\documentclass[12pt]{article}
\usepackage[utf8]{inputenc}
\usepackage{setspace}
\usepackage{geometry}
\usepackage[font=footnotesize,labelfont=bf,justification=justified,format=plain]{caption}
\usepackage{subcaption}
\usepackage{placeins}
\usepackage{graphicx}
\graphicspath{ {./graphs/} }
\usepackage{xcolor}
\usepackage{amsmath}
\usepackage{amssymb}
\usepackage{amsthm}
\usepackage{multirow}
\usepackage{booktabs}
\usepackage{tabularx}
\usepackage{threeparttable}
\usepackage{float}
\usepackage{makecell}
\usepackage[round]{natbib}
\usepackage[colorlinks=true,linkcolor=black,citecolor=blue,urlcolor=black]{hyperref}
\renewcommand{\floatpagefraction}{0.9}
\newcolumntype{L}{>{\raggedright\arraybackslash\hangindent=1em\hangafter=1}X}

\newtheorem{proposition}{Proposition}

\newcounter{appendixletter}
\renewcommand{\theappendixletter}{\Alph{appendixletter}}
\newcommand{\nextappendix}[1]{%
  \refstepcounter{appendixletter}%
  \setcounter{subsection}{0}%
  \renewcommand{\thesubsection}{\theappendixletter.\arabic{subsection}}%
  \renewcommand{\thefigure}{\theappendixletter.\arabic{figure}}%
  \renewcommand{\thetable}{\theappendixletter.\arabic{table}}%
  \setcounter{figure}{0}%
  \setcounter{table}{0}%
  \phantomsection
  \section*{\theappendixletter\quad #1}%
  \addcontentsline{toc}{section}{\protect\numberline{\theappendixletter}#1}%
  \suppressfloats[t]%
}

\title{\textbf{Floods and the Duplication of Production: \\ Evidence from the Global Car Industry}}
\author{Michele Fioretti${}^{\ast}$ \and Sophia Praetorius${}^{\dagger}$ \and Thierry Mayer${}^{\S}$}
\def\arxivid{}   %
\ifx\arxivid\empty \def\latestline{}\else \def\latestline{\\ \medskip {\small Latest version: \href{https://arxiv.org/abs/\arxivid}{arxiv.org/abs/\arxivid}}}\fi
\date{October 31, 2026 \\ \bigskip \textbf{Preliminary and Incomplete}\latestline}

\begin{document}
\maketitle
\thispagestyle{empty}

\begin{abstract}
\begin{spacing}{1.0}
Floods are becoming more frequent and can disrupt global supply chains. We study how multinational car makers adapt vehicle and key parts production to flood risk, and how their choices interact with insurance and public flood defences. Using plant-level data and a manually verified flood record in a staggered difference-in-differences design, we find that car makers duplicate production rather than move it away from flooded plants. Even more than five years after a flood, an engine is 19 percentage points more likely to be made in two or more plants, usually abroad and for cars assembled nearby, while the flooded plant's output shows no persistent decline beyond a temporary loss in the flood year. Firms duplicate only where adjustment costs are low, i.e. when a plant of theirs abroad already makes the same engine family. Flooded vehicle model versions are duplicated only temporarily, and transmission programs are not duplicated. Suggestive evidence that firms duplicate more where insurance is likely expensive and plants lack flood defences points to substitution between redundancy, insurance and public protection. \\[1em]
\end{spacing}
\end{abstract}

\vspace{1.5cm}
\begin{singlespace}
\rule{\textwidth}{0.05pt}\\
${}^{\ast}$ Bocconi and CEPR, \href{mailto:fioretti.m@unibocconi.it}{\color{blue}{fioretti.m@unibocconi.it}}\\
${}^{\dagger}$ University of Geneva, \href{mailto:sophia.praetorius@unige.ch}{\color{blue}{sophia.praetorius@unige.ch}}\\
${}^{\S}$ Sciences Po, CEPII, and CEPR, \href{mailto:thierry.mayer@sciencespo.fr}{\color{blue}{thierry.mayer@sciencespo.fr}}
\end{singlespace}

\clearpage
\setcounter{page}{1}
\onehalfspacing

\section{Introduction}
In October 2011, floods in Thailand inundated Honda's car plant in Ayutthaya and the industrial estates around it. Within a month, shortages of a small number of electronic components sourced from Thailand forced Honda to cut production by half at all six of its U.S. and Canadian vehicle plants.\footnote{Agence France-Presse, ``Thailand Floods Disrupt Honda's North America Production,'' \textit{IndustryWeek}, 1 November 2011. Honda Motor Co., ``About the Impact of Flooding in Thailand on Honda Operations (as of November 28, 2011),'' press release.} Due to climate change, floods like these are becoming more frequent and more intense \citep{IPCC2021}. How costly floods become depends not only on the damage they cause, directly and through supply chains, but also on how firms adapt and on what adaptation itself costs \citep{carleton2022,bilal2023}. In principle, a firm with plants in many countries can protect a site, insure it, move production away from it, or build a second source elsewhere in its network. Which of these adaptation margins firms use, and what limits them, determines how exposed global supply chains remain.

This paper asks how multinational firms adapt their production networks to flood risk, and how this private adaptation interacts with the insurance and flood protection that markets and governments provide. The interaction matters for policy. A firm that builds a second source weighs the loss of its own output, not the jobs and orders that a flood costs its workers and suppliers. Flood defences such as seawalls, by contrast, are built or subsidized by governments, and insurance depends on markets that governments regulate and that differ widely across countries in how competitive they are. Where firms build redundancy, and where it is too costly, therefore determines which risks private adaptation covers and which are left to public policy.

We study the global car industry, where cars, engines and transmissions differ greatly in how easily their production can be split across sites. We combine plant-level production data for vehicles, engines and transmissions in 2000--2024, including the location of every plant and the flows of components to assembly plants, with a record of plant floods that we collected by hand from news reports and company statements. We do not rely on public disaster databases, which are prone to error at the plant level \citep{patel2024}. We follow engine programs, transmission programs and car model generations, and compare each flooded product with products not yet flooded that were launched in the same years, were made in the same number of plants the year before the flood, and have their largest plant in another country, far enough away not to be affected by the flood. For the policy questions, we add each plant's flood hazard net of public protection, the concentration of its country's insurance market, and nearby flood defences.

The results provide new insights into the ways and costs of climate change adaptation, and consequently, the resilience of supply chains. First, average effects of floods on output can understate how firms adjust, since the main response we find is a change in the structure of production that leaves output at the flooded plant without a persistent decline, beyond a temporary loss in the year of the flood. Second, adaptation is uneven. Firms build redundancy where the adjustment costs of capital in their existing network are lowest, which means that the parts of supply chains with the largest economies of scale remain concentrated and exposed. Third, private redundancy and public protection may substitute for each other. If the suggestive evidence below holds, flood defences built by governments partly replace the second plants that firms would otherwise build, and the exposure that remains lies where plants are neither protected nor cheap to duplicate.

Our main finding is that firms respond to a flood by duplicating production rather than by moving it. Five to ten years after a documented flood, a flooded engine program is 19 percentage points more likely to be made in two or more plants than its controls. About half of the flooded programs were made in a single plant before the flood, so the effect is large. The original plants retain the programs, and their output shows no persistent decline. Flooded engine plants lose about 11\% of their output in the year of the flood, transmission and vehicle assembly plants less, and all return to their pre-flood paths the following year. 

The duplicated production is added at plants in other countries and serves markets close by. The added plants are small, and their volume goes to new model versions and new vehicle assembly plants, distinct from the original plants' customers. The floods we observe are short and cost the flooded programs little output in annual data. The flood changes where the program's additional volume is made, so that less of it depends on one site, and the firm gains a running line for the program elsewhere, an option that would pay off in a long disruption. 

Our second finding is that duplication occurs only when capital adjustment costs are sufficiently low. The response is driven entirely by engine programs whose owner already makes the same engine family at a plant in another country. For these, the probability of being made in two or more plants rises by 36 percentage points, with no effect for the others. This is in line with \citet{praetorius2026}, who shows that platforms, the production technologies that products are built on, are costly to deploy at a plant. Unless the platform already runs at a plant, starting a program there entails high capital adjustment costs. Priced with hand-collected public announcements of engine investments (Section~\ref{sec:price} and Appendix~\ref{app:defs}), a second plant after a flood costs \$26.8 million on average, about \$500 per engine of annual volume. At the frequency of documented floods, it pays off only as insurance against disruptions that last months, or if local supply is worth \$75 per engine a year (Table~\ref{tab:price}). Both thresholds are lower bounds, since we assume no extra unit cost at the second plant. 

Vehicle model versions and transmission programs are flooded too (421 versions and 121 transmission programs, against 216 engine programs), but they are not permanently duplicated. A flooded vehicle model version is made in a second plant three to four years after the flood, but this plant serves new markets rather than those of the flooded plant. This duplication vanishes by year five, and the version's output in the original plant remains unchanged. A plausible reason is that cars cost more to ship than engines, so that a car made in a second location needs sufficient local demand to be profitable. Yet the countries where the firm already builds the version's platform and the cost of duplication hence are low enough, account for almost none of the version's sales, suggesting low demand for these models in those locations. The flooded car version, however, expands more slowly than its controls into new export markets, and this shortfall persists. 
Automatic transmissions are held back from duplication by scale instead: their assembly lines are typically more than twice the size of an engine family line (205 against 91 thousand units a year), too large for the cars assembled near most plants.

Our third finding concerns the protection that markets and governments provide. Documented floods are too few to split by country, so for these tests we date the flood by the first flood signal at a plant, i.e., a flood seen by satellite within 5 km or a documented direct hit. We combine it with the flood defences mapped in OpenStreetMap close to the plants, such as seawalls, dykes and flood walls, and with the protection standards of FLOPROS \citep{scussolini2016}. Programs flooded at a plant with a flood defence mapped within 2 km add no second plant in the first six years after the flood, while programs flooded at unprotected plants duplicate their production in additional plants steadily, and the gap between the two reaches 12 percentage points in year 5. Similarly, programs flooded at high-hazard plants\footnote{We measure flood hazard as the residual probability of a flood at the plant, net of public flood protection. It combines the inundation maps of WRI Aqueduct Floods \citep{aqueduct2020}, which give the return periods at which a plant's location is flooded without defences, with the river flood protection standards of FLOPROS \citep{scussolini2016}. The residual probability is one over the larger of the shortest flooding return period and the protection standard, and we call a plant high-hazard when this probability is 1\% a year or more. Section~\ref{sec:otherdata} gives the details.} add a second plant more often where the insurance market is concentrated, so that insurance is likely to be expensive, than where it is competitive. Both patterns indicate a trade-off between private redundancy, market insurance and public protection, but they are only suggestive. The gap between protected and unprotected plants shrinks or reverses when defences farther from the plant are counted, and the insurance test measures the price of insurance only by how concentrated the national market is, with concentrated markets in five countries. 

Finally, three facts point to adaptation to flood risk, not to a demand-driven expansion or a repair. Before the flood, flooded and control programs add plants at the same rate and supply the same number of assembly countries. The effect is larger in years 8--10 than in years 2--4, instead of fading as control programs add plants, so the flood does not merely move planned expansions forward. The flooded plant recovers its output within a year, and a supply-chain stoppage, in which the plant stayed dry but a flood cut off its suppliers, destroys no capacity yet triggers about the same response as a direct hit (17.7 against 21.4 percentage points in years 5--10). The flood thus works through what the firm learns about the risk of producing in one place, and the response shows how production networks adapt as floods become more frequent.

We organize these findings with a simple model of the duplication decision. A firm that makes a component in one plant faces a risk of disruption, which it learns about from floods. It can also make the component at another of its plants, at a fixed cost that depends on the adjustment costs of capital, provided that demand nearby covers the minimum efficient scale of production there. A flood raises the perceived risk, and thus the value of redundancy, but redundancy is only built where its cost is low enough.

Our paper relates to four strands of the literature. First, a large literature exploits natural disasters and extreme weather to estimate the economic costs of climate change, with events ranging from temperature and precipitation shocks \citep{deschenes2007economic, dell2012temperature, burke2015global, burgess2017weather, somanathan2021impact} to hurricanes and cyclones \citep{hsiang2014climate, boustan2012moving, pelli2023costs} and floods \citep{leiter2009, guiteras2015satellites, desmet2018, kocornik2020flooded}. Micro-level evidence on how floods affect firms is hard to reconcile. Early studies find positive effects on employment and assets after floods in Europe, consistent with `creative destruction' \citep{leiter2009, noth2019}. More recent work finds large and persistent losses in output, capital and employment \citep{hossain2020}, or little effect on activity and location \citep{kocornik2020flooded, jia2022}. Part of this disagreement may come from measurement, since the public disaster databases used to assign floods to firms mostly capture proximity to large floods rather than floods at the firm (Section~\ref{sec:floods}). We therefore use a hand-verified record of plant floods, and we shift the focus from the average effect of a flood to how firms reorganize their production networks after it.

A second strand studies how disasters propagate through production networks. Disruptions at suppliers lower the output of their customers, especially when inputs are specific \citep{barrot2016, boehm2019, carvalho2021}, and the propagation of supply shocks in global value chains has received renewed attention since the pandemic \citep{lafrogne2022, baldwin2022}. Closest to us, several papers study how firms respond to climate risk in their supply chains. \citet{pankratz2023} show that heat and flood shocks at supplier locations lower the performance of suppliers and their customers, and that buyers terminate relationships when shocks exceed expectations. \citet{castro2024} show that Indian firms diversify their sourcing across suppliers in different regions to limit climate risk, at the cost of higher input prices, while \citet{balboni2023} find that flooded firms in Pakistan relocate and diversify their supplier base away from flood-prone locations. \citet{CV2024} studies the car industry and shows that firms shift production across their plants after floods and, in a quantitative model, that higher flood risk leads firms to build more and smaller plants. We provide direct evidence of a related margin, the duplication of production at plants the firm already runs, and show that its fixed cost, rather than the flood itself, determines where it occurs.

Third, a growing theoretical literature studies how firms and governments should balance efficiency and resilience in supply chains. \citet{GHL2023} ask whether policy should promote international diversification or reshoring, \citet{grossman2024} study optimal resilience in multi-tier supply chains, and \citet{EGL2022} show how the fragility of supply networks depends on the relationships firms choose to build. \citet{kopytov2024} model how uncertainty shapes the formation of production networks. In operations research, \citet{tomlin2006} compares mitigation through dual sourcing and inventories with contingency plans. These papers typically take the cost of an additional source as given. We show empirically that this cost varies greatly across products and plants with the firm's existing network, and that it governs where redundancy is built.

Finally, our paper relates to work on the location of multinational production, which emphasizes the trade-off between proximity to customers and the concentration of production to exploit economies of scale \citep{brainard1997, HMY2004}, as well as the role of export platforms \citep{tintelnot2017}, global sourcing \citep{AFT17, antras2024} and frictions in multinational production \citep{ARRY18, HM19}. The car industry has been a frequent laboratory for these questions \citep{HM19, HMMY2024}. \citet{gu2023} show that foreign direct investment responds to physical climate risk. We show that flood risk shifts the proximity-concentration trade-off at the margin, but only where a second plant fits within the firm's existing network and within local demand.

The rest of the paper is structured as follows. Section~\ref{sec:data} describes the setting and the data, and Section~\ref{sec:facts} presents stylised facts. Section~\ref{sec:model} develops the conceptual framework. Section~\ref{sec:design} explains the empirical strategy, and Section~\ref{sec:results} presents the results. Section~\ref{sec:concl} concludes.

\section{Setting and Data} \label{sec:data}
\subsection{Production Data} \label{sec:prod}
Our plant-level production data for the global automotive industry comes from S\&P Global Mobility (formerly IHS Markit), used among others by \citet{HM19}, \citet{CV2024}, \citet{HMMY2024}, and \citet{praetorius2026}. Our main source is S\&P's engine and transmission installation data (release of April 2025), which report, for 2000--2024, the annual production of each light vehicle model at each assembly plant, together with the plant that makes its engine and the plant that makes its transmission. The data include the manufacturer, platform, and body type of each model, its engine and transmission programs, and the location of each plant. In the sample, there are 1,196 vehicle assembly plants in 55 countries, 443 engine plants and 350 transmission plants. In a typical year, the data cover about 1,200 car models of about 100 brand owners, assembled in about 680 vehicle plants in 49 countries (Table~\ref{tab:sum_all} in Appendix~\ref{app:data}). We geocode every plant, which allows us to trace supply links between plants, and to measure distances between plants and to flood centroids.

\paragraph{{Products.}} We follow three kinds of products in automotive manufacturing. An \textit{engine program} is one engine design by an engine firm, and a \textit{transmission program} is defined in the same way. A \textit{vehicle model} is a nameplate of a brand owner, e.g., Ford Focus, and a \textit{model version} is one generation of it, i.e., the nameplate together with S\&P's vehicle program, e.g., Ford Focus C346. Our vehicle product is the model version. A new generation is engineered and tooled anew, often at other plants, so that a nameplate whose new generation starts at a second plant while the old one runs out at the first would look duplicated without being so. Below, ``vehicle model'' refers to the model version unless stated otherwise. For all three narrowly defined products, S\&P also reports the \textit{product family}, i.e., the platform of each program, which defines the engineering rules of the model architecture. Programs of the same family share most of their tooling and processes. We consider only models with at least 100 units of output per year. To describe sourcing, we also use the \textit{location}, i.e., a model version at an assembly plant, and, following \citet{HMMY2024}, the \textit{vehicle specification}, i.e., a vehicle together with the specification of its engine or transmission. Appendix~\ref{app:defs} gives the precise definition of every unit, threshold, and measure used in the paper.

\subsection{Documented Floods at Plants} \label{sec:floods}
Researchers interested in the economic effects of floods mostly rely on two public databases, the Dartmouth Flood Observatory archive \citep{DFO} and EM-DAT \citep{EMDAT}, and increasingly on satellite flood maps \citep{tellman2021}. These sources record large floods and the regions they affect, but not whether a given plant was flooded. Comparing them with the documented floods described below, we find that at the plant level they perform poorly. About 99\% of the plant-years they flag as flooded have no documented flood impact, while even the most inclusive rule misses a third of the floods documented at plants, and satellite maps recover almost none of them. 

\begin{figure}[t]
\centering
\caption{Plants with a Documented Flood, 2011} \label{fig:ra_map}
\includegraphics[width=0.9\textwidth]{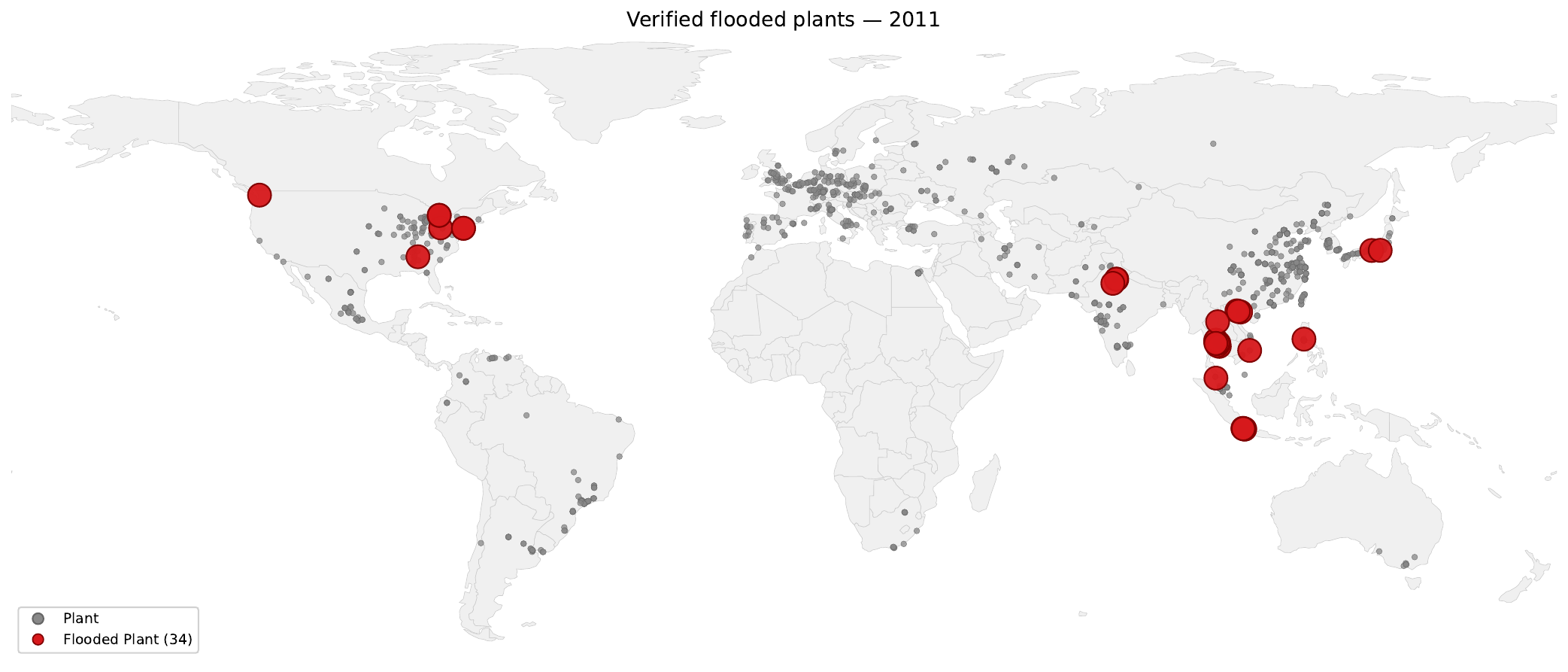}
\caption*{\textit{Notes:} Automotive plants with a documented flood in 2011, direct hits and stoppages. Documented floods are hand-collected from news reports and company statements.}
\end{figure}

We therefore use a record of floods at automotive plants that we collected by hand from local and international news, company press releases, and filings. For each case, the record gives the year and, where reported, the month, and it distinguishes two kinds of events. In a \textit{direct hit}, the plant closed because of a flood or storm at the plant. This does not require water in the plant: the plant also counts as hit when flooding around it cut its access and deliveries or when it closed as a precaution. In a \textit{stoppage}, production stopped because a flood elsewhere cut off the plant’s supplies while the plant itself stayed dry, as for Honda’s plants in the United States during the 2011 Thai floods. The record contains 195 flood plant-years, of which 112 are direct hits and 83 stoppages, and 186 fall in 2000--2024. It is concentrated in time, with 51 plant-years in 2011, the year of the Thai floods. Figure~\ref{fig:ra_map} shows the plants with a documented flood in 2011. We refer to both kinds of events as \textit{documented floods}. Appendix~\ref{app:strict} restricts direct hits to floods that submerged a plant. Conversely, absence from the record does not prove that a plant was unaffected, since small disruptions may go unreported.
A product is considered treated in the year of the first documented flood at a plant that made it in the year before. This gives us 216 flooded engine programs of 14 owners, 121 flooded transmission programs of 15 owners, and 421 flooded vehicle model versions of 28 owners. For 21 of the engine programs, other public reports describe the first event as something other than a flood at the plant, mostly the earthquake of March 2011. We keep the record as collected, and Appendix~\ref{app:strict} rechecks it and drops such events, with a similar main estimate.
\subsection{Other Data} \label{sec:otherdata}
For the extensions in Section~\ref{sec:ext}, we measure the residual flood hazard at each plant, i.e., the probability of a flood net of public flood protection, and we classify a plant as high-hazard when this probability is 1\% a year or more. Flood hazard comes from the WRI Aqueduct Floods maps \citep{aqueduct2020}, which give the inundation depth of an undefended flood for return periods of 2 to 1,000 years on a grid of about 1 km, and protection from the river flood protection standards of FLOPROS \citep{scussolini2016}. The residual riverine probability is $1/\max(rp_{\min},S)$, where $rp_{\min}$ is the shortest return period at which the plant's cell is flooded and $S$ is the protection standard. Coastal floods are not netted, as FLOPROS has no coastal layer. We measure how expensive insurance is likely to be with the concentration of the country's non-life insurance market, the market share of the five largest insurers (CR5) in non-life gross written premiums. No public cross-country dataset exists, so we collected CR5 from insurance supervisors, insurance associations, IMF and World Bank financial sector assessments, and EIOPA, for the 26 countries with flooded plants over 2003--2020. It is computed on a company basis except for the United States and Canada (group basis), and we take the value for the year nearest $F-1$. Finally, we use OpenStreetMap to check whether a dyke, levee, flood wall, or seawall is mapped within 2 km of a plant (within 5 km in Appendix~\ref{app:def}). We also use S\&P's sales data by destination country, and public announcements of investments in engine capacity, which we collected by hand (Appendix~\ref{app:defs}).

\section{Stylised Facts} \label{sec:facts}
This section describes the setting in which firms decide whether to respond to flood risk by adding a plant. We ask how exposed a product is to a flood at one of its plants, what a flood does to the firm, and what a second plant would need in order to be built. Table~\ref{tab:org} describes how products, plants and markets are linked, with measures that follow \citet{HMM2022}, who use the same S\&P data to study the sourcing of car parts. Table~\ref{tab:floods} turns to the flooded products. Appendix~\ref{app:defs} defines every unit and measure, and Appendix~\ref{app:sum} reports the full distributions by world region.

\begin{table}[t]
\centering
\footnotesize
\caption{Plants, Products and Sources} \label{tab:org}
\setlength{\tabcolsep}{8pt}
\renewcommand{\arraystretch}{1.12}
\begin{tabularx}{\textwidth}{@{}Lccc@{}}
\toprule
 & Vehicles & Engines & Transmissions\\
\midrule
\multicolumn{4}{@{}l}{\textit{A. Plants and products, 2000--2024}}\\
\quad Products per plant, mean & 3.19 & 5.40 & 4.43\\
\quad Plants per product, mean & 1.51 & 1.38 & 1.37\\
\quad Plants per product, output-weighted mean & 2.71 & 2.91 & 2.38\\
\quad Product-years made in one plant (\%) & 76.1 & 79.0 & 78.2\\
\quad Product-years made in one country (\%) & 79.1 & 82.3 & 84.5\\
\addlinespace[4pt]
\multicolumn{4}{@{}l}{\textit{B. Links between plants, 2000--2024 (per component plant unless stated)}}\\
\quad Assembly plants supplied, mean &  & 5.66 & 8.68\\
\quad Carmaker groups supplied, mean &  & 2.58 & 3.58\\
\quad Component plants supplying an assembly plant, mean & 5.25 & 2.55 & 3.01\\
\addlinespace[4pt]
\multicolumn{4}{@{}l}{\textit{C. Supplied by two or more plants (\%), 2000--2024}}\\
\quad Vehicles, share of vehicle-years & 15.0 & 39.0 & 52.8\\
\quad Vehicles, share of units & 48.0 & 54.9 & 68.6\\
\quad Specifications, share of specification-years & 11.0 & 0.7 & 2.2\\
\quad Specifications, share of units & 25.6 & 1.8 & 4.6\\
\quad Locations, share of location-years &  & 41.3 & 56.2\\
\quad Locations, the same program, share of location-years &  & 4.8 & 4.0\\
\addlinespace[4pt]
\multicolumn{4}{@{}l}{\textit{D. Sources of a model version in a market, 2000--2023}}\\
\quad Plants supplying the version in the market, mean & 1.07 & 2.17 & 2.45\\
\quad Supplied by two or more plants (\%) & 6.1 & 67.5 & 74.4\\
\quad Markets a plant sells in, mean & 11.1 & 16.8 & 21.0\\
\quad Markets a plant sells in, median & 3 & 6 & 11\\
\addlinespace[4pt]
\multicolumn{4}{@{}l}{\textit{E. Documented floods, 2000--2024}}\\
\quad Plants & 1,196 & 443 & 350\\
\quad Plants with a documented flood & 132 & 47 & 30\\
\quad World car output, plant flooded that year (\%) & 1.08 & 0.78 & 0.58\\
\quad Flooded vehicle specifications & 2,244 & 1,041 & 754\\
\quad Flooded specifications from one plant in $t-1$ (\%) & 80.5 & 97.3 & 97.9\\
\quad Flooded specifications from two or more plants in $t+3$ (\%) & 21.0 & 0.8 & 2.9\\
\quad Never-flooded specifications from two or more plants (\%) & 10.4 & 0.7 & 2.2\\
\bottomrule
\end{tabularx}

\caption*{\textit{Notes:} Columns refer to the kind of plant (assembly, engine or transmission) and of product (model version, engine program or transmission program). For vehicles, the plants in panels C and E are assembly plants. A plant makes a product, and supplies another plant, with 100 units or more in the year, and a vehicle or vehicle specification counts every plant supplying any units. Empty cells: not defined, as an assembly plant supplies no other plant and belongs to one carmaker group, and for a car the market (D) takes the place of a location. Appendix~\ref{app:defs} defines every unit, threshold and measure, and Appendix~\ref{app:sum} gives the tables by region. Own calculations based on S\&P Global Mobility data. }
\end{table}

\begin{figure}[t]
\centering
\includegraphics[width=\textwidth]{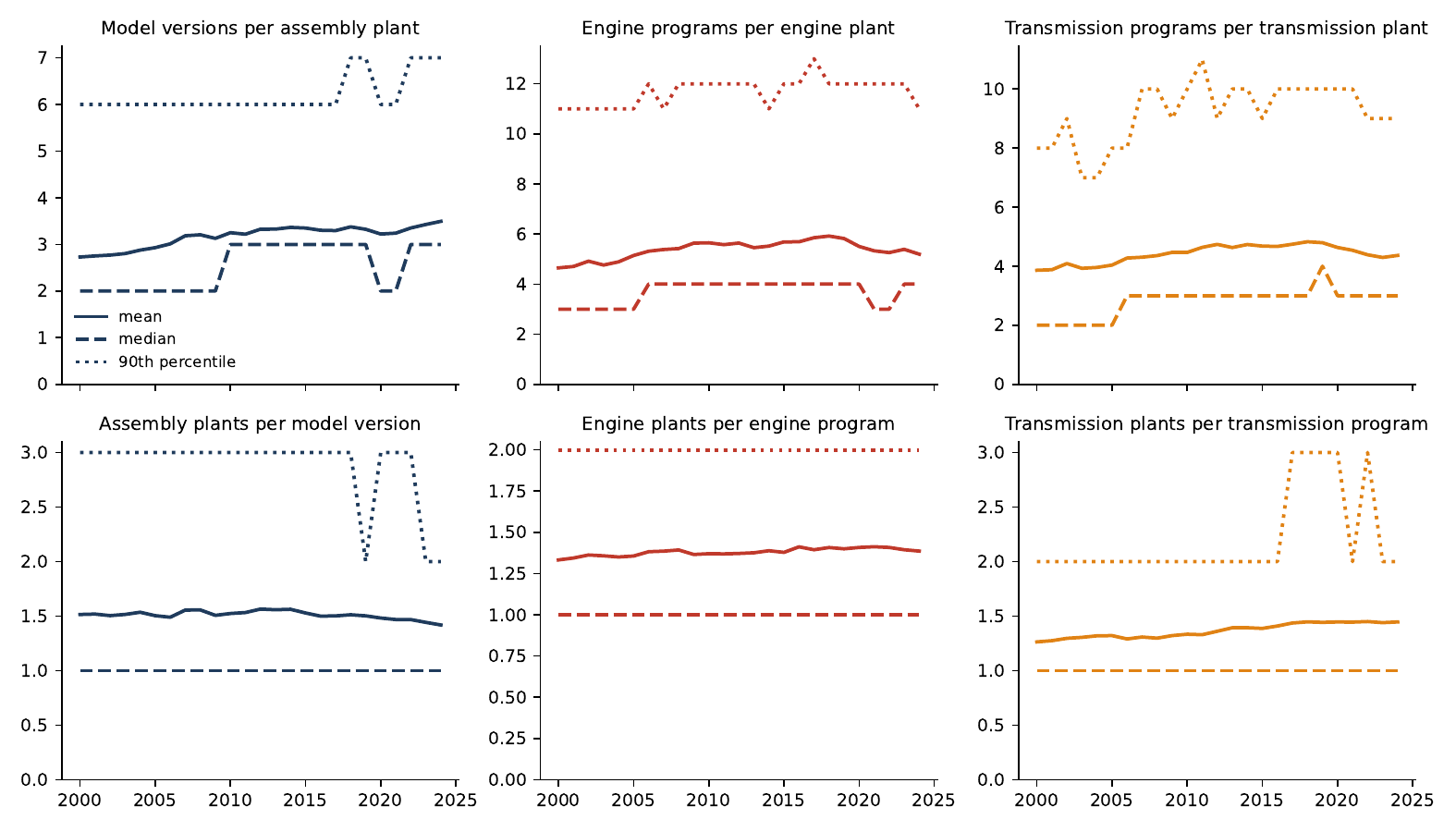}
\caption{Products per Plant and Plants per Product, 2000--2024} \label{fig:trend}
\caption*{\textit{Notes:} Mean (solid), median (dashed) and 90th percentile (dotted) by year. Top: model versions per assembly plant, engine programs per engine plant and transmission programs per transmission plant, over plant-years. Bottom: plants per model version, engine program and transmission program, over product-years. A plant makes a product with 100 units or more in the year. Percentiles are the lower of the two neighbouring values, i.e., actual counts. By region in Figure~\ref{fig:regions}. Own calculations based on S\&P Global Mobility data. }
\end{figure}

\begin{figure}[!htp]
\centering
\includegraphics[width=\textwidth]{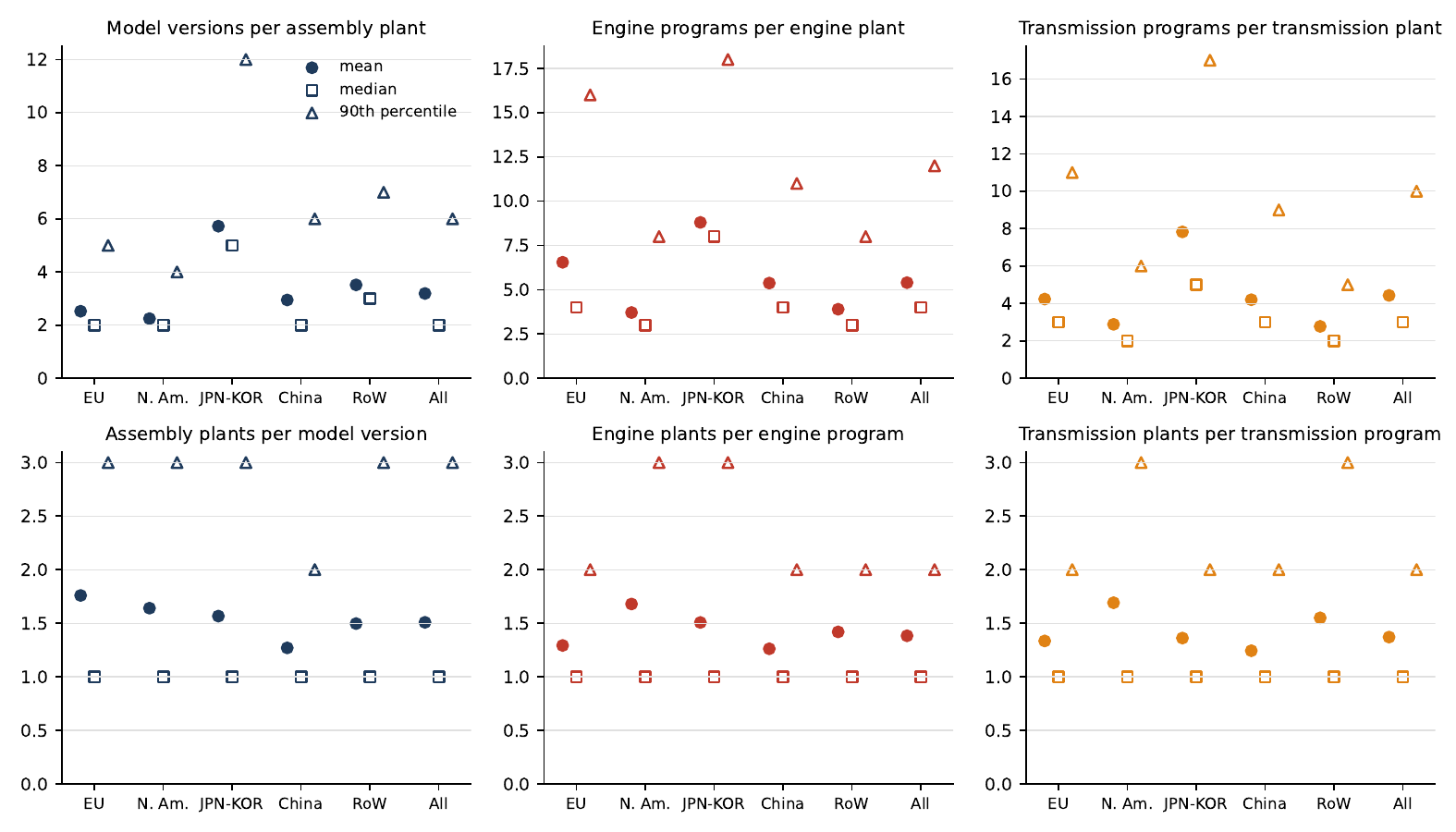}
\caption{Products per Plant and Plants per Product, by Region, 2000--2024} \label{fig:regions}
\caption*{\textit{Notes:} Mean (filled circle), median (square) and 90th percentile (triangle), pooling 2000--2024. Top: model versions per assembly plant, engine programs per engine plant and transmission programs per transmission plant, over plant-years, by region of the plant. Bottom: plants per model version, engine program and transmission program, over product-years, by region of the product's largest plant. A plant makes a product with 100 units or more in the year. Percentiles are actual counts. Numbers in Tables~\ref{tab:plants_v}--\ref{tab:plants_t}.}
\end{figure}

\paragraph{{Production of each product is concentrated in a single plant.}} The typical product is made in a single plant. This holds for 79\% of engine program-years, 78\% of transmission program-years and 76\% of vehicle model-version-years, and the median is one plant in every year and every region (Figures~\ref{fig:trend} and~\ref{fig:regions}). Pooling the years, even the 90th percentile is only two plants for engine and transmission programs and three for model versions. The products made in two or more countries are the larger ones and account for about half of output (Table~\ref{tab:floods}). Concentration has not fallen over time. The mean number of plants per engine program rose only from 1.33 to 1.39 between 2000 and 2024, and model versions became more concentrated, from 1.51 plants to 1.42. Multiple sourcing does not soften this exposure. Although 39\% of vehicles get their engines from two or more plants, this reflects engine variants, as \citet{HMM2022} show. Once the vehicle is defined together with the specification of its engine, only 0.7\% of vehicle specifications get it from two or more plants (2.2\% for transmissions), and only 4.8\% of the locations get the same engine program from two or more plants. A flood at the plant that makes an engine thus stops every car fitted with it, and this is the exposure that a second plant could reduce.

\paragraph{{Component plants are shared hubs close to assembly.}} Each component plant, in contrast, makes many products. The median engine plant makes 4 engine programs and the 90th percentile 12, against 2 and 6 vehicle models at assembly plants, and in Japan and Korea the median engine plant makes 8 programs. An engine plant supplies 5.7 assembly plants on average and a transmission plant 8.7, and most of what they make is fitted nearby. In China, Japan, Korea and India, more than 93\% of the engines fitted in cars are made in the country itself, and the engine and the transmission both come from the region of assembly for 83\% of the cars assembled in the EU and 70\% of those assembled in North America (Table~\ref{tab:origin_e} and Figure~\ref{fig:config} in Appendix~\ref{app:sum}). Two features of a cheap second plant follow. A plant that already makes programs of the same family has the tooling and the workers to take on one more, which keeps the fixed cost $F_B$ of the framework low. The engines made at a second plant can also be fitted in vehicles assembled nearby, which gives the second plant the local demand $D_B$ that keeps it busy in normal times.

\paragraph{{Vehicles are sold close to where they are assembled.}} Vehicle models differ in this respect. In 94\% of the version-market-years, a model version is sold in a country from a single assembly plant, and about three quarters of an assembly plant's sales are in the market where it stands, counting the EU as one market (77\% on average across plants, 74\% of all sales). An assembly plant sells in 11 countries on average but in only 3 at the median. A second plant for a car model would therefore need buyers of that model in its own market, whereas a second plant for an engine program only needs cars assembled nearby.

\begin{table}[t]
\centering
\footnotesize
\caption{Flooded Models} \label{tab:floods}
\setlength{\tabcolsep}{5pt}
\renewcommand{\arraystretch}{1.12}
\begin{tabularx}{\textwidth}{@{}Lccccc@{}}
\toprule
 & Vehicles & Engines & \multicolumn{3}{c}{Transmissions}\\
\cmidrule(l){4-6}
 & & & All & Manual & Automatic\\
\midrule
\multicolumn{6}{@{}l}{\textit{A. All models, 2000--2024}}\\
\quad Made in two or more countries, share of model-years & 0.21 & 0.18 & 0.15 &  & \\
\quad Made in two or more countries, share of output & 0.48 & 0.55 & 0.49 &  & \\
\addlinespace[4pt]
\multicolumn{6}{@{}l}{\textit{B. Exposure: flooded models, the year before the flood}}\\
\quad Made in one plant, share & 0.58 & 0.48 & 0.53 &  & \\
\quad Largest plant's share of output & 0.86 & 0.80 & 0.83 &  & \\
\addlinespace[4pt]
\multicolumn{6}{@{}l}{\textit{C. The flood year}}\\
\quad Output relative to controls & $-0.009$ & $+0.003$ & $-0.036$ &  & \\
 & (0.036) & (0.049) & (0.049) &  & \\
\addlinespace[4pt]
\multicolumn{6}{@{}l}{\textit{D. Made in one plant, share made in two or more plants five years later}}\\
\quad Flooded models & 0.115 & 0.128 & 0.038 &  & \\
\quad Never-flooded models & 0.025 & 0.023 & 0.035 &  & \\
\addlinespace[4pt]
\multicolumn{6}{@{}l}{\textit{E. A place for a second plant: flooded models made in one plant}}\\
\quad A plant abroad makes the family, share & 0.57 & 0.60 & 0.28 & 0.50 & 0.03\\
\addlinespace[4pt]
\multicolumn{6}{@{}l}{\textit{F. Scale and shipping: all families}}\\
\quad Family line, median peak output (thousands) &  & 91 & 114 & 77 & 205\\
\quad Elasticity of shipments to distance &  & $-1.10$ & $-1.04$ & $-0.99$ & $-1.12$\\
 &  & (0.09) & (0.08) & (0.09) & (0.16)\\
\bottomrule
\end{tabularx}

\caption*{\textit{Notes:} Manual: manual gearboxes. Automatic: automatic, CVT and dual-clutch transmissions. All models: model-years 2000--2024. Exposure: flooded models, the year before the first documented flood. The flood year: output at all plants over output in the year before, flooded models minus controls (Section~\ref{sec:design}), owner-bootstrap standard errors in parentheses. Made in one plant: models made in one plant in a base year, share made in two or more plants five years later, never-flooded models evaluated in the base years of the flooded models. A place for a second plant: family for engines and transmissions, platform of the same group for model versions. Scale and shipping: peak yearly output of a family at a plant, all family lines. Elasticity: Poisson regression of the units a plant sends to an assembly plant on their log distance, among families made in two or more plants, with fixed effects for assembly plant, family and year and for supplying plant, family and year. Empty cells: not available by kind of transmission, or not defined for vehicle models. }
\end{table}

\paragraph{{Flooded models depend on the flooded plant.}} Floods are rare at the plant level. Over 2000--2024, 132 of the 1,196 assembly plants, 47 of the 443 engine plants and 30 of the 350 transmission plants have a documented flood. Cars assembled at a plant in a year of a documented flood there make up 1.1\% of the world's cars over the period, and cars fitted with an engine or a transmission from such a plant 0.8\% and 0.6\% (Table~\ref{tab:org}). When a flood occurs, however, the model at stake typically depends on the flooded plant. About half of the flooded models were made in that plant alone in the year before, and the largest plant made over 80\% of their output on average (Table~\ref{tab:floods}).

\paragraph{{A flood reveals exposure rather than destroying output.}} In the flood year, the output of flooded models does not fall relative to their controls, for any of the three models. The flooded plant as a whole loses output only in the flood year, by about a tenth for engine plants, and returns to its pre-flood path in the following year (Section~\ref{sec:engines}). The two estimates use different units, the model's output at all its plants and the plant's output of all its models (including models flooded before at other plants), and both are imprecise, with overlapping intervals. Many documented floods stop a plant for days or weeks, and these stoppages are small in annual data or are made up later in the year. A flood thus matters mainly through what it tells the firm about the risk. In the framework of Section~\ref{sec:model}, it raises the perceived probability of a disruption $\hat\pi$, and with it the value of a second plant.

\paragraph{{After a flood, engine programs are duplicated, but cars do not obtain a second engine source.}} Among engine programs made in a single plant, 12.8\% of the flooded programs are made in two or more plants five years later, against 2.3\% of never-flooded programs in the same years. For transmission programs, the two shares are the same, at 3.8\% and 3.5\%. For model versions, the raw comparison also shows a gap, 11.5\% against 2.5\%, but Section~\ref{sec:products} shows that, against similar versions, it is a second plant for a few years only, gone by year five. The second engine plant does not supply the vehicles that the flooded plant supplied. Of the 1,041 vehicle specifications whose engine plant has its first documented flood, 97.3\% got the engine only from that plant the year before, and three years later 0.8\% get it from two or more plants, close to the 0.7\% of specifications never supplied by a flooded plant (Table~\ref{tab:org}). The added plant thus supplies other cars, a point Section~\ref{sec:receiving} develops. In principle, it reduces the exposure of the program as a whole, since a future flood at one plant would then affect only part of its output.

\paragraph{{A second plant requires existing capacity and local demand.}} The products differ in how cheaply a second plant can be added. For 60\% of the flooded engine programs made in a single plant, a plant of the same owner in another country already makes the same engine family, which is the low fixed cost of the hubs described above. For transmission programs, the share is only 28\%, and the gap is entirely due to automatic transmissions (3\%), while manual gearboxes look like engines (50\%). For model versions, suitable capacity exists, as a plant of the group abroad builds the same platform for 57\% of the flooded versions made in one plant, but local demand is lacking, since vehicles are sold close to where they are assembled. Automatic transmissions lack suitable capacity for a different reason, namely scale. Their family lines are more than twice the size of an engine family line, while their shipments are as sensitive to distance (Table~\ref{tab:floods}), so that a second line is too large for the cars assembled near most plants, i.e., its minimum scale $m$ exceeds local demand $D_B$ (Section~\ref{sec:products}).

In this setting, each product is concentrated in a single plant, a flood reveals this exposure without causing large output losses, and a second plant is inexpensive only where the firm already makes the family at another plant and vehicle assembly near that plant can absorb the output of a family line. The next section formalizes the decision, and the rest of the paper tests it.

\section{Conceptual Framework} \label{sec:model}
We develop a simple model of the decision to duplicate production. Its purpose is to organize the empirical analysis and to derive testable predictions, not to quantify the value of redundancy.

\paragraph{{Setup.}} A firm makes a component, e.g., an engine program, at a single plant $A$ and fits it in cars assembled downstream. Each year, plant $A$ is disrupted with probability $\pi$, for instance by a flood at the plant or at its suppliers. A disruption stops the plant for part of the year, and every unit of the component that is not delivered costs the firm a margin $\lambda$ on the cars that cannot be assembled. If the program's yearly volume is $Q$, the expected loss is $\pi \lambda Q$ per year. The firm can also make the component at another of its plants, $B$, located in another country. Production at $B$ costs a fixed amount $F_B$ per year, i.e., the annualized cost of plant, tooling, qualification of suppliers and training. Its output $q_B$ is fitted in cars assembled near $B$, which would otherwise use another component or the output of $A$. Producing there costs $\Delta c$ more per unit than at $A$, and production at $B$ is only viable at a minimum efficient scale $m$. The output of $B$ is thus bounded by the local demand $D_B$ for components of this kind, and production at $B$ is feasible only if $D_B \geq m$.

With two sites, a disruption at $A$ stops only the part of the volume that $A$ makes. For simplicity, we assume that the two plants are not disrupted at the same time and that $A$ keeps its volume $Q$, while $B$ serves the additional local demand $q_B$. The expected loss avoided by the second plant is then proportional to the volume that $B$ can deliver in a disruption, which rises with $q_B$. We write it as $\pi \lambda \phi q_B$, where $\phi \in (0,1]$ captures how much of the volume that $B$ delivers would otherwise be exposed to a disruption at $A$, either because $B$ can redirect output to $A$'s customers or because $B$ makes volume that $A$ would otherwise have made. The firm adds plant $B$ if
\begin{equation}
\hat\pi \, \lambda \phi \, q_B \;\geq\; F_B + \Delta c \, q_B, \qquad m \leq q_B \leq D_B, \label{eq:dup}
\end{equation}
where $\hat\pi$ is the firm's belief about the probability of a disruption.

\paragraph{{Learning from floods.}} The firm does not know $\pi$ and learns about it from experience. With a Beta prior with parameters $(a,b)$, after $T$ years with $k$ disruptions, the belief is $\hat\pi = (a+k)/(a+b+T)$. A flood raises the belief by $(1-\hat\pi)/(a+b+T+1)$, and by $1/(a+b+T+1)$ relative to a year without one. Since the flood reveals information about the risk of the site, the update does not depend on whether the plant was physically damaged. A stoppage caused by a flood that cut off the plant's suppliers reveals a risk to the program in the same way as a direct hit.

\paragraph{{The cost of a second plant.}} The firm holds a portfolio of technologies, here engine families, each with its own capacity at the plants that run it, as in models of firms with technology portfolios \citep{FHT2024}. The fixed cost $F_B$ depends on what plant $B$ already makes. If $B$ already makes the same family, the component can share its tooling, processes, suppliers and workers, and $F_B$ is low. Deploying a platform at a plant that does not yet run it is costly \citep{praetorius2026}, so that $F_B$ is higher if $B$ makes engines of another family, and highest if it makes no engines at all. The local demand $D_B$ depends on the product. For an engine, $D_B$ is the volume of engines of this kind fitted in the cars assembled near $B$, which can switch to the program's engine. For a vehicle model version, $D_B$ is the version's sales in $B$'s country, since a car assembled at $B$ has to be sold somewhere and shipping it back defeats the purpose of local production. For an automatic transmission, the minimum efficient scale $m$ is large relative to the transmissions fitted in any one country.

\begin{proposition} \label{prop:dup}
Suppose that, before a flood, condition~\eqref{eq:dup} does not hold for any plant $B$. Then:
\begin{enumerate}
\item[(i)] A flood raises $\hat\pi$ and thus the probability that the product is duplicated in the following years.
\item[(ii)] The response does not depend on whether the flood damaged the plant, as long as it reveals a risk to the product's supply.
\item[(iii)] Production is added rather than moved, i.e., plant $A$ keeps its output, and the second plant serves local demand at $B$.
\item[(iv)] The response is concentrated among products for which a plant $B$ with a low fixed cost $F_B$ exists, i.e., a plant that already makes the same family, and among plants $B$ with high local demand $D_B$.
\item[(v)] Products for which $D_B < m$ at every plant, because local demand is small (vehicle models) or because the minimum efficient scale is large (automatic transmissions), do not respond.
\item[(vi)] Other ways of managing the risk that lower the loss borne by the firm, such as market insurance (lower $\lambda$), or the probability of disruption, such as flood defences (lower $\pi$), lower the response.
\end{enumerate}
\end{proposition}

Parts (i)--(iii) follow directly from condition~\eqref{eq:dup} and the updating rule. Part (iv) follows because a flood moves $\hat\pi$ by the same amount for all products of the same history, so that the products that cross the threshold are those with the lowest $F_B/q_B$. Part (v) follows because the constraint $m \leq q_B \leq D_B$ cannot be met. Part (vi) follows because insurance and defences lower the left-hand side of condition~\eqref{eq:dup}, which is the standard substitution between market insurance, self-insurance and self-protection in \citet{ehrlich1972}.

The framework also clarifies what we cannot observe. Condition~\eqref{eq:dup} is about the value of redundancy during a disruption, which yearly data cannot show directly. The two channels of $\phi$ pay off differently. Volume placed at $B$ that would otherwise depend on $A$ is protected in any disruption, while the ability to raise output at $B$ for $A$'s customers matters mostly when $A$ is out for long, so that a short stoppage need not trigger it. Nor do we observe the firm's beliefs. We therefore test the predictions on the timing and the pattern of the response, not on the size of the risk.

\section{Empirical Strategy} \label{sec:design}
For each flooded product, we take the change in the outcome between the year before the flood and $q$ years after it, and subtract the average change over the same years among its controls. The estimate $\hat\theta_q$ is the average over flooded products,
\begin{align}
\hat\theta_{q} &= \frac{1}{N_{q}} \sum_{i:\,F_i<\infty}
   \Big[ \big(Y_{i,F_i+q}-Y_{i,F_i-1}\big)
   - \frac{1}{|\mathcal{C}_i(q)|} \sum_{j\in\mathcal{C}_i(q)} \big(Y_{j,F_i+q}-Y_{j,F_i-1}\big) \Big], \label{eq:theta}\\
\mathcal{C}_i(q) &= \big\{\, j:\ F_j > F_i-1,\ \ E_{j,F_i-1} > F_i+q,\ \ g_j=g_i,\ \ c_j\neq c_i \,\big\}. \notag
\end{align}
$Y_{it}$ equals one if product $i$ is made in two or more plants in year $t$, and $F_i$ is the year of its first documented flood. A control $j$ is a product of the same kind that (i) is not flooded yet, (ii) has plants, as of year $F_i-1$, that stay dry through year $F_i+q$, where $E_{j,F_i-1}$ is the first flood at those plants, (iii) is in the same cell $g$ as the flooded product, and (iv) has its largest plant in another country, $c$, so that it is not caught by the same flood or by what the flood does to its neighbours. For $q<0$, the same differences are taken before the flood and should be zero.

\paragraph{{What is matched.}} The cell $g$ is the product's launch cohort (its first year of output, in three-year bins) times the number of plants that made it in $F_i-1$ (one, two, three, or four or more). Both are measured in the year before the flood, for the flooded product and for its controls, so that the cell cannot reflect the response. The matching is exact. Each flooded product is compared with the plain average of the controls in its cell, and no covariate enters a regression. The number of plants is the part that matters. A product made in more plants has more plants that can be flooded, so that floods hit multi-plant products disproportionately, and these are, on average, in the phase of their life in which plants are added. Compared with all not-yet-flooded products, or with those of the same launch cohort only, flooded products are therefore already adding plants before the flood. Within a cell, a flooded product and its controls start from the same number of plants, and $\hat\theta_q$ measures how the two diverge from that common starting point.

\paragraph{{Estimator and inference.}} This is the estimator of \citet{CH2024} with not-yet-treated controls, restricted to exact cells, rather than a two-way fixed-effects regression, which is biased in staggered designs with heterogeneous effects. It can also be written as a regression. Stacking the long differences of every flooded product and its controls, with a fixed effect for each flooded product and horizon and a weight of one over the number of controls in the cell, an OLS regression on event-time dummies reproduces every point estimate. Off-the-shelf implementations such as \texttt{did\_multiplegt\_dyn} condition on the launch cohort but not on the number of plants in $F_i-1$, which changes over time. Without that condition, the pre-flood test fails, as flooded products add plants faster than their controls before the flood, by 0.16 plants for engine programs and 0.31 for model versions between years $-4$ to $-2$ and year $-1$. Matching on the exact launch year alone fails in the same way, while matching on the exact launch year times plants gives the main estimates (Table~\ref{tab:design} in Appendix~\ref{app:est}). We draw owners with replacement to compute standard errors (499 draws), since the decision to duplicate is taken by the owner and floods hit several programs of the same owner at once. Standard errors clustered by owner in the stacked regression are about half as large, which reflects the few flooded owners, and we report the bootstrap throughout. In every event-study figure, year $-1$ is zero by construction and the bars show 95\% confidence intervals. We summarize the dynamics with averages over years $-4$ to $-2$ (pre-trends), years 0--4 and years 5--10, and, where the response is short-lived, over years 0--2 and 3--4.

\paragraph{{Robustness of the design.}} Table~\ref{tab:design} reports the main outcomes under alternative control sets. The estimates are the same with controls of any country and with controls of other owners only, so that the main estimates do not rest on comparing products within or across owners. Controls of the product's own owner are too few, and they differ before the flood. For the outcomes on markets, we also refine the cell by the number of export markets the product served in $F_i-1$ (none, 1--2, 3--5, 6--10, or 11 or more, and no recorded sales). This compares, for instance, a flooded version sold in three export markets with versions sold in three to five. The estimates keep their sign and size or grow, with wider intervals.

Two features of the design deserve comment. First, requiring controls from other countries protects against floods that hit many plants in one country at once, but it also excludes the products that may respond to the same news. We return to this in Section~\ref{sec:engines}. Second, the support of the estimates thins out over time. Eleven owners remain in year 5 for engine programs, so that the long-run estimates rest on few decision makers.

\section{Results} \label{sec:results}
\subsection{Production Duplication} \label{sec:engines}
\begin{figure}[t]
\centering
\includegraphics[width=0.85\textwidth]{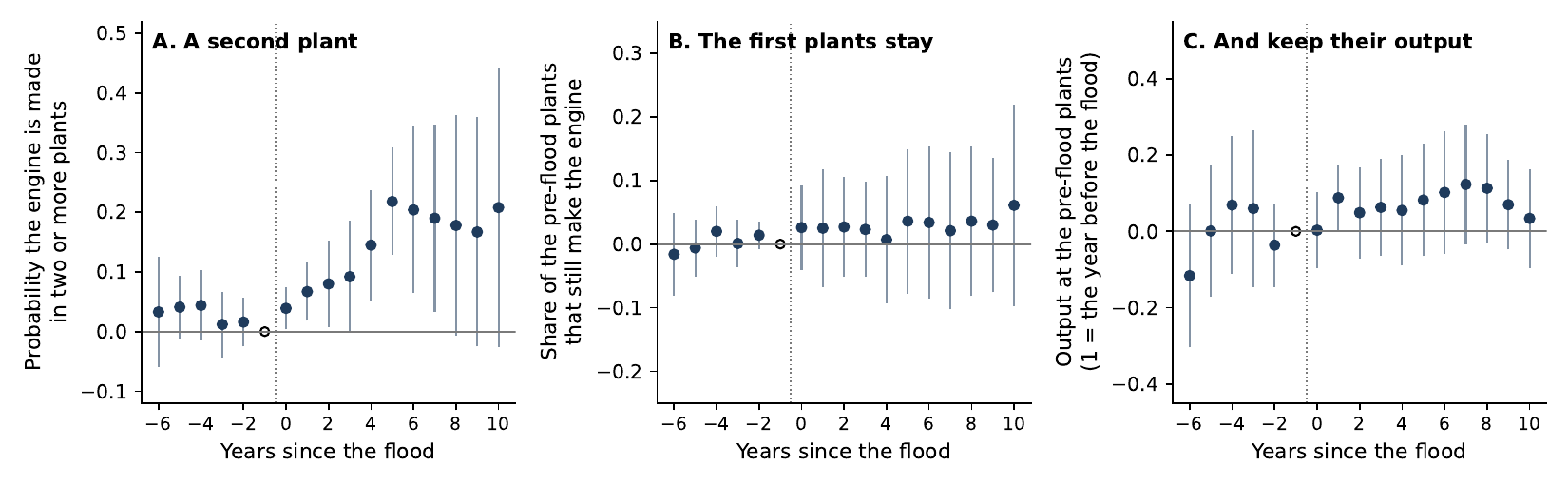}
\caption{Engine Programs after a Documented Flood} \label{fig:engines}
\caption*{\textit{Notes:} One outcome per panel. A: the program is made in two or more plants (0 or 1). B: of the plants that made the program the year before the flood, the share that still make it (0 if the program has ended). C: the program's output at those plants, divided by its output the year before the flood (0 if it has ended). Each dot is $\hat\theta_q$, the change since the year before the flood, flooded programs minus their controls. 95\% intervals, owner bootstrap. }
\end{figure}

We compare the 216 flooded engine programs with their controls. Panel A of Figure~\ref{fig:engines} shows the main outcome. Panels B and C ask what happens to the plants that made the program before the flood. If the second plant replaced the first, these plants would lose the program (B) or its output (C).

Flooded engine programs are made in two or more plants more often than their controls, by 0.085 (s.e.\ 0.027) in years 0--4 and by 0.194 (s.e.\ 0.061) in years 5--10, from a flat path before the flood (0.024, s.e.\ 0.024, in years $-4$ to $-2$). Since about half of the flooded programs were made in one plant before the flood, the effect is large. It is the same against controls of any country (0.180, s.e.\ 0.064), against controls of the same size (0.178, s.e.\ 0.073), without any Japanese program (0.178, s.e.\ 0.079) and without the floods of 2011 (0.176, s.e.\ 0.093). After the flood, the program is also less concentrated, with the largest plant's share of output falling by 0.074 (s.e.\ 0.036).

\paragraph{{No relocation.}} The plants that made the program before the flood keep it (panel B), and their output does not fall (panel C, $+0.087$ of the output of the year before, s.e.\ 0.066, in years 5--10). Programs do not end up being made only at added plants. Production is thus added rather than moved, as part (iii) of Proposition~\ref{prop:dup} predicts.

\paragraph{{No contraction of the flooded plant.}} The same holds for the flooded engine plant as a whole, not only for the flooded program. We compare each engine plant with a first documented flood (43 plants of 15 owners) with engine plants not yet flooded, in other countries, of the same cohort and with the same number of engine family lines the year before, using the same estimator (Appendix~\ref{app:plants}). The engine plant loses output in the flood year, by 0.112 (s.e.\ 0.061) of its output of the year before, and recovers the next year ($+0.066$, s.e.\ 0.070). There is no lasting change afterwards, neither in its output ($+0.123$, s.e.\ 0.123, in years 5--10) nor in its output relative to the vehicle output of the assembly countries it supplies ($-0.017$, s.e.\ 0.134). The estimators of \citet{CH2024}, \citet{CS2021} and \citet{BJS2024}, which cannot match on the plant's family lines, give a flood-year drop of 0.11 to 0.16 log points and the same recovery (Table~\ref{tab:plants_estimators}). The flood affects the composition of the engine plant's output more than its level. Over years 5--10, the engine family lines it ran before the flood lose output ($-0.086$, s.e.\ 0.113, of the plant's output of the year before), while other engine family lines gain ($+0.191$, s.e.\ 0.115; each outcome is capped at three times the output of the year before, so that the two need not add up to the plant's output). A larger share of the engine programs it made before the flood is made only at other engine plants ($+0.107$, s.e.\ 0.054), and fewer end altogether ($-0.136$, s.e.\ 0.055). The flooded engine plant thus keeps its output, while some of its engine programs leave and other engine family lines take their place.

\paragraph{{Other customers in normal times.}} We also look at the 78 additions after a flood, i.e., cases in which a plant makes a flooded engine program for the first time, from 42 programs of 10 owners. Of these, 63 are in another country, at an engine plant the owner already had. They send only 5\% of their volume to the assembly plants that the flooded plants supplied, so that in normal times they do not serve the flooded plant's customers. Instead, 37\% of their volume goes into model versions that were already built at the destination with another engine, and 57\% into nameplates built there before, counting earlier generations. Assembly plants near the added plant thus switch part of their production to the flooded program's engine. The output of the plants added after a flood is small ($+0.062$, s.e.\ 0.043, of the program's output in $F-1$ in years 5--10), and it goes to model versions and assembly plants new to the program. Supplier switches, i.e., deliveries to model versions that the program already supplied from another plant, fall ($-0.002$, s.e.\ 0.001).

\paragraph{{More assembly countries served.}} The second plant also extends the program to new locations. We count the assembly countries to which an engine program supplies 100 units or more, relative to their number in $F-1$. Flooded programs supply more of them than their controls, by 0.121 (s.e.\ 0.031) in years 0--2, 0.293 (s.e.\ 0.070) in years 3--4 and 0.361 (s.e.\ 0.100) in years 5--10, from $-0.027$ (s.e.\ 0.048) in years $-4$ to $-2$. Countries new to the program account for most of the rise (0.214, s.e.\ 0.076, in years 3--4). Of this, 0.138 (s.e.\ 0.054) are supplied by the plants of $F-1$ and 0.076 (s.e.\ 0.046) only by added plants, and the latter grows to 0.179 (s.e.\ 0.096) in years 5--10. The estimates are the same when the cell also matches the number of foreign assembly countries supplied in $F-1$ (0.263, s.e.\ 0.104, in years 3--4, and 0.216, s.e.\ 0.115, for countries new to the program, Table~\ref{tab:design}). Since the markets of an engine are the countries where cars fitted with it are assembled, this also reflects where carmakers assemble those cars.

\paragraph{{A response to risk.}} Two facts suggest that firms respond to risk rather than to growth or damage. First, the response starts with the flood, with flat trends before it. Second, a stoppage, in which the plant stayed dry and production stopped because a flood cut off its suppliers, triggers about the same response as a direct hit, 0.177 (s.e.\ 0.096) against 0.214 (s.e.\ 0.087) in years 5--10. With direct hits restricted to floods that put water in the plant, the two cannot be told apart either (Appendix~\ref{app:strict}). This is part (ii) of Proposition~\ref{prop:dup}.

\paragraph{{A persistent response.}} A flood could also bring forward additions that the program would have made later in any case. The effect would then peak early and fade as control programs add plants. The data do not show this pattern. The effect on being made in two or more plants is 0.106 (s.e.\ 0.038) in years 2--4 and 0.184 (s.e.\ 0.078) in years 8--10, and the share of programs that have ever added a plant rises by a further 0.074 (s.e.\ 0.037) between the two windows. First-time additions remain more frequent than among controls in years 5--10 (summed, $+0.105$, s.e.\ 0.025). The pattern is the same for the 93 programs flooded up to 2014, for which year 10 is observed. The rise is specific to the flooded program, however. The engine family's later programs are not made in more plants ($-0.025$, s.e.\ 0.133).

\paragraph{{The second plant as insurance.}} These results fit a second plant that the firm starts as a backup, in case it is needed. In normal times it serves its own markets, so that the program's additional volume does not depend on the flooded site, and since the original plants keep their output, it does not take volume away from that site. Its value as insurance lies in a long disruption: the program then already runs at a second site, with tooling, suppliers and workers in place, where the firm could raise output for the customers of the flooded plant. This is the first channel of $\phi$ in condition~\eqref{eq:dup}, and it is used only when a disruption lasts. The floods we observe are short. Programs made in a single plant do not lose output in the flood year relative to their controls ($+0.069$, s.e.\ 0.068), and the confidence interval rules out losses larger than about 6\% of a year's output, i.e., about three weeks of production. There is thus little for a second plant to cover in the floods we observe. Long disruptions are too rare to test it directly: in 2011, the year with the most documented floods, direct hits concern 20 engine programs of three firms. The evidence for the insurance motive comes instead from the response itself. It follows the revealed risk rather than physical damage, it persists, and it is concentrated where a second plant is cheap (Section~\ref{sec:cheap}) and, more tentatively, where market insurance is expensive (Section~\ref{sec:ext}).

\paragraph{{Controls from the flooded country.}} With controls from the flooded program's own country instead of other countries, the estimate falls to 0.064 (s.e.\ 0.070) in years 5--10, although the same 64 programs give 0.206 (s.e.\ 0.075) against controls of any country (Table~\ref{tab:controls}). Never-flooded engine programs of the flooded country lean the same way (0.050, s.e.\ 0.039), so they are not a clean comparison. This could reflect learning within the country, as firms learn from floods at their neighbours' plants, or a country-wide shock, mostly in Japan in 2011. Controls of other owners give the same estimate (0.208, s.e.\ 0.063, in years 5--10). Only 6--7\% of the matched controls belong to the flooded program's owner, so the result does not come from owners shifting programs among their own plants.

\begin{table}[t]
\centering
\footnotesize
\caption{Engine Programs, Made in Two or More Plants: Alternative Controls} \label{tab:controls}
\setlength{\tabcolsep}{5pt}
\begin{tabularx}{\textwidth}{@{}Lccc@{}}
\toprule
 & Years 0--4 & Years 5--10 & Programs [owners]\\
\midrule
Controls from other countries (main) & $+0.085$ (0.027) & $+0.194$ (0.061) & 99 [11]\\
Controls from the same country & $+0.056$ (0.057) & $+0.064$ (0.070) & 64 [11]\\
The same 64 programs, controls of any country & $+0.075$ (0.031) & $+0.206$ (0.075) & 64 [11]\\
Never-flooded programs of the flooded country & $+0.022$ (0.015) & $+0.050$ (0.039) & 299 [21]\\
\bottomrule
\end{tabularx}
\caption*{\textit{Notes:} The last row treats the never-flooded programs of a flooded country as if they were flooded, against programs abroad. Owner-bootstrap standard errors in parentheses. Programs and owners in year 5. }
\end{table}

\subsection{Role of Capital Adjustment Costs} \label{sec:cheap}
Starting an engine at a plant that already makes engines of the same family needs no new plant and little new tooling. Bringing the platform to a plant that does not run it is costly \citep{praetorius2026}, so that without such a plant a new program cannot easily move elsewhere. If the cost of a second plant is what decides, as in part (iv) of Proposition~\ref{prop:dup}, a flood should bring a second plant only to programs whose owner already makes the family in another country.

\begin{figure}[t]
\centering
\includegraphics[width=0.85\textwidth]{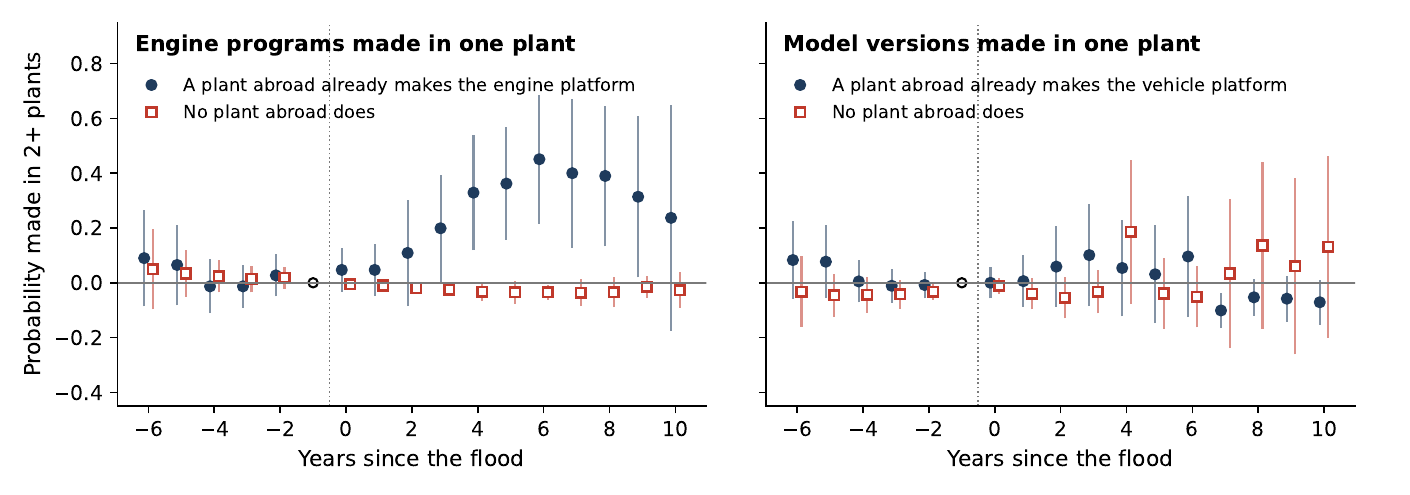}
\caption{Products Made in One Plant: Made in Two or More Plants, by Whether a Plant Abroad Makes the Platform} \label{fig:cheap}
\caption*{\textit{Notes:} Products made in one plant the year before the flood. Left: engine programs, by whether a plant in another country already makes the engine family. Right: model versions, by whether a plant of the group in another country builds the vehicle platform. Controls of the same status. 95\% intervals, owner bootstrap. }
\end{figure}

We keep the flooded engine programs made in one plant the year before the flood (104 of the 216) and split them into those for which a plant in another country makes the same family in that year (62 programs) and those for which no plant does (42). Each group is compared with controls of the same status, so that the left panel of Figure~\ref{fig:cheap} compares flooded and not-yet-flooded programs that both have such a plant. The right panel does the same for model versions, where the plant abroad builds the version's platform.

Engine programs with a plant abroad that makes the family get a second plant, by 0.146 (s.e.\ 0.069) in years 0--4 and by 0.359 (s.e.\ 0.105) in years 5--10. The others do not ($-0.030$, s.e.\ 0.018). The additions after a flood are consistent with this pattern. Of these, 59\% are at a plant that already made the family, 32\% at a plant that made another engine family of the owner, and 9\% at a plant new to engines. Diversification thus builds on the firm's existing production network. The left panel, however, rests on 20 programs of 10 owners in year 5, and having such a plant abroad is not random, as it goes with large owners and with engine families made worldwide, which is why the controls have the same status.

\subsection{The Price of Redundancy} \label{sec:price}
The framework does not quantify the value of redundancy, but its cost can be priced. S\&P reports output and no costs, so we price the capacity behind the output with the investments that firms announce. This gives what a second plant costs, how much capital floods add, and what a disruption must cost the firm for a second plant to pay.

\begin{table}[t]
\centering
\footnotesize
\caption{The Price of a Second Plant for Engine Programs} \label{tab:price}
\setlength{\tabcolsep}{4pt}
\renewcommand{\arraystretch}{1.12}
\begin{tabularx}{\textwidth}{@{}Lcccc@{}}
\toprule
\multicolumn{5}{@{}l}{\textit{A. An addition: capital in million US dollars}}\\
 & After a flood & Ordinary & Difference & \makecell[b]{Median,\\after a flood}\\
\cmidrule(l){2-5}
\quad Capital at announced prices & 26.8 & 39.7 & $-12.9$ & 8.1\\
 & & & (3.5) & \\
\quad Capital per unit of yearly volume (\$) & 501 & 572 & $-71$ & 590\\
 & & & (23) & \\
\quad Capital with a \$164 million line & 136.7 & 144.0 & $-7.4$ & 164.0\\
 & & & (5.0) & \\
\quad Volume (thousand engines a year) & 53.7 & 65.2 & $-11.4$ & 19.1\\
 & & & (11.5) & \\
\addlinespace[6pt]
\multicolumn{5}{@{}l}{\textit{B. Capital added since the year before the flood, per flooded program, million US dollars}}\\
 & Years $-4$ to $-2$ & Years 0--4 & Years 5--10 & Programs\\
\cmidrule(l){2-5}
\quad Announced prices, all programs & $+4.5$ & $-3.0$ & $-8.0$ & 203, 99\\
 & (7.7) & (3.6) & (12.8) & \\
\quad Announced prices, one plant in $F-1$ & $+0.0$ & $+2.3$ & $+9.1$ & 95, 30\\
 & (0.0) & (4.1) & (18.0) & \\
\quad \$164 million line, all programs & $+6.5$ & $+1.0$ & $+22.2$ & 203, 99\\
 & (6.3) & (13.7) & (39.4) & \\
\quad \$164 million line, one plant in $F-1$ & $-0.1$ & $+14.1$ & $+82.3$ & 95, 30\\
 & (1.0) & (8.3) & (33.9) & \\
\addlinespace[6pt]
\multicolumn{5}{@{}l}{\textit{C. Break-even of condition~\eqref{eq:dup} with $\Delta c=0$ (lower bounds), $r=0.15$, US dollars}}\\
 & \makecell[b]{$K$\\(million)} & \makecell[b]{$rK/q_B$\\(per engine,\\a year)} & \makecell[b]{$rK/\hat\pi$\\(million)} & \makecell[b]{$\lambda\phi$\\(per engine)}\\
\cmidrule(l){2-5}
\quad At announced prices & 26.8 & 75 & 343 & 6,394\\
\quad With a \$164 million line & 136.7 & 382 & 1,751 & 32,613\\
\bottomrule
\end{tabularx}
\caption*{\textit{Notes:} An addition is a plant that starts making an engine program after the program's launch year. A: the 78 additions within ten years of a documented flood (42 programs of 10 owners) and the 534 additions of never-flooded programs of the same owners. Capital from equation~\eqref{eq:price}, or a new line of \$164 million for every addition that needs capacity. Volume: the addition's peak output in its first five years. Per engine: capital over volume, averaged over additions. B: estimator~\eqref{eq:theta} with the capital of the plants added since $F-1$ as the outcome, averaged over each window (years $-4$ to $-2$ are close to zero by construction for programs made in one plant in $F-1$). Programs with controls in year 0 and in year 5, 11 owners in year 5. C: $K$ and $q_B$, the mean capital and volume of an addition after a flood (panel A). $\hat\pi=0.0117$, the share of the flooded plant's active years with another documented flood after the program's first (23 of 1,965); at any plant that makes the program, $\hat\pi=0.0265$ and $\lambda\phi$ at announced prices is 2,827. $r$: the annual capital charge, assumed, to which every value in panel C is proportional. Owner-bootstrap standard errors in parentheses (499 draws). Current US dollars of the announcements, not deflated. }
\end{table}

\paragraph{{Prices from announced investments.}} We collected 363 public announcements of investments in engine production capacity from company press releases, stock-exchange filings, union contract summaries and news (Appendix~\ref{app:defs}). Eighty of them concern engines only and state the capacity of the project. On these, we estimate
\begin{equation}
\ln C_j = a_{s(j)} + \beta \ln Q_j + \varepsilon_j, \label{eq:costfn}
\end{equation}
where $C_j$ is the amount of announcement $j$ in millions of current US dollars, $Q_j$ the capacity of the project in engines a year, and $s(j)$ its kind, i.e., a new plant, a new line at an existing engine plant, an expansion, or a retooling for a new engine family. The elasticity is $\hat\beta=1.04$ (s.e.\ 0.11, robust to heteroskedasticity), so that capital is about proportional to capacity. At 200,000 engines a year, a new line costs \$615 per engine of annual capacity and a new plant \$1,033. Below 79,500 engines a year, the tenth percentile of announced capacity, we keep the cost proportional to capacity, at \$590 and \$991 per engine.

\paragraph{{The price of an addition.}} An addition is a plant that makes an engine program for the first time, after the program's first year. We price it by what the receiving plant $p$ made the year before, with $C_{\text{line}}$ and $C_{\text{plant}}$ from equation~\eqref{eq:costfn}:
\begin{equation}
K_{ip}=\begin{cases}
0 & \text{if } x_{fp}>0 \text{ and } x_{fp}+v_{ip}\le \bar x_{fp},\\
C_{\text{line}}(x_{fp}+v_{ip})-C_{\text{line}}(\bar x_{fp}) & \text{if } x_{fp}>0 \text{ and } x_{fp}+v_{ip}>\bar x_{fp},\\
C_{\text{line}}(X_{fp})\,v_{ip}/X_{fp} & \text{if } x_{fp}=0 \text{ and } p \text{ makes engines},\\
C_{\text{plant}}(X_{p})\,v_{ip}/X_{p} & \text{if } p \text{ makes no engines}.
\end{cases} \label{eq:price}
\end{equation}
S\&P gives output only, so we measure the capacity that an addition needs by $v_{ip}$, the peak output of program $i$ at plant $p$ in its first five years there. Here, $x_{fp}$ is the output of the program's family $f$ at $p$ the year before, and $\bar x_{fp}$ its peak over the five years before, which we take as the capacity of the family line (the platform line of Section~\ref{sec:receiving}, measured by its recent rather than its lifetime peak). A plant that makes the family thus adds the program at no cost if its line has room, and pays for the capacity above the line's past peak if not. A plant that makes only other families pays for a new family line, and a plant that makes no engines for a new plant. Both are sized at their peak output $X$ in their first five years and shared by volume among the programs that start on them. Since small lines may cost more than equation~\eqref{eq:costfn} implies, we also charge every addition that needs capacity the median announced new line, \$164 million, whatever its size. We compare the 78 additions after a documented flood with the 534 additions of never-flooded programs of the same owners, and we apply the estimator of equation~\eqref{eq:theta} to the capital of the plants added since $F-1$.

\paragraph{{Second plants are cheap.}} An addition after a flood costs \$26.8 million on average and \$8.1 million at the median, or \$501 per engine of annual volume (Table~\ref{tab:price}, panel A). This is \$12.9 million (s.e.\ 3.5) less than the same owners' ordinary additions. Additions after a flood are smaller, with a median of 19,100 engines a year against 30,900, and cheaper per engine, at \$501 against \$572, and 17\% of them join a family line with room and cost nothing, against 12\% of ordinary ones. The capital that floods add cannot be told apart from zero at announced prices (panel B). In years 5--10, it is \$9.1 million per program (s.e.\ 18.0) for programs made in one plant before the flood, and $-\$8.0$ million (s.e.\ 12.8) for all flooded programs. Only if every addition is charged a full new line does the capital added to programs made in one plant stand out from zero, at \$82.3 million (s.e.\ 33.9).

\paragraph{{What a disruption must cost.}} Condition~\eqref{eq:dup} turns these prices into the loss that a disruption at plant $A$ must cause. With $F_B=rK$, where $r$ is the annual capital charge for depreciation and return, and with $\Delta c=0$, the second plant pays if
\begin{equation}
\lambda\phi \;\geq\; \frac{rK}{\hat\pi\, q_B}. \label{eq:breakeven}
\end{equation}
After a program's first documented flood, the flooded plant is flooded again in 1.17\% of its active years (23 of 1,965 plant-years), which we take as $\hat\pi$. With $r=0.15$ and the mean capital and volume of an addition after a flood, \$26.8 million and 53,702 engines a year, a disruption must cost the firm \$343 million, or \$6,394 for each engine of the second plant's yearly volume (panel C). Since $\phi\le 1$, the margin $\lambda$ lost per engine not delivered must be at least as large. The flood-year loss of programs made in one plant is at most about 6\% of a year's output, about three weeks of production (Section~\ref{sec:engines}). In a disruption of three weeks, the second plant protects at most three weeks of its volume, $\phi\le 0.06$, and $\lambda$ would have to exceed \$100,000 per engine. With the flooded engine plant's loss in the flood year instead, 0.112 of its output (Section~\ref{sec:engines}), $\phi\le 0.11$ and $\lambda$ would still have to exceed \$57,000. At the observed frequency of floods, short stoppages thus cannot pay for a second plant. It pays as insurance only against disruptions that stop the first plant for months, the role that Section~\ref{sec:engines} gives it, or through the markets it serves. Without any flood, the second plant pays for itself if making the engine near the cars assembled around $B$ is worth \$75 per engine a year, its capital charge $rK/q_B$.

\paragraph{{A lower bound.}} These prices are lower bounds on the cost of a second plant, since they set $\Delta c=0$, i.e., they assume that plant $B$ makes the engine at the same unit cost as plant $A$. The added lines are small, with a median of 19,100 engines a year against a peak of 91 thousand for the median engine family line (Section~\ref{sec:facts}). Of the additions after a flood, 85\% are smaller than nine in ten announced projects, and most of them share the receiving plant with its other lines. A lower unit cost at $B$ than at $A$ is thus unlikely. Each dollar of $\Delta c$ adds $q_B$ dollars a year to the right-hand side of condition~\eqref{eq:dup}, so that the thresholds in panel C are lower bounds as well. The loss per engine must be at least $(rK/q_B+\Delta c)/\hat\pi$, which rises by about \$8,500 for every \$100 of $\Delta c$, and local supply must be worth at least \$75 plus $\Delta c$ per engine a year. The prices also leave out the tooling of an addition to a line with room, which we price at zero, so that they understate $F_B$ as well. Measuring capacity by peak output is about right on average: at 42 announced projects that we can link to an S\&P engine plant, the peak output of the plant's engine families in the first five years is a median 1.05 of the announced capacity (1.06 weighted by capacity), although single projects vary widely (Appendix~\ref{app:defs}). Finally, the amounts are current dollars of the announcement years, not deflated, and $r$ is an assumption.

\subsection{Recipient Plants} \label{sec:receiving}
The framework predicts that the program goes to a plant with a low fixed cost and high local demand. To test this, we take the 78 additions after a documented flood, i.e., cases in which a plant makes a flooded engine program for the first time within ten years of the flood, and, as a benchmark, the 912 additions of never-flooded programs after their launch year. For each, we look at the receiving plant the year before the addition, namely the engines it made, by family and by engine type (fuel, i.e., spark ignition, diesel or electric, times the cylinder layout, such as inline four or V6), and whether an assembly plant or a transmission plant stands within 2 km (Table~\ref{tab:recv}).

\begin{figure}[t]
\centering
\includegraphics[width=0.85\textwidth]{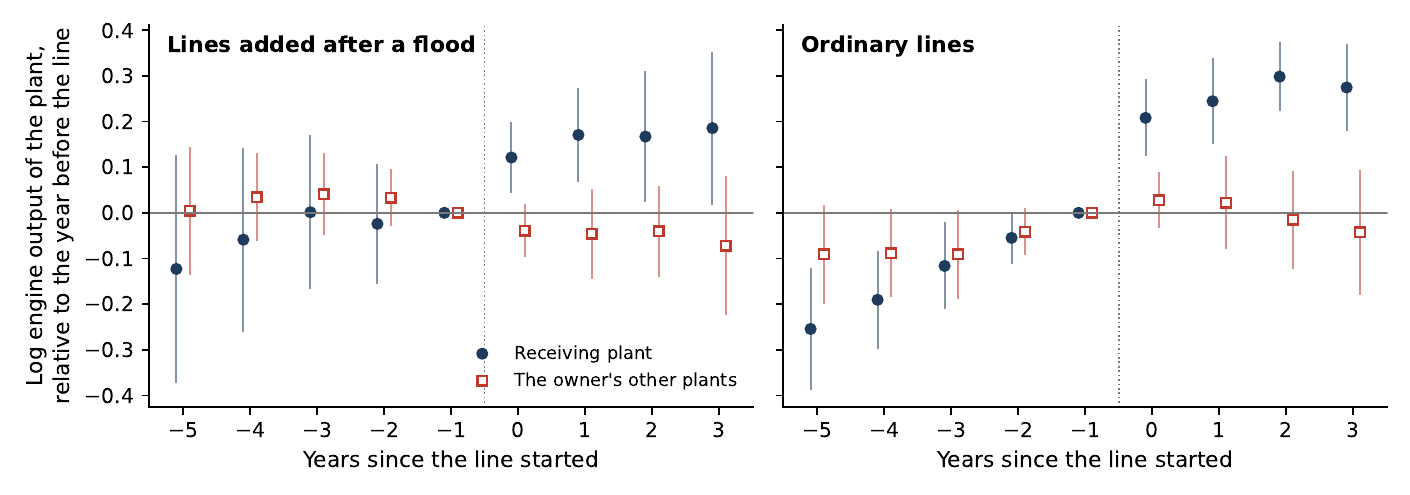}
\caption{Engine Output of the Receiving Plant and of the Owner's Other Engine Plants} \label{fig:recv}
\caption*{\textit{Notes:} Engine output of the plant that receives an engine program and of the owner's other engine plants that could have received it, in log points relative to the year before the addition. Left: additions after a documented flood (70 additions, 10 owners). Right: ordinary additions (615 additions, 24 owners). Additions in 2005 or later. Means over additions, 95\% intervals, owner bootstrap. }
\end{figure}

We then compare it with the plants the firm could have chosen, i.e., the owner's other engine plants that did not make the program in that year, about 19 per addition. To see whether engines behave differently, we do the same for the additions of transmission programs and model versions. We look at how the plants' output moved before and after the addition (Figure~\ref{fig:recv} for engines) and at their slack. The data have no capacity figures, so we measure capacity from the platform lines a plant runs. A platform line is a platform at a plant, and its capacity is the largest yearly output it ever has there, over 2000--2024. A plant's slack is one minus its output over the summed capacity of the platform lines it runs. Because the lifetime peak can come after the addition, we also measure slack against the peaks reached so far. To capture demand, we measure the plant's access to the product's markets the year before, namely the distance to them, the share of them in the plant's country, in contiguous countries, in countries with a common language and in countries with a trade agreement, and the applied tariff from the plant's country, each weighted by the market's size. The markets of an engine or transmission program are the assembly plants that fit it, weighted by units, and those of a vehicle model are its sales by destination country (Figure~\ref{fig:recv_summary} and Table~\ref{tab:recv2} in Appendix~\ref{app:recv}). Finally, a conditional logit \citep{mcfadden1974} gives how each trait changes the odds that a plant is chosen,
\begin{equation}
\Pr\big(p_\ell=p \,\big|\, \mathcal S_\ell\big)=\frac{\exp(x_{p\ell}'\beta)}{\sum_{p'\in\mathcal S_\ell}\exp(x_{p'\ell}'\beta)},\qquad
\mathrm{slack}_{pb}=1-\frac{\sum_{f\in\mathcal F_{pb}}q_{fpb}}{\sum_{f\in\mathcal F_{pb}}\max_{2000\le s\le 2024}q_{fps}}, \label{eq:clogit}
\end{equation}
where $\ell$ is an addition, $\mathcal S_\ell$ the owner's plants of the same kind that make its products in the year $b$ before the addition but not this product, with the receiving plant among them, $q_{fpb}$ the output of platform line $f$ at plant $p$ in year $b$, $\mathcal F_{pb}$ the platform lines the plant runs in $b$, and $x_{p\ell}$ the plant's traits in Table~\ref{tab:clogit}. A coefficient $\beta$ on a binary trait multiplies the odds of being chosen by $e^{\beta}$. Standard errors are again based on drawing owners with replacement (499 draws).

\begin{table}[t]
\centering
\footnotesize
\caption{The Plant That Receives an Engine Program} \label{tab:recv}
\setlength{\tabcolsep}{5pt}
\renewcommand{\arraystretch}{1.12}
\begin{tabularx}{\textwidth}{@{}Lcccc@{}}
\toprule
 & \makecell[b]{After a\\flood} & \makecell[b]{Started\\by $F+4$} & Ordinary & Difference\\
\midrule
\multicolumn{5}{@{}l}{\textit{A. The year before the addition}}\\
\quad Made the program's family & 0.59 & 0.62 & 0.52 & $+0.07$ (0.08)\\
\quad Made the same engine type, another family & 0.21 & 0.22 & 0.22 & $-0.02$ (0.04)\\
\quad Made only engines of other types & 0.12 & 0.15 & 0.09 & $+0.03$ (0.06)\\
\quad Made no engines & 0.09 & 0.02 & 0.17 & $-0.08$ (0.04)\\
\quad Its output of the program's type, share & 0.74 & 0.73 & 0.67 & $+0.07$ (0.08)\\
\quad An assembly plant on the site & 0.49 & 0.51 & 0.34 & $+0.15$ (0.08)\\
\quad A transmission plant on the site & 0.33 & 0.36 & 0.44 & $-0.11$ (0.06)\\
\quad Engines only & 0.40 & 0.38 & 0.42 & $-0.02$ (0.06)\\
\addlinespace[4pt]
\multicolumn{5}{@{}l}{\textit{B. The year of the addition and the two years after}}\\
\quad Output of its other engines over the year before & 1.07 & 1.06 & 1.21 & $-0.13$ (0.07)\\
\quad Share of the added program's volume fitted on the site & 0.34 & 0.33 & 0.15 & $+0.19$ (0.08)\\
\midrule
\quad Additions & 78 & 55 & 912 & \\
\bottomrule
\end{tabularx}
\caption*{\textit{Notes:} Means over additions. After a flood: the plant starts a flooded program within ten years of the flood (started by $F+4$: within four). Ordinary: additions of never-flooded programs after their launch year. Engine type: fuel times cylinder layout. Site: within 2 km. Rows on the plant's output: plants that made engines the year before. The year of the addition and the two years after: additions by 2022. Difference: after a flood minus ordinary, owners drawn with replacement. }
\end{table}

\begin{figure}[t]
\centering
\includegraphics[width=\textwidth]{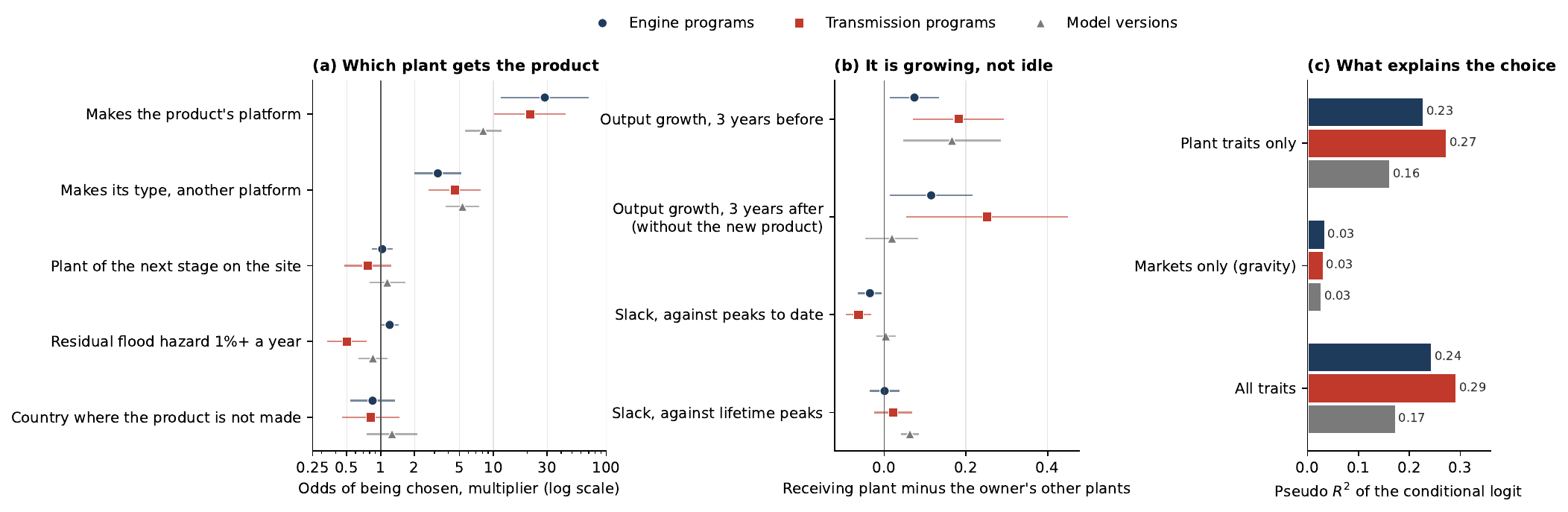}
\caption{Which of the Owner's Plants Receives the Product} \label{fig:recv_summary}
\caption*{\textit{Notes:} Additions of engine programs, transmission programs and model versions, 2004--2024, after a flood and in ordinary times pooled. Choice set: the owner's plants of the same kind that make its products the year before the addition but not this product. (a) Conditional logit \citep{mcfadden1974}, coefficients of binary traits shown as odds multipliers $e^{\beta}$ on a log scale. (b) The receiving plant minus the mean of the other plants of its choice set. Growth: symmetric rate. Slack: one minus the plant's output over the summed capacity of the platform lines it runs. (c) Pseudo $R^2$ of the conditional logit with all traits, with the plant's own traits only, and with the market terms only. 95\% intervals, owner bootstrap with 499 draws. Full estimates, also separately after a flood and in ordinary times, in Tables~\ref{tab:recv2} and~\ref{tab:clogit} of Appendix~\ref{app:recv}. }
\end{figure}

\paragraph{{The receiving plant makes the same kind of engine.}} In six additions out of ten, the receiving plant already made the program's family, in another two the same engine type in another family, and three quarters of its output is of the program's type (Table~\ref{tab:recv}). Plants new to engines are rare after a flood (2\% of the additions within four years). The family is the decisive factor. In the logit, a plant that makes it is chosen with nearly 30 times the odds of one that does not ($e^{3.35}$, Figure~\ref{fig:recv_summary}, panel a), and a plant that makes the engine type in another family with about 3 times the odds ($e^{1.17}$). After a flood the estimate for the family is of the same size but imprecise ($+2.85$, s.e.\ 1.56). Transmissions and model versions follow the same pattern, so that the plant's products, not its location, explain most of the choice. Plant traits alone yield a pseudo $R^2$ of 0.23 for engines, the market terms alone 0.03.

\paragraph{{The receiving plant is expanding rather than underused.}} The two measures of slack are consistent once read together. Against the peaks its platform lines have reached so far, the receiving plant has less slack than the owner's other plants ($-0.035$, s.e.\ 0.015, for engines), and so does its family line for the program ($-0.074$, s.e.\ 0.022). Against the peaks its platform lines reach over their whole life, it has as much slack ($+0.001$, s.e.\ 0.019, and $+0.018$, s.e.\ 0.017, Figure~\ref{fig:recv_summary}, panel b). Its platform lines are thus at their past high the year before the addition, but below the high they reach later, as in a plant that is expanding. Its output growth confirms this. It grows faster than the owner's other plants over the three years before the addition (by $+0.07$, s.e.\ 0.03), and, leaving the new program's own units out, over the three years after it (by $+0.12$, s.e.\ 0.05), with its family line for the program growing faster still ($+0.30$, s.e.\ 0.08). The same holds for transmissions and model versions, whose receiving plants grow faster before the addition by 0.18 and 0.17 (s.e.\ 0.06 and 0.06). In the logit, which holds the platform and the plant's markets fixed, growth before the addition raises the odds of being chosen ($+0.016$ per 10 points, s.e.\ 0.005, for engines), while slack against the lifetime peaks does not matter, neither for engines ($-0.008$, s.e.\ 0.035) nor for the plant as a whole for model versions ($+0.018$, s.e.\ 0.020). For model versions, only the slack of the platform assembly line draws the addition ($+0.091$, s.e.\ 0.014), a platform assembly line that has yet to reach its peak. After a flood the receiving plant grows by about as much as in ordinary times (by $+0.12$ before the addition and $+0.28$ on the family line after it), but with 62 additions, those of 2004--2024 with every trait measured, the differences are imprecise (s.e.\ 0.10 and 0.18, Table~\ref{tab:recv2}).

\paragraph{{The receiving plant is close to vehicle assembly and to the program's markets.}} After a flood, though not in ordinary times, the program also goes to a plant next to an assembly plant. The receiving plant is more likely than the owner's other plants to have one on its site ($+0.18$, s.e.\ 0.09, against $+0.00$, s.e.\ 0.03, in ordinary times, Table~\ref{tab:recv2}), and a third of the added program's volume is fitted there (Table~\ref{tab:recv}). This is the local demand $D_B$ of the framework. More broadly, the receiving plant is closer to the program's markets, about 1,000 km closer on average, with more of them in its own country and lower tariffs to them. In the logit, the odds of being chosen fall with distance ($-0.18$ per log point, s.e.\ 0.05) and with the tariff. One additional percentage point lowers the odds by about 12\% for engines and 16\% for transmissions, and by 3\% for model versions. Taking on the program does not reduce the engine plant's other output. Against engine plants that start a never-flooded program in the same year, run as many engine family lines and stand in other countries, its output without the new engine programs does not change ($+0.021$, s.e.\ 0.125, in years 0--2, and $+0.055$, s.e.\ 0.180, in years 3--6), and more of the engine programs it made before stay there ($+0.197$, s.e.\ 0.078, in years 3--6) rather than ending ($-0.247$, s.e.\ 0.059, Table~\ref{tab:plants_receiving}). A raw comparison with ordinary additions of all years shows slower growth of the plant's other engines (1.07 times their output of the year before in the year of the addition and the two years after, against 1.21), but this reflects the years in which programs are added after floods, not the addition itself. Nor is the receiving engine plant a safer one, as its flood hazard is, if anything, higher ($+0.19$, s.e.\ 0.10), whereas new transmission programs do avoid plants at high hazard ($-0.69$, s.e.\ 0.21). After a flood, the program's second plant is typically a growing plant that already makes the family and is located close to the vehicles it will supply, and the program mostly adds new capacity there rather than filling spare capacity, as only 17\% of the additions fit on a family line with room (Section~\ref{sec:price}). The additions after a flood are few: 62 for engines and 29 for transmissions, whose estimates after a flood rest on 5 owners, with 364 of 499 draws converging.

\subsection{Vehicle Assembly and Transmissions} \label{sec:products}
\begin{figure}[t]
\centering
\includegraphics[width=0.85\textwidth]{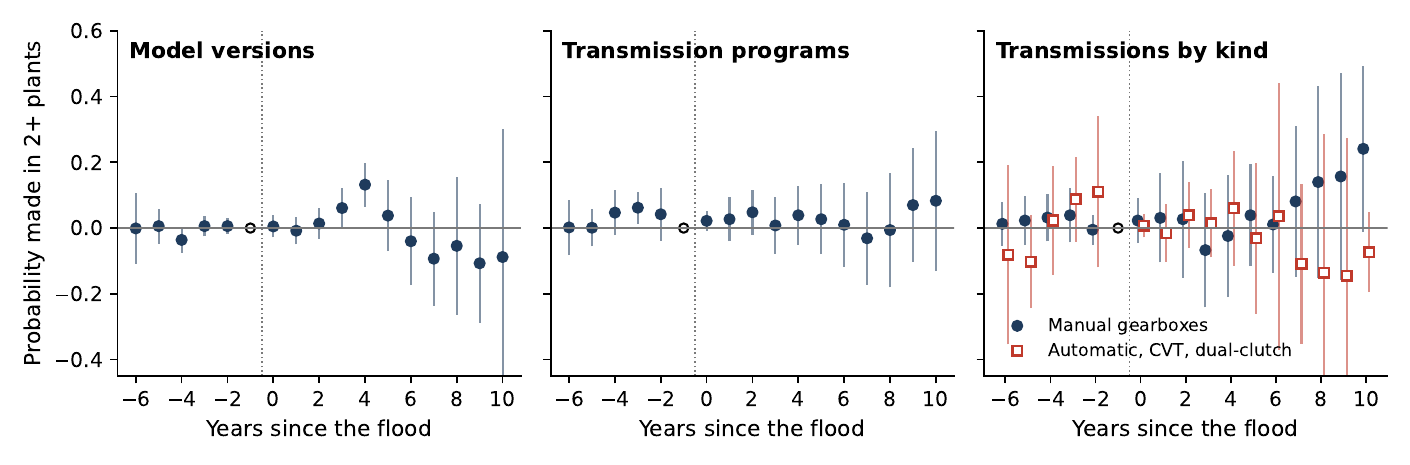}
\caption{Model Versions and Transmission Programs after a Documented Flood} \label{fig:products}
\caption*{\textit{Notes:} Made in two or more plants. Right: transmission programs by kind, controls of the same kind. 95\% intervals, owner bootstrap. }
\end{figure}

We apply the same design to the 421 flooded vehicle model versions, where the flood is at an assembly plant that built the version, and to the 121 flooded transmission programs. A flooded version is compared with not-yet-flooded versions launched in the same three years, made in the same number of plants the year before the flood and assembled in other countries. Panel C of Figure~\ref{fig:products} splits the transmission programs by kind, with controls of the same kind.

\paragraph{{A temporary second plant for model versions.}} Model versions are flooded more often than engine programs, and their output losses in the flood year are equally small. Flooded versions are made in two or more plants more often than their controls three to four years after the flood, by 0.096 (s.e.\ 0.028), from a flat path before it ($-0.008$, s.e.\ 0.012, in years $-4$ to $-2$). The second plant is short-lived, however. The difference is 0.004 (s.e.\ 0.018) in years 0--2, and it disappears by year five ($-0.057$, s.e.\ 0.061, in years 5--10, Table~\ref{tab:est}). The pattern is the same with controls of any country or of other owners only, with the exact launch year times plants, and with the cell refined by export markets (0.076, s.e.\ 0.036, in years 3--4, Table~\ref{tab:design}).

\paragraph{{Not a generational changeover.}} A second plant for a few years could be the old version running out at one plant while its successor starts at another. Defining the product as the version rather than the nameplate already separates the two generations, and the rise comes from versions without a successor launched since the flood (0.070, s.e.\ 0.046, in years 3--4, against 0.027, s.e.\ 0.035, for versions with one). Flooded versions are not more likely to get a successor ($+0.038$, s.e.\ 0.039). When they get one, it starts at a plant of the predecessor at least as often as after ordinary changeovers (60\% against 55\%).

\paragraph{{A second plant for other buyers.}} In normal times, the added plant serves buyers other than the flooded plant's. The share of the version's markets of $F-1$ (destination countries with 100 units or more) supplied from two or more plants does not change ($+0.012$, s.e.\ 0.017, in years 3--4). Seventeen flooded versions made in one plant are made in two plants in year 3 or 4. At the median, their added plants hold none of the version's sales in the markets of $F-1$. Six of them get a second plant within Japan after the floods of 2018. Most of the others start assembly in countries such as Russia, Kazakhstan, Malaysia, Vietnam or Brazil, for buyers there. The version's output ($+0.018$, s.e.\ 0.056) and sales ($+0.041$, s.e.\ 0.070) do not change in years 3--4. Nor does the flooded assembly plant shrink. Its output falls only in the flood year, and only slightly ($-0.022$, s.e.\ 0.049, or $-0.064$, s.e.\ 0.037, with the estimator of \citet{CH2024}), and its platform assembly lines and model versions do not change. An assembly plant that receives a model version after a flood does not give up any of its other output (Appendix~\ref{app:plants}).

\paragraph{{Slower expansion into new markets.}} The flood has a longer-lasting effect on where the version is sold. Flooded versions enter fewer new destination countries than their controls, by 0.164 (s.e.\ 0.056) of their number of markets in $F-1$ in years 0--2, 0.185 (s.e.\ 0.101) in years 3--4 and 0.086 (s.e.\ 0.043) in years 5--10, from $-0.090$ (s.e.\ 0.097) in years $-4$ to $-2$. The shortfall is in countries that the plants of $F-1$ would have supplied ($-0.120$, s.e.\ 0.031, in years 0--2), and the added plants do not make up for it ($-0.044$, s.e.\ 0.033). Flooded versions do not drop more of the markets they already served ($-0.016$, s.e.\ 0.019), and a larger share of their sales goes to those markets ($+0.023$, s.e.\ 0.008, in years 0--2). A flood thus delays the version's roll-out to new countries, while its output and its existing markets hold. This is not because flooded versions were exporters at a different stage. When the cell also matches the number of export markets in $F-1$, the shortfall is larger ($-0.401$, s.e.\ 0.198, in years 0--2, and $-0.194$, s.e.\ 0.091, through the plants of $F-1$).

\paragraph{{Local demand is lacking.}} Suitable capacity for a second plant exists, since for 57\% of the versions made in one plant a plant of the group abroad builds the same platform. However, these versions get no lasting second plant either ($-0.026$, s.e.\ 0.031, in years 5--10, Table~\ref{tab:est}). A likely reason is local demand. Using S\&P's sales by country, we compute for each of these versions the share of its sales, in the year before the flood, in the countries where such a plant is located. It is zero at the median, and zero for 52\% of the versions. An engine made in a second country goes into cars that are built there anyway, whereas a car made in a second country has to be sold there. In the terms of the framework, $D_B$ is close to zero. Where a second plant is added, it is for a new market, and it does not last.

\paragraph{{Transmissions.}} Transmission programs are not duplicated either ($+0.026$, s.e.\ 0.066). Here, suitable capacity is lacking. A plant abroad makes the family for 28\% of the programs made in one plant, against 60\% for engines, and the gap is entirely due to automatic, CVT and dual-clutch transmissions (3\%), while manual gearboxes look like engines (50\%). After a flood, manual gearbox programs respond like engine programs ($+0.112$, s.e.\ 0.094) and the others do not ($-0.076$, s.e.\ 0.067). The difference between the two is not distinguishable from zero ($+0.188$, s.e.\ 0.120). These estimates should be read with care, since flooded transmission programs already differ from their controls before the flood ($+0.050$, s.e.\ 0.023, in years $-4$ to $-2$), as much as after it, and the series for automatic transmissions is noisy.

\begin{figure}[t]
\centering
\includegraphics[width=0.85\textwidth]{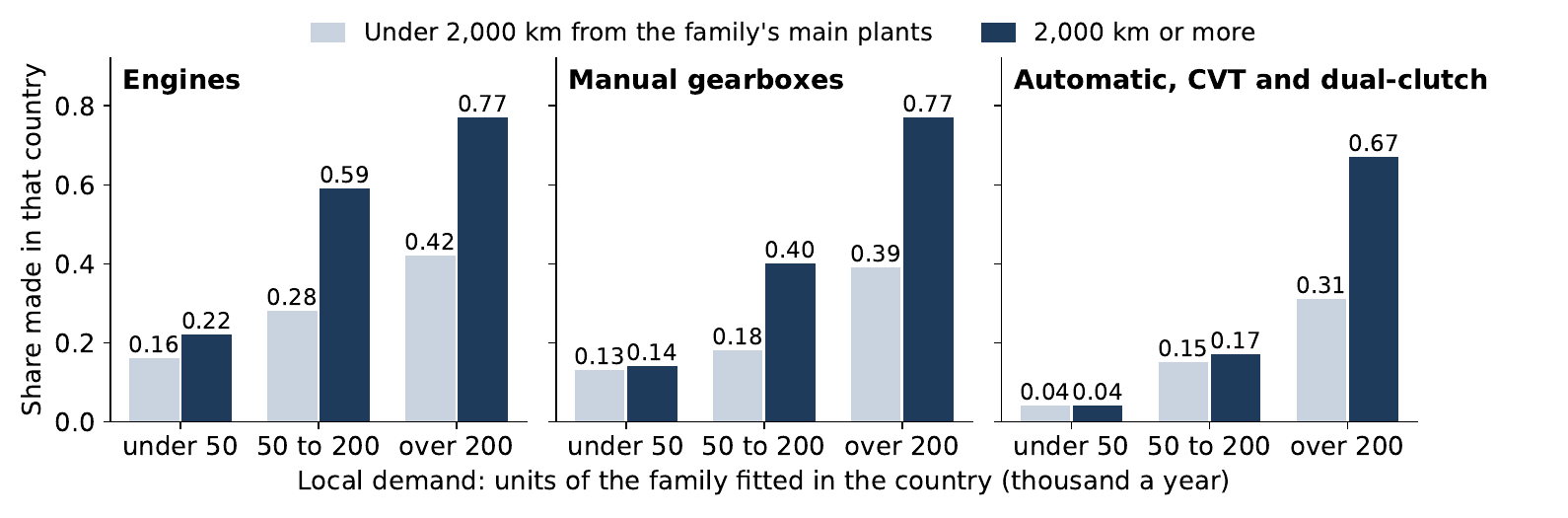}
\caption{Families Made in a Country Where They Are Fitted, by Local Demand and Distance} \label{fig:scale}
\caption*{\textit{Notes:} Share of families made in a country where they are fitted, by local demand (units of the family fitted in cars assembled there) and by distance from the family's main plants. Family $\times$ country $\times$ year, countries other than the family's main country. Not an event study. }
\end{figure}

\paragraph{{Why automatic transmissions are made in so few places.}} Either their transport costs are low, or the minimum efficient scale of a family line exceeds the demand of a single market. The two act at different margins, and three pieces of evidence tell them apart. First, for families made in two or more plants, the fixed costs are sunk, so how an assembly plant splits its purchases among those plants shows what distance costs. A Poisson regression of the units a plant sends to an assembly plant on their log distance, with fixed effects for what each assembly plant needs and what each plant makes, gives an elasticity of $-1.10$ (s.e.\ 0.09) for engines and $-1.12$ (s.e.\ 0.16) for automatic transmissions. Transport costs are thus not lower. Second, family lines are large. The median peak output of a family at a plant is 205 thousand units a year for automatic transmissions, against 91 thousand for engines. Third, we look at where a family line opens. Figure~\ref{fig:scale} takes every family and every country where it is fitted in cars, other than the family's main country, and asks whether the family is also made there. Below about 200 thousand units of local demand, distance does not bring a local automatic transmission family line (far minus near, $-0.01$ under 50 thousand and $+0.02$ from 50 to 200 thousand, against $+0.06$ and $+0.31$ for engines). Above it, distance brings a local family line as often as for engines ($+0.36$ against $+0.35$, far minus near). This is part (v) of Proposition~\ref{prop:dup}, where the minimum efficient scale $m$ exceeds local demand $D_B$.

\paragraph{{A transmission program crowds out the plant's other output.}} The scale of transmission production also shows at the plant that takes on a program. A transmission plant that starts a transmission program after a flood (21 plants of 11 owners) loses output on the rest of the plant, against transmission plants that start a never-flooded program in the same year, run as many transmission family lines and stand in other countries. Before the program starts, the two move together ($+0.078$, s.e.\ 0.138, in years $-4$ to $-2$). After it, the plant's output without the new transmission programs is lower by 0.404 (s.e.\ 0.174) in years 0--2 and by 0.355 (s.e.\ 0.243) in years 3--6, also relative to the vehicle output of the assembly countries it supplies ($-0.329$, s.e.\ 0.170, in years 0--2), and more of its other transmission programs move to another plant ($+0.101$, s.e.\ 0.052, in years 3--6). Against all transmission plants starting a program that year, whatever their country and number of transmission family lines, the drop is about half as large and imprecise ($-0.226$, s.e.\ 0.176, in years 0--2). No such crowding out appears at engine plants or assembly plants that receive a product after a flood. This suggests that transmission production comes in units that are large relative to what a plant can absorb next to its other products.

\subsection{Role of Inventories} \label{sec:stock}
An alternative explanation for why transmissions are not duplicated is that inventories already protect assembly plants against disruptions. S\&P records engines and transmissions only once they are fitted in a car, so stocks of components cannot be observed. We can observe how far the components travel to the assembly plant, which determines the stock in transit. A transmission shipped across an ocean is weeks on the way, and the assembly plant holds weeks of it. We split the flooded programs by how they were shipped in $F-1$. They are \textit{fitted nearby} if half of their volume or more goes to assembly plants within 100 km, and \textit{shipped} otherwise. Each group is compared with controls of the same group, adding the group to the cell $g$ in equation~\eqref{eq:theta}. If stocks in transit stand in for a second plant, programs shipped far should keep their output in the flood year, add a second plant less often, and account for the gap between engines and transmissions.

\begin{figure}[t]
\centering
\includegraphics[width=0.85\textwidth]{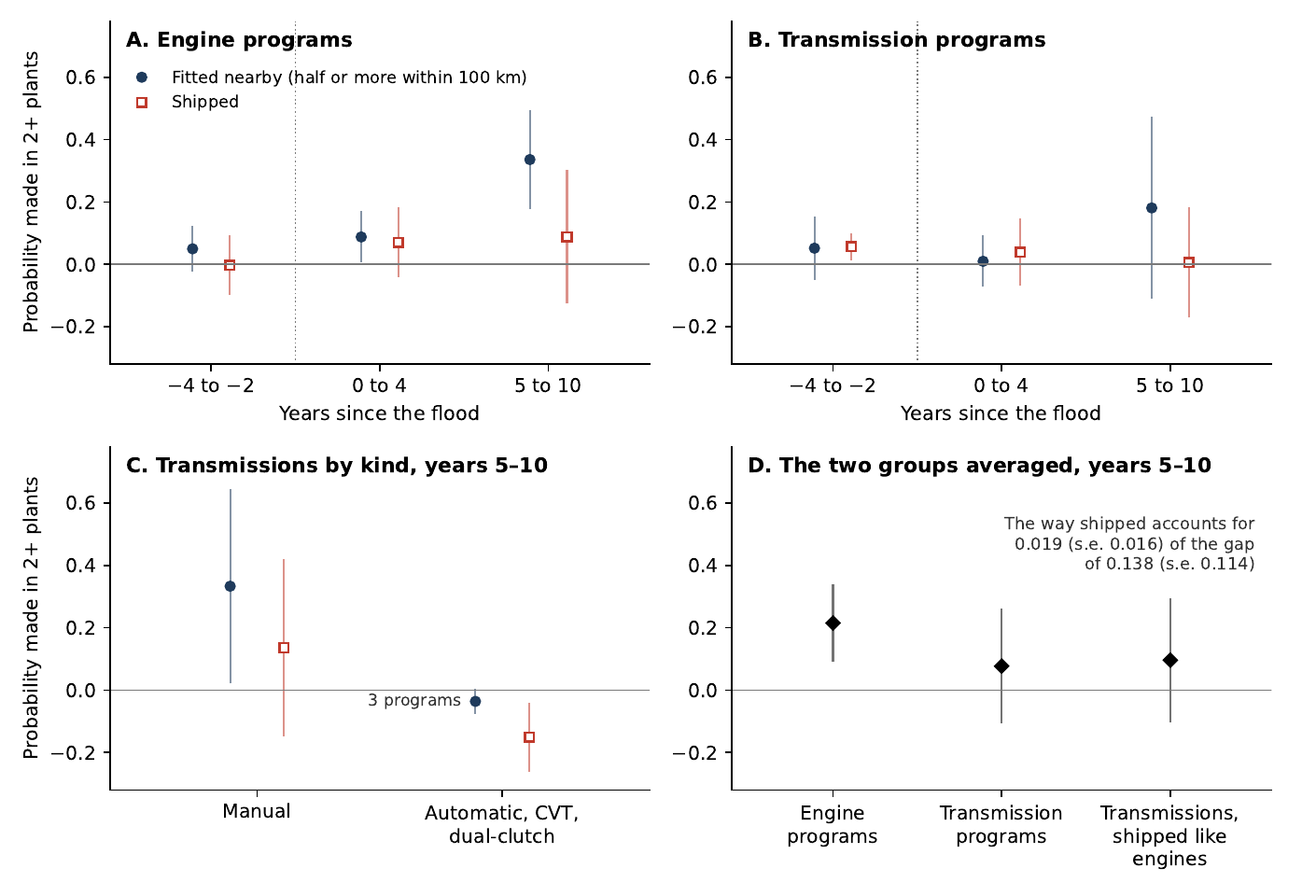}
\caption{Flooded Programs by How They Are Shipped} \label{fig:stock}
\caption*{\textit{Notes:} One outcome in every panel: made in two or more plants, flooded programs minus controls from other countries shipped the same way. Fitted nearby: half of the program's volume or more fitted within 100 km of the component plant in $F-1$. Shipped: otherwise. A and B: by window of years since the flood. C: transmission programs by kind, years 5--10. D: each product's two groups averaged with its own shares of programs fitted nearby (0.51 for engine programs, 0.40 for transmission programs), and the transmission groups averaged with the engines' shares, years 5--10. 95\% intervals, owner bootstrap. Estimates in Table~\ref{tab:est}. }
\end{figure}

Transmissions travel farther. Of the flooded programs, 40\% of the transmission programs are fitted nearby, against 51\% of the engine programs, and the median distance to the assembly plant is 1,197 km against 562 km (933 km for manual gearboxes and 1,596 km for automatic transmissions). Programs shipped far add a second plant less often, for both products (panels A and B of Figure~\ref{fig:stock}). However, the way they are shipped explains little of why transmissions are not duplicated. Averaging each product's two groups with its own shares gives $+0.215$ for engines and $+0.077$ for transmissions (made in two or more plants, years 5--10), and giving transmissions the engines' shares raises theirs only to $+0.096$ (panel D). The way of shipping thus accounts for 0.019 (s.e.\ 0.016) of a gap of 0.138 (s.e.\ 0.114). Manual gearboxes respond like engines, whether fitted nearby or shipped, while automatic transmissions do not respond either way (panel C).

These results come with caveats. Output in the flood year does not fall even for programs fitted nearby ($-0.002$, s.e.\ 0.065, of the output of $F-1$ for engine programs and $-0.009$, s.e.\ 0.084, for transmission programs), so that yearly data hardly show the stoppage a stock of components would have to cover (the flooded plant's own output shows only a temporary loss in the flood year, Section~\ref{sec:engines}). The stoppage shows instead in the stock of finished vehicles, as we see below. Only three automatic programs fitted nearby remain by year 5, so that the comparison that would separate stocks in transit from the size of a family line has no sample. Programs fitted nearby are not a random group, as they are more often at integrated sites, in Japan, and made in one plant. Shipped transmission programs, like transmission programs as a whole, already differ from their controls before the flood ($+0.057$, s.e.\ 0.022, in years $-4$ to $-2$). Finally, with controls of the same group, the two transmission groups average to more than the transmission estimate of Section~\ref{sec:products} ($+0.077$ against $+0.026$).

A second stock stands between any stoppage upstream and a lost sale, namely finished vehicles waiting to be sold. Unlike the stock of components, it can be measured, as the cumulated world production of a vehicle program minus its cumulated world sales. We use the programs whose lifetime sales are within 5\% of their lifetime production, which make up 65\% of production, and among them the programs that have ended, which make up about 30\% of output. A small mismatch between the two sources accumulates over a program's life. For programs that have ended the true final stock is zero, so we rescale their sales to lifetime production. Vehicles are sold from about 55 days of stock in mid-life, most of it on the way to the buyer: 16 days for programs exported less than 5\% and 90 days for programs exported more than 75\%. This stock is the same behind both components. The vehicles an engine program goes into hold 57 days of sales, and those a transmission program goes into 58 (output-weighted medians, 2004--2023), so thicker stocks downstream do not explain why transmissions are not duplicated. It is also where a stoppage shows. When a model's plant is stopped by a flood at a supplier, its sales hold and the stock falls by 25 days (s.e.\ 10) in the flood year, and it is not rebuilt ($-36$ days, s.e.\ 12, in years 2--4). It does not tell which flooded engine programs add a plant, as programs whose vehicles hold less stock than the median respond as those holding more ($-0.05$, s.e.\ 0.15, made in two or more plants, years 5--10). The stock is measured by program rather than by plant, because the sales data attribute vehicles to plants by a rule of their own, and the corrected stock covers about 30\% of output. Its level is not comparable across countries. In China and India, where sales appear to be recorded on delivery to the dealer, it is 12 to 15 days.

\subsection{Role of Insurance and Flood Defences} \label{sec:ext}
A firm can also insure the plant or protect the site. If these are substitutes for a second plant, as in part (vi) of Proposition~\ref{prop:dup}, a flood should bring fewer second plants where insurance is cheap and where the plant is protected.

\paragraph{{A different flood measure.}} Documented floods are too few to split by country characteristics, so these tests date the flood by the first \textit{flood signal} at a plant of the program, i.e., a flood seen by satellite within 5 km or a documented direct hit (1,138 engine programs, 2000--2021). Controls come from the same World Bank region for insurance, since country cells leave too few programs, and from the same country for defences. For defences, the comparison of interest is between programs flooded at defended and at undefended plants. Same-country controls hold the country's flood policy and insurance market fixed. Learning within the country, which makes such controls a poor comparison for the main estimate (Section~\ref{sec:engines}), shifts the controls of both groups alike and largely cancels in their difference. With controls from other countries of the same region, the gap is larger (Appendix~\ref{app:def}).

\begin{figure}[t]
\centering
\includegraphics[width=0.85\textwidth]{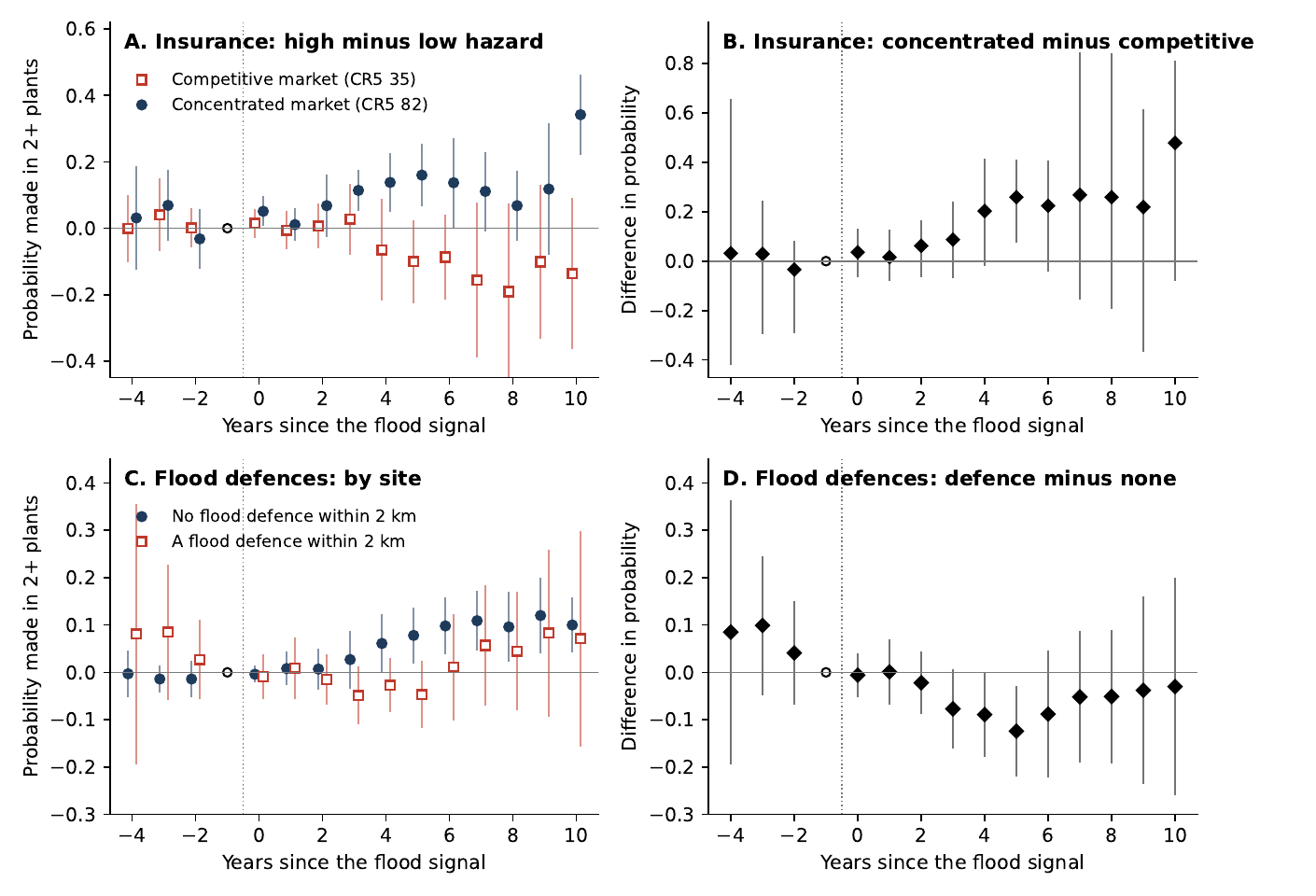}
\caption{Insurance and Flood Defences after a Flood Signal} \label{fig:ext}
\caption*{\textit{Notes:} Each row has two groups on the left and their difference on the right. A: engine programs, the response of programs flooded at high-hazard plants minus that of programs at low-hazard plants, from equation~\eqref{eq:cr5}, at a competitive (CR5 of 35) and at a concentrated (CR5 of 82) insurance market, 95\% intervals from country-clustered standard errors. B: the difference between the two series of panel A, 95\% intervals from the wild cluster bootstrap. C: engine and transmission programs made in two or more plants, by whether a flood defence is mapped within 2 km of the flooded plant (within 5 km in Figure~\ref{fig:def_5km}). D: their difference, 95\% intervals, owner bootstrap. Years $-4$ to 10. }
\end{figure}

\paragraph{{Insurance.}} We ask whether firms rely more on a second plant where insurance is expensive. We measure two things at the flooded plant, namely its flood risk ($\mathrm{High}_i$, a residual flood hazard of 1\% a year or more, which applies to a fifth of the programs) and how expensive insurance is likely to be, proxied by the concentration of the country's non-life insurance market ($\mathrm{CR5}$). The test has three steps. First, for each flooded program and each year $q$, we compute its own effect, i.e., its change since the year before the signal minus the average change of its controls, chosen as in Section~\ref{sec:design} but within the same region,
\begin{equation}
\hat e_{iq} = \big(Y_{i,F_i+q}-Y_{i,F_i-1}\big) - \frac{1}{|\mathcal{C}_i(q)|}\sum_{j\in\mathcal{C}_i(q)}\big(Y_{j,F_i+q}-Y_{j,F_i-1}\big). \label{eq:eiq}
\end{equation}
Second, we regress these effects across programs, year by year,
\begin{equation}
\hat e_{iq} = b_{0q} + b_{1q}\,\mathrm{High}_i + b_{2q}\,\mathrm{CR5}_{c(i)} + b_{3q}\,\mathrm{High}_i\times\mathrm{CR5}_{c(i)} + \gamma_q' X_i + u_{iq}, \label{eq:cr5}
\end{equation}
with $\mathrm{CR5}$ in tens of points and $X_i$ holding income per head, the year of the signal, and the program's plants and output in the year before, and with standard errors clustered by country. If a second plant substitutes for insurance, the extra response of high-hazard programs should be larger where insurance is dearer, i.e., $b_{3q}>0$. Third, we evaluate the regression at a competitive market ($\mathrm{CR5}=35$, the 25th percentile of the programs) and at a concentrated one ($\mathrm{CR5}=82$, the 75th percentile). Panel A of Figure~\ref{fig:ext} plots, for each year, how much more programs flooded at high-hazard plants respond than programs at low-hazard plants, $b_{1q}+b_{3q}\,\mathrm{CR5}$, at the two levels. Panel B plots the difference between the two lines. The programs sit in two clusters of countries, with $\mathrm{CR5}$ around 25--35 and around 75--86, so that the regression in effect compares these two groups.

Before the signal, the two lines of panel A coincide. After it, in concentrated markets, programs flooded at high-hazard plants add a second plant more often than those at low-hazard plants, by 0.160 (s.e.\ 0.048) in year 5. In competitive markets they do not ($-0.100$, s.e.\ 0.064, in year 5). Over years 5--10, the difference (panel B) is 0.214 (s.e.\ 0.078, with a wild-bootstrap interval from 0.012 to 0.413), which corresponds to $b_3=0.046$ per ten points ($p=0.041$). In years 0--4, it is $b_3=0.014$ per ten points ($p=0.025$).

This evidence is fragile. There are 89 high-hazard programs in 23 countries in years 5--10, and those in concentrated markets come from five countries. With comparison groups inside the country, the coefficient is 0.038 (s.e.\ 0.027, $p=0.32$), and four groups instead of continuous concentration give 0.222 (s.e.\ 0.154). Transmissions have the opposite sign, and the depth of the insurance market shows nothing. Stacking all event years in one regression with country clusters across years, the pre-trend is flat (joint test of years $-4$ to $-2$, wild-bootstrap $p=0.93$), but the joint test that every coefficient after the signal is zero is not rejected (wild-bootstrap $p=0.52$ for years 0--10 and $p=0.18$ for years 5--10). Weighting program-years instead of programs, the coefficient on years 5--10 is 0.056 (s.e.\ 0.021, wild-cluster-bootstrap $p=0.060$).

\paragraph{{Flood defences.}} For each flooded plant, we check whether a dyke, levee, flood wall or seawall is mapped in OpenStreetMap within 2 km, close enough to protect the plant. This is the case for 31 of the 354 flooded plants and 154 of the 1,756 flooded programs. Panel C of Figure~\ref{fig:ext} shows the response with and without such a defence, against controls of the same country, and panel D shows their difference. Programs flooded at a plant with no defence nearby add plants steadily, by 0.100 (s.e.\ 0.029) in years 5--10. Those flooded at a defended plant add none in the first six years after the signal, and the gap between the two groups reaches $-0.124$ (s.e.\ 0.049) in year 5. Over years 0--4 the difference is $-0.038$ (s.e.\ 0.030), and over years 5--10 it is $-0.064$ (s.e.\ 0.069), as the defended programs add some plants after year 6. Appendix~\ref{app:def} shows that the gap comes from the flooded programs themselves rather than from their controls, explains the late rise, and reports the comparison within 5 km and under other definitions of protection, under which the gap is smaller or reversed. A second fact points the same way. The additions of engine programs after documented floods go to plants with a lower public protection standard than the flooded plant, with 63\% of them going to a worse-protected plant and 26\% to a better-protected one.

Both signs fit a trade-off, since firms duplicate more where insurance is expensive and where no defence protects the plant. Neither result is robust, however, and both rest on satellite flood signals.

\section{Conclusion} \label{sec:concl}
This paper studies how multinational firms adapt their production networks to flood risk, using the global car industry and a hand-verified record of plant floods. Firms respond to floods by duplicating production rather than by moving it. After a flood, engine programs are added at a second plant, usually abroad, while the flooded plant's output, after a temporary loss in the flood year, shows no persistent decline. The second plant supplies the vehicles assembled nearby, which keeps it busy in normal times and gives the firm a second site that already makes the engine, a backup in case it is needed.

This form of adaptation is inexpensive only in some parts of the supply chain, and firms use it only there. Engine programs can draw on existing plants and on existing demand within the firm, as the second plant is typically a growing plant that already makes the engine family, close to the cars it will supply. A vehicle model version has suitable capacity but lacks local demand. After a flood, it obtains a second plant only for a few years, for new markets, and it expands more slowly into new countries. An automatic transmission lacks suitable capacity, because a single family line exceeds the demand near most plants and serves several countries. Stocks in transit explain little of these differences, and the stock of finished vehicles is the same behind engines and transmissions.

These findings have three implications. First, studies that measure the cost of floods by their effect on output may miss the main adjustment, a change in the structure of production. Second, the costs of adaptation, not only the risk, determine where it happens, so that the most scale-intensive parts of supply chains remain concentrated and exposed. Third, private redundancy interacts with the protection that markets and governments provide: firms add fewer second plants where a flood defence protects the plant and more where insurance is likely to be expensive. If these patterns hold, public flood defences and private second plants address the same risk, so that policies that protect sites and policies that encourage a second source should be designed together, and the exposure that remains lies where plants are neither protected nor cheap to duplicate.

Our evidence has limits. The long-run estimates rest on few owners, the floods of 2011 play a large role, and annual data cannot show whether the second site covers for the first during a stoppage of a few weeks. The tests on insurance and flood defences rest on satellite flood signals, and the defences mapped in OpenStreetMap are not dated. Monthly production data, and data on insurance and protection at the plant, would let us address these questions directly.

\clearpage
\bibliographystyle{plainnat}
\bibliography{references,references_methods,references_duplication}

\begin{thebibliography}{55}
\providecommand{\natexlab}[1]{#1}
\providecommand{\url}[1]{\texttt{#1}}
\expandafter\ifx\csname urlstyle\endcsname\relax
  \providecommand{\doi}[1]{doi: #1}\else
  \providecommand{\doi}{doi: \begingroup \urlstyle{rm}\Url}\fi

\bibitem[Antr{\`a}s et~al.(2017)Antr{\`a}s, Fort, and Tintelnot]{AFT17}
Pol Antr{\`a}s, Teresa~C Fort, and Felix Tintelnot.
\newblock The margins of global sourcing: Theory and evidence from us firms.
\newblock \emph{American Economic Review}, 107\penalty0 (9):\penalty0
  2514--2564, 2017.

\bibitem[Antr{\`a}s et~al.(2024)Antr{\`a}s, Fadeev, Fort, and
  Tintelnot]{antras2024}
Pol Antr{\`a}s, Evgenii Fadeev, Teresa~C Fort, and Felix Tintelnot.
\newblock Exporting, global sourcing, and multinational activity: Theory and
  evidence from the united states.
\newblock \emph{Review of Economics and Statistics}, pages 1--48, 2024.

\bibitem[Arkolakis et~al.(2018)Arkolakis, Ramondo, Rodr{\'\i}guez-Clare, and
  Yeaple]{ARRY18}
Costas Arkolakis, Natalia Ramondo, Andr{\'e}s Rodr{\'\i}guez-Clare, and Stephen
  Yeaple.
\newblock Innovation and production in the global economy.
\newblock \emph{American Economic Review}, 108\penalty0 (8):\penalty0
  2128--2173, 2018.

\bibitem[Balboni et~al.(2024)Balboni, Boehm, and Waseem]{balboni2023}
C.~Balboni, J.~Boehm, and M.~Waseem.
\newblock Firm adaptation in production networks: evidence from extreme weather
  events in pakistan.
\newblock \emph{PEDL Working Paper}, 2024.

\bibitem[Baldwin and Freeman(2022)]{baldwin2022}
Richard Baldwin and Rebecca Freeman.
\newblock Risks and global supply chains: What we know and what we need to
  know.
\newblock \emph{Annual Review of Economics}, 14:\penalty0 153--180, 2022.

\bibitem[Barrot and Sauvagnat(2016)]{barrot2016}
Jean-Noël Barrot and Julien Sauvagnat.
\newblock Input specificity and the propagation of idiosyncratic shocks in
  production networks.
\newblock \emph{The Quarterly Journal of Economics}, 131\penalty0 (3):\penalty0
  1543--1592, 2016.
\newblock \doi{10.1093/qje/qjw018}.

\bibitem[Bilal and Rossi-Hansberg(2023)]{bilal2023}
Adrien Bilal and Esteban Rossi-Hansberg.
\newblock Anticipating climate change across the united states.
\newblock Technical report, National Bureau of Economic Research, 2023.

\bibitem[Boehm et~al.(2019)Boehm, Flaaen, and Pandalai-Nayar]{boehm2019}
Christoph~E. Boehm, Aaron Flaaen, and Nitya Pandalai-Nayar.
\newblock Input linkages and the transmission of shocks: Firm-level evidence
  from the 2011 tōhoku earthquake.
\newblock \emph{The Review of Economics and Statistics}, 101\penalty0
  (1):\penalty0 60--75, 2019.
\newblock \doi{10.1162/rest_a_00739}.

\bibitem[Borusyak et~al.(2024)Borusyak, Jaravel, and Spiess]{BJS2024}
Kirill Borusyak, Xavier Jaravel, and Jann Spiess.
\newblock Revisiting event-study designs: robust and efficient estimation.
\newblock \emph{Review of Economic Studies}, 91\penalty0 (6):\penalty0
  3253--3285, 2024.

\bibitem[Boustan et~al.(2012)Boustan, Kahn, and Rhode]{boustan2012moving}
Leah~Platt Boustan, Matthew~E Kahn, and Paul~W Rhode.
\newblock Moving to higher ground: Migration response to natural disasters in
  the early twentieth century.
\newblock \emph{American Economic Review}, 102\penalty0 (3):\penalty0 238--244,
  2012.

\bibitem[Brainard(1997)]{brainard1997}
S.~Lael Brainard.
\newblock An empirical assessment of the proximity-concentration trade-off
  between multinational sales and trade.
\newblock \emph{American Economic Review}, 87\penalty0 (4):\penalty0 520--544,
  1997.

\bibitem[Burgess et~al.(2017)Burgess, Deschênes, Donaldson, and
  Greenstone]{burgess2017weather}
Robin Burgess, Olivier Deschênes, Dave Donaldson, and Michael Greenstone.
\newblock Weather and death in india: Mechanisms and implications for climate
  change.
\newblock \emph{American Economic Journal: Applied Economics}, 6\penalty0
  (3):\penalty0 1--36, 2017.

\bibitem[Burke et~al.(2015)Burke, Hsiang, and Miguel]{burke2015global}
Marshall Burke, Solomon~M. Hsiang, and Edward Miguel.
\newblock Global non-linear effect of temperature on economic production.
\newblock \emph{Nature}, 527:\penalty0 235--239, 2015.

\bibitem[Callaway and Sant’Anna(2021)]{CS2021}
Brantly Callaway and Pedro~HC Sant’Anna.
\newblock Difference-in-differences with multiple time periods.
\newblock \emph{Journal of econometrics}, 225\penalty0 (2):\penalty0 200--230,
  2021.

\bibitem[Carleton et~al.(2022)Carleton, Jina, Delgado, Greenstone, Houser,
  Hsiang, Hultgren, Kopp, McCusker, Nath, Rising, Rode, Seo, Viaene, Yuan, and
  Zhang]{carleton2022}
Tamma Carleton, Amir Jina, Michael Delgado, Michael Greenstone, Trevor Houser,
  Solomon Hsiang, Andrew Hultgren, Robert~E Kopp, Kelly~E McCusker, Ishan Nath,
  James Rising, Ashwin Rode, Hee~Kwon Seo, Arvid Viaene, Jiacan Yuan, and
  Alice~Tianbo Zhang.
\newblock Valuing the global mortality consequences of climate change
  accounting for adaptation costs and benefits*.
\newblock \emph{The Quarterly Journal of Economics}, 137\penalty0 (4):\penalty0
  2037--2105, 04 2022.
\newblock ISSN 0033-5533.
\newblock \doi{10.1093/qje/qjac020}.
\newblock URL \url{https://doi.org/10.1093/qje/qjac020}.

\bibitem[Carvalho et~al.(2021)Carvalho, Nirei, Saito, and
  Tahbaz-Salehi]{carvalho2021}
Vasco~M. Carvalho, Makoto Nirei, Yukiko~U. Saito, and Alireza Tahbaz-Salehi.
\newblock Supply chain disruptions: Evidence from the {G}reat {E}ast {J}apan
  {E}arthquake.
\newblock \emph{Quarterly Journal of Economics}, 136\penalty0 (2):\penalty0
  1255--1321, 2021.

\bibitem[Castro-Vincenzi(2024)]{CV2024}
Juanma Castro-Vincenzi.
\newblock Climate hazards and resilience in the global car industry.
\newblock Working paper, Harvard University, 2024.
\newblock Version of 4 June 2024. Revise and resubmit, \textit{American
  Economic Review}.

\bibitem[Castro-Vincenzi et~al.(2024)Castro-Vincenzi, Khanna, Morales, and
  Pandalai-Nayar]{castro2024}
Juanma Castro-Vincenzi, Gaurav Khanna, Nicolas Morales, and Nitya
  Pandalai-Nayar.
\newblock Weathering the storm: Supply chains and climate risk.
\newblock NBER Working Paper 32218, National Bureau of Economic Research, 2024.
\newblock Revise and resubmit, \textit{Review of Economics and Statistics}.

\bibitem[De~Chaisemartin and d'Haultfoeuille(2024)]{CH2024}
Cl{\'e}ment De~Chaisemartin and Xavier d'Haultfoeuille.
\newblock Difference-in-differences estimators of intertemporal treatment
  effects.
\newblock \emph{Review of Economics and Statistics}, pages 1--45, 2024.

\bibitem[Delforge et~al.(2025)Delforge, Wathelet, Below, Sofia, Tonnelier, van
  Loenhout, and Speybroeck]{EMDAT}
Damien Delforge, Valentin Wathelet, Regina Below, Cinzia~Lanfredi Sofia, Margo
  Tonnelier, Joris~AF van Loenhout, and Niko Speybroeck.
\newblock Em-dat: the emergency events database.
\newblock \emph{International Journal of Disaster Risk Reduction}, page 105509,
  2025.

\bibitem[Dell et~al.(2012)Dell, Jones, and Olken]{dell2012temperature}
Melissa Dell, Benjamin~F. Jones, and Benjamin~A. Olken.
\newblock Temperature shocks and economic growth: Evidence from the last half
  century.
\newblock \emph{American Economic Journal: Macroeconomics}, 4\penalty0
  (3):\penalty0 66--95, 2012.

\bibitem[Deschênes and Greenstone(2007)]{deschenes2007economic}
Olivier Deschênes and Michael Greenstone.
\newblock The economic impacts of climate change: Evidence from agricultural
  output and random fluctuations in weather.
\newblock \emph{American Economic Review}, 97\penalty0 (1):\penalty0 354--385,
  2007.

\bibitem[Desmet et~al.(2021)Desmet, Kopp, Kulp, Nagy, Oppenheimer,
  Rossi-Hansberg, and Strauss]{desmet2018}
Klaus Desmet, Robert~E. Kopp, Scott~A. Kulp, Dávid~Krisztián Nagy, Michael
  Oppenheimer, Esteban Rossi-Hansberg, and Benjamin~H. Strauss.
\newblock Evaluating the economic cost of coastal flooding.
\newblock \emph{American Economic Journal: Macroeconomics}, 13\penalty0
  (2):\penalty0 444–86, April 2021.
\newblock \doi{10.1257/mac.20180366}.
\newblock URL \url{https://www.aeaweb.org/articles?id=10.1257/mac.20180366}.

\bibitem[Ehrlich and Becker(1972)]{ehrlich1972}
Isaac Ehrlich and Gary~S. Becker.
\newblock Market insurance, self-insurance, and self-protection.
\newblock \emph{Journal of Political Economy}, 80\penalty0 (4):\penalty0
  623--648, 1972.

\bibitem[Elliott et~al.(2022)Elliott, Golub, and Leduc]{EGL2022}
Matthew Elliott, Benjamin Golub, and Matthew~V. Leduc.
\newblock Supply network formation and fragility.
\newblock \emph{American Economic Review}, 112\penalty0 (8):\penalty0
  2701--2747, 2022.

\bibitem[Fioretti et~al.(2024)Fioretti, He, and Tamayo]{FHT2024}
Michele Fioretti, Junnan He, and Jorge Tamayo.
\newblock Capacity, technology portfolios, and the paradox of concentration.
\newblock arXiv:2407.03504, 2024.

\bibitem[Grossman et~al.(2023)Grossman, Helpman, and Lhuillier]{GHL2023}
Gene~M. Grossman, Elhanan Helpman, and Hugo Lhuillier.
\newblock Supply chain resilience: Should policy promote international
  diversification or reshoring?
\newblock \emph{Journal of Political Economy}, 131\penalty0 (12):\penalty0
  3462--3496, 2023.

\bibitem[Grossman et~al.(2024)Grossman, Helpman, and Sabal]{grossman2024}
Gene~M Grossman, Elhanan Helpman, and Alejandro Sabal.
\newblock Optimal resilience in multitier supply chains.
\newblock \emph{The Quarterly Journal of Economics}, 139\penalty0 (4):\penalty0
  2377--2425, 2024.

\bibitem[Gu and Hale(2023)]{gu2023}
Grace~Weishi Gu and Galina Hale.
\newblock Climate risks and fdi.
\newblock \emph{Journal of International Economics}, 146:\penalty0 103731,
  2023.
\newblock \doi{10.1016/j.jinteco.2023.103731}.

\bibitem[Guiteras et~al.(2015)Guiteras, Jina, and
  Mobarak]{guiteras2015satellites}
Raymond Guiteras, Amir Jina, and A~Mushfiq Mobarak.
\newblock Satellites, self-reports, and submersion: exposure to floods in
  bangladesh.
\newblock \emph{American Economic Review}, 105\penalty0 (5):\penalty0 232--236,
  2015.

\bibitem[Head and Mayer(2019)]{HM19}
Keith Head and Thierry Mayer.
\newblock Brands in motion: How frictions shape multinational production.
\newblock \emph{American Economic Review}, 109\penalty0 (9):\penalty0
  3073--3124, 2019.

\bibitem[Head et~al.(2023)Head, Mayer, and Melitz]{HMM2022}
Keith Head, Thierry Mayer, and Marc Melitz.
\newblock The unintended consequences of high regional content requirements.
\newblock In Lili~Yan Ing and Gene~M. Grossman, editors, \emph{Local Content
  Requirements: Promises and Pitfalls}, Routledge-ERIA Studies in Development
  Economics, chapter~4. Routledge, London, 2023.
\newblock \doi{10.4324/9781003415794-4}.

\bibitem[Head et~al.(2026)Head, Mayer, Melitz, and Yang]{HMMY2024}
Keith Head, Thierry Mayer, Marc Melitz, and Chenying Yang.
\newblock Industrial policies for multi-stage production: The battle for
  battery-powered vehicles.
\newblock NBER Working Paper 34884, National Bureau of Economic Research, 2026.

\bibitem[Helpman et~al.(2004)Helpman, Melitz, and Yeaple]{HMY2004}
Elhanan Helpman, Marc~J. Melitz, and Stephen~R. Yeaple.
\newblock Export versus {FDI} with heterogeneous firms.
\newblock \emph{American Economic Review}, 94\penalty0 (1):\penalty0 300--316,
  2004.

\bibitem[Hossain(2020)]{hossain2020}
Farzana Hossain.
\newblock Creative destruction or just destruction? effects of floods on
  manufacturing establishments in india.
\newblock \emph{SSRN Working Paper}, 2020.

\bibitem[Hsiang and Jina(2014)]{hsiang2014climate}
Solomon~M. Hsiang and Amir~S. Jina.
\newblock The causal effect of environmental catastrophe on long-run economic
  growth: Evidence from 6,700 cyclones.
\newblock \emph{NBER Working Paper}, 20352, 2014.

\bibitem[Jia et~al.(forthcoming)Jia, Ma, and Xie]{jia2022}
Ruixue Jia, Xiao Ma, and Victoria~Wenxin Xie.
\newblock Expecting floods: Firm entry, employment, and aggregate implications.
\newblock Technical report, forthcoming.

\bibitem[Kettner et~al.(2021)Kettner, Brakenridge, Schumann, and Shen]{DFO}
AJ~Kettner, G~Robert Brakenridge, Guy~JP Schumann, and X~Shen.
\newblock Dfo—flood observatory.
\newblock In \emph{Earth observation for flood applications}, pages 147--164.
  Elsevier, 2021.

\bibitem[Kocornik-Mina et~al.(2020)Kocornik-Mina, McDermott, Michaels, and
  Rauch]{kocornik2020flooded}
Adriana Kocornik-Mina, Thomas K.~J. McDermott, Guy Michaels, and Ferdinand
  Rauch.
\newblock Flooded cities.
\newblock \emph{American Economic Journal: Applied Economics}, 12\penalty0
  (2):\penalty0 35--66, 2020.

\bibitem[Kopytov et~al.(2024)Kopytov, Mishra, Nimark, and
  Taschereau-Dumouchel]{kopytov2024}
Alexandr Kopytov, Bineet Mishra, Kristoffer Nimark, and Mathieu
  Taschereau-Dumouchel.
\newblock Endogenous production networks under supply chain uncertainty.
\newblock \emph{Econometrica}, 92\penalty0 (5):\penalty0 1621--1659, 2024.
\newblock \doi{10.3982/ECTA20629}.

\bibitem[Lafrogne-Joussier et~al.(2022)Lafrogne-Joussier, Martin, and
  Mejean]{lafrogne2022}
Raphael Lafrogne-Joussier, Julien Martin, and Isabelle Mejean.
\newblock Supply shocks in supply chains: Evidence from the early lockdown in
  china.
\newblock Technical Report DP16813, Centre for Economic Policy Research, 2022.
\newblock URL \url{https://ssrn.com/abstract=4026681}.

\bibitem[Leiter et~al.(2009)Leiter, Oberhofer, and Raschky]{leiter2009}
Andrea~M. Leiter, Harald Oberhofer, and Paul~A. Raschky.
\newblock Creative disasters? flooding effects on capital, labor and
  productivity within european firms.
\newblock \emph{Environmental and Resource Economics}, 43\penalty0
  (3):\penalty0 333--350, 2009.

\bibitem[Masson-Delmotte et~al.(2021)Masson-Delmotte, Zhai, Pirani, Connors,
  Péan, Berger, Caud, Chen, Goldfarb, Gomis, Huang, Leitzell, Lonnoy,
  Matthews, Maycock, Waterfield, Yelekçi, Yu, and Zhou]{IPCC2021}
V.~Masson-Delmotte, P.~Zhai, A.~Pirani, S.L. Connors, C.~Péan, S.~Berger,
  N.~Caud, Y.~Chen, L.~Goldfarb, M.I. Gomis, M.~Huang, K.~Leitzell, E.~Lonnoy,
  J.B.R. Matthews, T.K. Maycock, T.~Waterfield, O.~Yelekçi, R.~Yu, and
  B.~Zhou.
\newblock \emph{Climate Change 2021: The Physical Science Basis. Contribution
  of Working Group I to the Sixth Assessment Report of the Intergovernmental
  Panel on Climate Change}.
\newblock Cambridge University Press, Cambridge, United Kingdom and New York,
  NY, USA, 2021.
\newblock \doi{10.1017/9781009157896}.

\bibitem[McFadden(1974)]{mcfadden1974}
Daniel McFadden.
\newblock Conditional logit analysis of qualitative choice behavior.
\newblock In Paul Zarembka, editor, \emph{Frontiers in Econometrics}, pages
  105--142. Academic Press, New York, 1974.

\bibitem[Noth and Rehbein(2019)]{noth2019}
Felix Noth and Oliver Rehbein.
\newblock Badly hurt? natural disasters and direct firm effects.
\newblock \emph{Finance Research Letters}, 28:\penalty0 254--258, 2019.

\bibitem[Pankratz and Schiller(2024)]{pankratz2023}
Nora M.~C. Pankratz and Christoph~M. Schiller.
\newblock Climate change and adaptation in global supply-chain networks.
\newblock \emph{The Review of Financial Studies}, 37\penalty0 (6):\penalty0
  1729--1777, 2024.
\newblock \doi{10.1093/rfs/hhad027}.

\bibitem[Patel(2024)]{patel2024}
Dev Patel.
\newblock Floods.
\newblock \emph{Working Paper}, 2024.

\bibitem[Pelli et~al.(2023)Pelli, Tschopp, Bezmaternykh, and
  Eklou]{pelli2023costs}
Martino Pelli, Jeanne Tschopp, Natalia Bezmaternykh, and Kodjovi~M. Eklou.
\newblock In the eye of the storm: Firms and capital destruction in india.
\newblock \emph{Journal of Urban Economics}, 134:\penalty0 103529, 2023.
\newblock ISSN 0094-1190.
\newblock \doi{https://doi.org/10.1016/j.jue.2022.103529}.
\newblock URL
  \url{https://www.sciencedirect.com/science/article/pii/S009411902200105X}.

\bibitem[Praetorius(2026)]{praetorius2026}
Sophia Praetorius.
\newblock Technology, licensing and multinational production.
\newblock \emph{Job Market Paper}, 2026.

\bibitem[Scussolini et~al.(2016)Scussolini, Aerts, Jongman, Bouwer, Winsemius,
  de~Moel, and Ward]{scussolini2016}
Paolo Scussolini, Jeroen C. J.~H. Aerts, Brenden Jongman, Laurens~M. Bouwer,
  Hessel~C. Winsemius, Hans de~Moel, and Philip~J. Ward.
\newblock {FLOPROS}: An evolving global database of flood protection standards.
\newblock \emph{Natural Hazards and Earth System Sciences}, 16:\penalty0
  1049--1061, 2016.
\newblock \doi{10.5194/nhess-16-1049-2016}.

\bibitem[Somanathan et~al.(2021)Somanathan, Somanathan, Sudarshan, and
  Tewari]{somanathan2021impact}
E.~Somanathan, R.~Somanathan, A.~Sudarshan, and M.~Tewari.
\newblock The impact of temperature on productivity and labor supply: Evidence
  from indian manufacturing.
\newblock \emph{Journal of Political Economy}, 129\penalty0 (6):\penalty0
  1797--1827, 2021.

\bibitem[Tellman et~al.(2021)Tellman, Sullivan, Kuhn, Kettner, Doyle,
  Brakenridge, Erickson, and Slayback]{tellman2021}
Beth Tellman, Jonathan~A Sullivan, Catherine Kuhn, Albert~J Kettner, Colin~S
  Doyle, G~Robert Brakenridge, Tyler~A Erickson, and Daniel~A Slayback.
\newblock Satellite imaging reveals increased proportion of population exposed
  to floods.
\newblock \emph{Nature}, 596\penalty0 (7870):\penalty0 80--86, 2021.

\bibitem[Tintelnot(2017)]{tintelnot2017}
Felix Tintelnot.
\newblock Global production with export platforms.
\newblock \emph{The Quarterly Journal of Economics}, 132\penalty0 (1):\penalty0
  157--209, 2017.

\bibitem[Tomlin(2006)]{tomlin2006}
Brian Tomlin.
\newblock On the value of mitigation and contingency strategies for managing
  supply chain disruption risks.
\newblock \emph{Management Science}, 52\penalty0 (5):\penalty0 639--657, 2006.

\bibitem[Ward et~al.(2020)Ward, Winsemius, Kuzma, et~al.]{aqueduct2020}
Philip~J. Ward, Hessel~C. Winsemius, Samantha Kuzma, et~al.
\newblock Aqueduct floods methodology.
\newblock Technical note, World Resources Institute, Washington, DC, 2020.

\end{thebibliography}

\clearpage
\setcounter{topnumber}{4}
\setcounter{totalnumber}{4}
\renewcommand{\floatpagefraction}{0.5}
\nextappendix{Data} \label{app:data}

\begin{table}[!htp]
\centering
\caption{Production Data, Counts per Year} \label{tab:sum_all}
\footnotesize
\begin{tabular}{lrrrrrr}
\toprule
 & Mean & SD & P25 & P50 & P75 & Max\\
\midrule
Car models & 1,235 & 267 & 1,003 & 1,226 & 1,500 & 1,695\\
Brand owners & 102 & 10.9 & 92 & 100 & 112 & 121\\
Plant operators (vehicle) & 158 & 9.7 & 152 & 156 & 165 & 176\\
Plant operators (engine) & 74.2 & 5.5 & 69 & 74 & 79 & 83\\
Plant operators (transmission) & 108 & 6.8 & 103 & 109 & 113 & 116\\
Production countries (vehicle) & 49.4 & 1.2 & 48 & 49 & 50 & 51\\
Production countries (engine) & 35.7 & 1.1 & 35 & 36 & 37 & 37\\
Production countries (transmission) & 33.7 & 1.9 & 32 & 33 & 36 & 36\\
Plants (vehicle) & 677 & 76.3 & 609 & 682 & 749 & 781\\
Plants (engine) & 296 & 28.4 & 272 & 292 & 324 & 341\\
Plants (transmission) & 228 & 15.4 & 217 & 226 & 242 & 248\\
Platforms & 460 & 43.5 & 419 & 483 & 492 & 516\\
Output per vehicle plant (1,000) & 111 & 10.1 & 102 & 114 & 120 & 127\\
Share of output assembled under license & 0.07 & 0.02 & 0.04 & 0.07 & 0.08 & 0.09\\
\bottomrule
\end{tabular}

\caption*{\textit{Notes:} Statistics across the 25 years 2000--2024 of yearly counts in the April 2025 release, plant-model-years with positive output. Brand owners: firms owning the brands produced. Plant operators: firms operating the plants. Plants: vehicle assembly, engine and transmission plants with positive output in the year. Share of output assembled under license: output of models assembled at a plant of another firm than the brand owner, over total output. Own calculations based on S\&P Global Mobility data.}
\end{table}

\begin{figure}[!htp]
\centering
\caption{Automotive Plants, 2021} \label{fig:plants}
\includegraphics[width=0.9\textwidth]{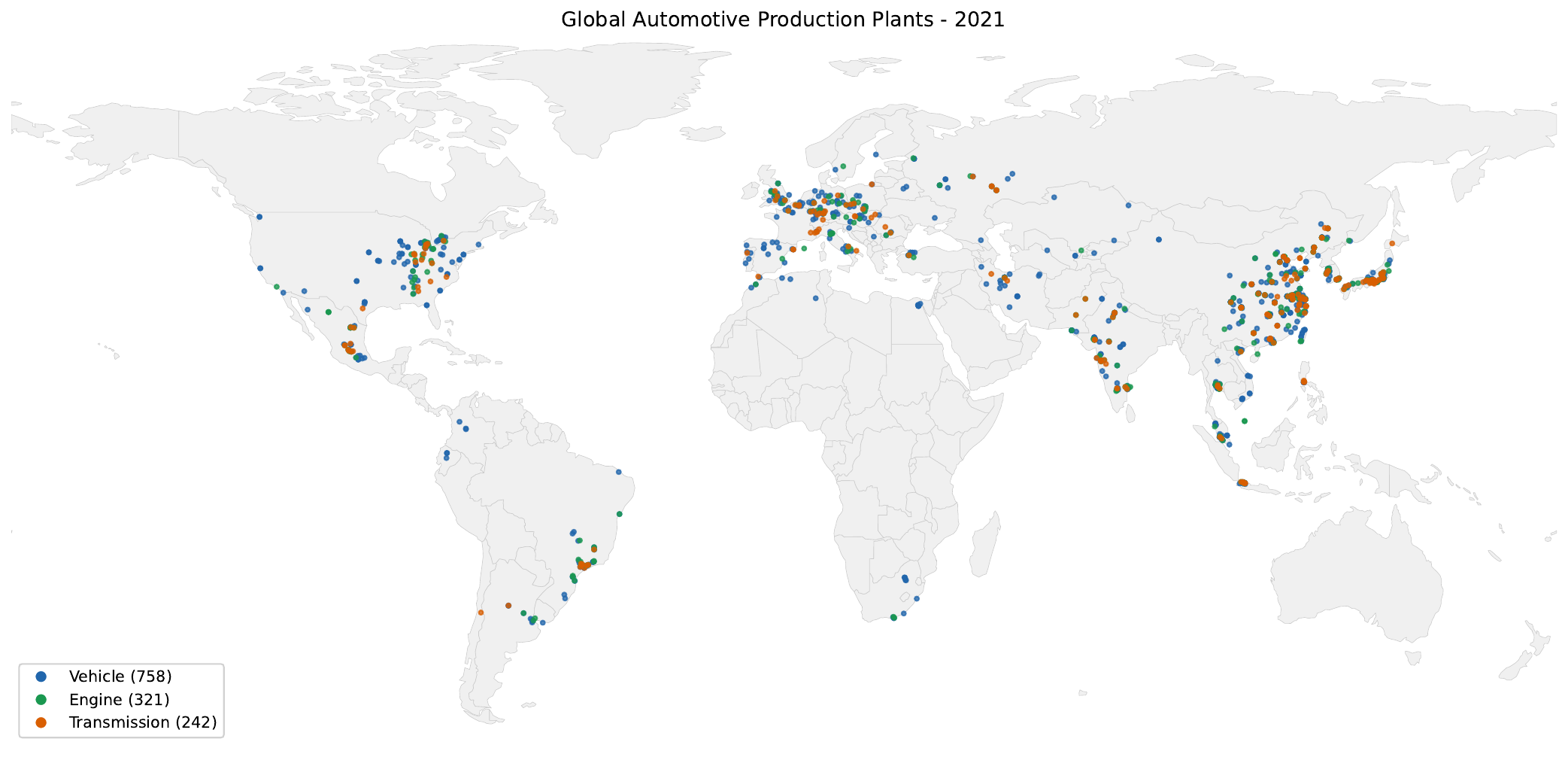}
\caption*{\textit{Notes:} Vehicle assembly, engine and transmission plants active in 2021. Own calculations based on S\&P Global Mobility data.}
\end{figure}

\clearpage
\nextappendix{Definitions} \label{app:defs}
This appendix defines the units, thresholds and measures used in Section~\ref{sec:facts}, in Appendix~\ref{app:sum} and in the analysis.

\subsection{Samples}
\paragraph{{Production.}} S\&P's engine and transmission installation data, release of April 2025. A row gives the output of a vehicle at an assembly plant in a year, together with the plant that made its engine and the plant that made its transmission. We use the rows with positive output in 2000--2024, i.e., 261,987 rows and 1,895 million vehicles. Battery-electric vehicles have no engine, and S\&P records them with a placeholder engine and transmission and no plant (2.3\% of the vehicles for engines, 2.5\% for transmissions). These placeholders are not counted as engines, transmissions or plants. This leaves 443 engine plants and 350 transmission plants.

\paragraph{{Sales.}} S\&P's sales data, which give the units of each vehicle model sold in each destination country in a year, together with the assembly plant and country it came from. We use the rows with positive sales in 2000--2023 and a known plant, origin country and destination country, i.e., 676,361 rows and 1,744 million vehicles.

\paragraph{{Engine investments.}} Public announcements of investments in engine production capacity, i.e., new plants, new lines, expansions and retooling for a new engine family, 1998--2024, which we collected from company press releases, stock-exchange filings, union contract summaries and news. There are 363 announcements, each with its source and a short quote. For 360, we saved a copy of the source and checked the amount, its currency, the capacity and the year against it. Amounts are converted into current US dollars with annual averages of the ECB reference rates and are not deflated. We keep the 188 amounts that were seen in the source and are for engine capacity at one plant, without vehicles, transmissions, development, components alone, heavy-duty engines or press estimates, and without duplicates. Of these, 80 state the capacity of the project, i.e., the capacity added or, for a retooling, the capacity of the retooled line, and enter equation~\eqref{eq:costfn}: 38 new plants, 19 new lines, 14 expansions and 9 retoolings, of 30,000 to 520,000 engines a year. The median new line costs \$164 million (35 announcements). To check that peak output measures capacity, we link 61 of the 80 announcements to an S\&P engine plant by firm group, country and city, and compare the rise in the peak output of the plant's engine families, from the five years before production starts (the announced year, or else two years after the announcement) to the five years after, with the announced capacity. Coverage is incomplete, as most Asian engine plants are announced within the amount for a vehicle plant, and Toyota, Honda, Nissan and Hyundai do not publish amounts for their home-country engine lines.

\subsection{Units}
\begin{itemize}
\item \textit{Plant.} A manufacturing group, plant name and country, so that a plant that changes owner group becomes a new plant. Assembly, engine and transmission plants are separate plants, even when they share a site.
\item \textit{Vehicle model.} A brand and its global nameplate, e.g., Ford Focus. A model spans its generations and all its assembly plants. There are 3,813 vehicle models, 4,390 generations (S\&P vehicle programs) and 1,123 vehicle platforms.
\item \textit{Model version.} A vehicle model together with its S\&P vehicle program, i.e., one generation of the nameplate, e.g., Ford Focus C346. A version spans all its assembly plants. There are 5,763 model versions, as a program can carry several nameplates. The model version is the vehicle product in the analysis.
\item \textit{Engine program.} An engine code of the firm that designed it (S\&P's design parent and engine program), e.g., Honda K24. There are 3,538 engine programs.
\item \textit{Engine family.} The engine architecture a program is built on (design parent and S\&P's engine platform), e.g., Honda K. There are 544 families, with 6.5 programs per family on average, and 99.6\% of the programs belong to a single family.
\item \textit{Transmission program} and \textit{transmission family.} Defined in the same way as for engines. There are 1,831 transmission programs and 543 families.
\item \textit{Product.} An engine program, a transmission program or a model version.
\item \textit{Owner.} The design parent for engine and transmission programs, and the manufacturer group for model versions (the group with most of the version's output). Standard errors are bootstrapped by owner.
\item \textit{Location.} A model version at an assembly plant.
\item \textit{Vehicle.} As in \citet{HMM2022}, a brand, model, platform, generation, body type and manufacturer at an assembly plant in a year. A vehicle can be fitted with several engines and transmissions.
\item \textit{Vehicle-engine specification.} A vehicle together with the displacement, fuel type, propulsion system and horsepower of its engine.
\item \textit{Vehicle-transmission specification.} A vehicle together with the design, number of forward gears, drive type and torque of its transmission.
\item \textit{Vehicle across assembly plants.} In the vehicles column of Table~\ref{tab:org}, a vehicle without its assembly plant and without S\&P's manufacturer, which differs across the plants of a model version in 71\% of the version-years made in two or more plants, so that a car assembled at two plants is one vehicle. Its specification adds the specifications of both its engine and its transmission, and its sources are the assembly plants that assemble any units of it.
\item \textit{Market.} A destination country in the sales data. With the EU as one market, the 28 countries of the EU, the United Kingdom included in every year, form one market, and these countries also count as one origin.
\item \textit{Region.} EU (as above), North America (United States, Canada and Mexico), Japan and Korea, China, and the rest of the world. A plant belongs to the region of its country. A vehicle model, engine program or transmission program belongs to the region of the plant with its largest output that year, a vehicle and a vehicle specification to the region of their assembly plant, and a market to the region of its country.
\end{itemize}

\subsection{Links and Thresholds}
\begin{itemize}
\item A plant \textit{makes} a product in a year if it makes 100 units or more of it. A product is \textit{made in two or more plants} (countries) if two or more plants (in different countries) make it. The 100-unit threshold removes pilot runs and the last units of programs that are ending.
\item An engine or transmission plant \textit{supplies} a vehicle or a vehicle specification if it supplies any units to it in the year, as in \citet{HMM2022}.
\item A plant \textit{supplies} another plant, a vehicle model or a carmaker group in a year if 100 units or more go from the one to the other in that year. These are the links of panel B of Table~\ref{tab:org} and of Table~\ref{tab:links}.
\item A location \textit{gets} a component from a plant if 100 units or more of it go from that plant to the location in the year. Location-years with fewer than 1,000 units of the component are left out.
\item An assembly plant, or an origin country, \textit{sells} a model in a market if it sells 100 units or more of the model there in the year.
\item In the vehicles column of panel B of Table~\ref{tab:org}, the component plants supplying an assembly plant are its engine and transmission plants together, and a plant that supplies both counts once.
\item In the engines and transmissions columns of panel D of Table~\ref{tab:org}, the plants supplying a model version in a market are the engine (transmission) plants that supplied 100 units or more of the version, of any engine (transmission), to the assembly plants selling it there in the year. Sales rows are matched to production by version, plant name and country, which matches every selling plant for 75\% of the version-market-years, and the panel uses those. A component plant sells in a market if 100 units or more of its engines (transmissions) are in cars sold there, an assembly plant's sales of a version being split over its suppliers by their shares of the version.
\end{itemize}

\subsection{Measures}
\paragraph{{Counts per unit.}} Tables~\ref{tab:plants_v}--\ref{tab:plants_t}, \ref{tab:links} and~\ref{tab:served} describe a count per unit-year, such as the number of engine programs an engine plant makes in a year, by:
\begin{itemize}
\item the mean, the 25th, 50th, 75th and 90th percentiles and the maximum over unit-years,
\item \% one, the share of unit-years with a count of one,
\item \% top, the mean over unit-years of the share of the unit's output that comes from its largest partner, e.g., the share of an engine plant's output that is its largest program,
\item the units-weighted mean, i.e., the mean count weighted by the unit's output,
\item N, the number of unit-years.
\end{itemize}

\paragraph{{Sources per chooser.}} Following \citet{HMM2022}, Tables~\ref{tab:veh_e}, \ref{tab:veh_t} and~\ref{tab:markets} describe the number of sources of a \textit{chooser}, i.e., the unit that buys, in a year. The chooser is a vehicle, a vehicle specification or a model in a market, and the source is a plant or a country. The measures are:
\begin{itemize}
\item the mean number of sources,
\item \% top, the mean share of the chooser's units that come from its largest source,
\item \% $n>1$, \% $n>2$ and \% $n>3$, the shares of chooser-years with two or more, three or more and four or more sources,
\item the maximum number of sources,
\item \% units $n>1$, the share of all units that go to chooser-years with two or more sources,
\item the number of chooser-years.
\end{itemize}

\begin{figure}[tp]
\centering
\includegraphics[width=0.6\textwidth]{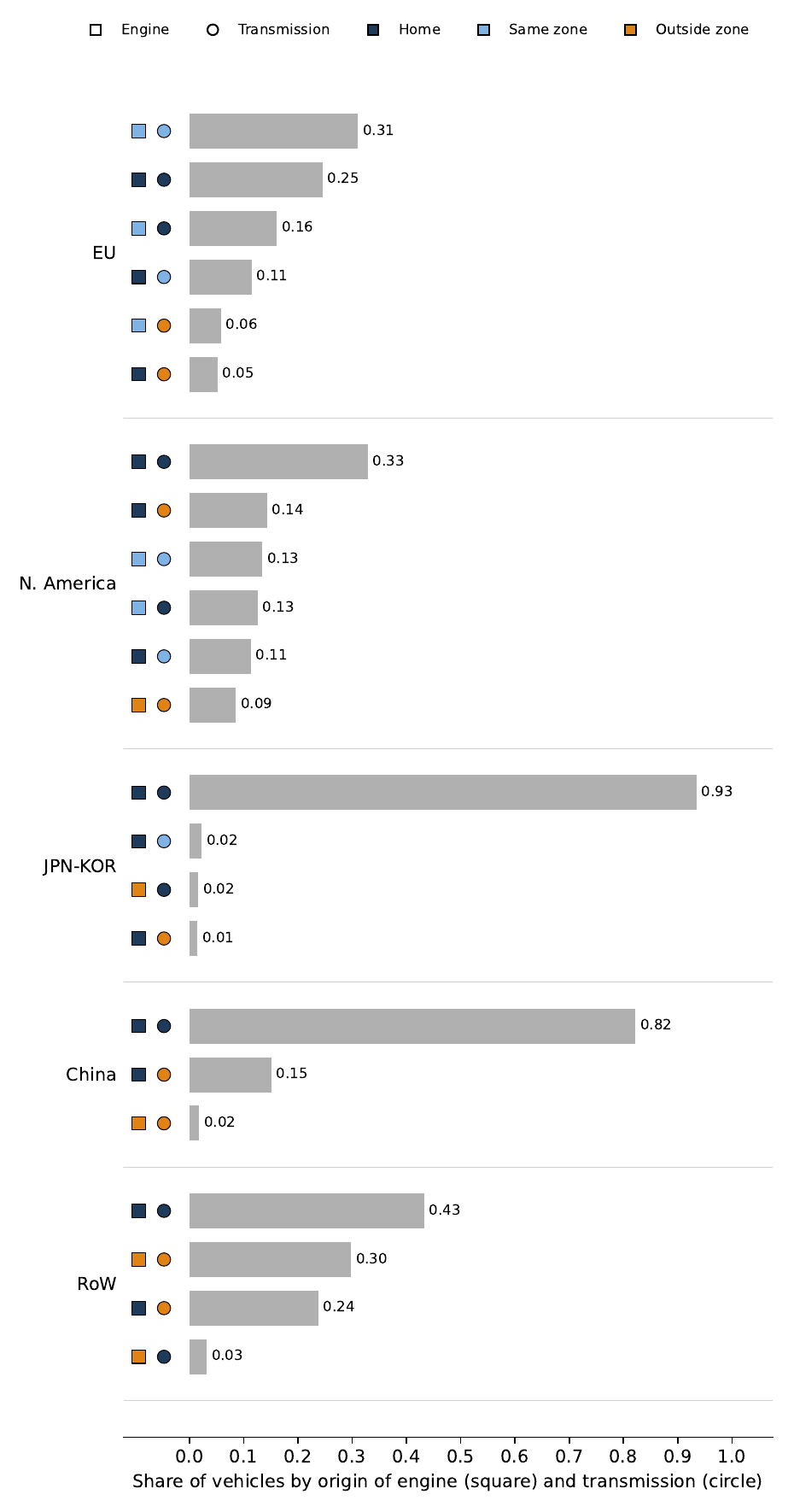}
\caption{Vehicles by Where the Engine and the Transmission Were Made, 2000--2024} \label{fig:config}
\caption*{\textit{Notes:} Share of the vehicles assembled in each region, by where the engine (square) and the transmission (circle) were made, namely in the country of assembly (home), in another country of the region (same zone) or outside the region. Combinations with 1\% or more of the vehicles, up to six per region. As \citet{HMM2022}, Figure 5.}
\end{figure}

\paragraph{{Origin shares.}} Tables~\ref{tab:origin_e} and~\ref{tab:origin_t} give, for the ten countries with the largest vehicle output, the share of the engines (transmissions) fitted in cars assembled there that were made in each country of origin. Figure~\ref{fig:config} classifies every vehicle by where its engine and its transmission were made, namely in the country of assembly (home), in another country of its region (same region), or outside its region.

\subsection{Floods}
\begin{itemize}
\item A \textit{documented flood} is a plant-year in the hand-collected record, either a \textit{direct hit}, where the plant closed because of a flood or storm at the plant or in its own area, with or without water in the plant, or a \textit{stoppage}, where production stopped because a flood elsewhere cut off its supplies while the plant stayed dry. Of the record's flood plant-years, 186 fall in 2000--2024, and 183 of these match a plant in the production data. A site with an assembly plant and an engine or transmission plant under the same name has the flood at each of them. This is the case for 28 flood plant-years (9 with all three kinds of plant), so that the counts by kind of plant in Table~\ref{tab:floods_plants} add up to 220.
\item A product is \textit{flooded} in year $F$, the year of the first documented flood at a plant that made it in $F-1$. Later floods at its plants do not change $F$.
\item A vehicle specification (or a model version, engine program or transmission program, in Table~\ref{tab:floods_choosers}) is flooded in year $t$ if a plant that supplied (made) it in $t-1$ has a documented flood in $t$, for the first such $t$. Shares in $t+k$ are among those still made in $t+k$. Specifications never supplied by a flooded plant give the comparison share, pooled over all their years. In the vehicles column of Table~\ref{tab:org}, the specification is that of a vehicle across assembly plants, and the plant is its assembly plant.
\item World car output with a plant flooded is the share of all vehicles in 2000--2024 that were assembled at, or fitted with an engine or transmission from, a plant in a year with a documented flood there.
\end{itemize}

\clearpage
\nextappendix{Summary Statistics by Region} \label{app:sum}
The tables in this appendix give the full distributions behind Table~\ref{tab:org}, by region. Definitions are in Appendix~\ref{app:defs}. All are own calculations based on S\&P Global Mobility data.

\begin{table}[tp]
\centering
\caption{Model Versions and Assembly Plants, 2000--2024} \label{tab:plants_v}
\footnotesize\setlength{\tabcolsep}{4pt}
\begin{tabular}{lcccccccccc}
\toprule
Zone & mean & p25 & median & p75 & p90 & max & \% one & \% top & units-weighted mean & N\\
\midrule
\multicolumn{11}{l}{\emph{Model versions per assembly plant}}\\
\quad EU & 2.52 & 1 & 2 & 3 & 5 & 11 & 29.5 & 73.4 & 3.09 & 3,038\\
\quad N. America & 2.25 & 1 & 2 & 3 & 4 & 18 & 32.7 & 76.9 & 2.52 & 2,329\\
\quad JPN-KOR & 5.72 & 3 & 5 & 7 & 12 & 23 & 10.8 & 55.2 & 8.10 & 1,305\\
\quad China & 2.95 & 1 & 2 & 4 & 6 & 20 & 32.0 & 74.7 & 4.61 & 4,004\\
\quad RoW & 3.52 & 2 & 3 & 5 & 7 & 38 & 24.2 & 67.3 & 4.90 & 5,909\\
\quad All & 3.19 & 1 & 2 & 4 & 6 & 38 & 27.2 & 70.6 & 4.51 & 16,585\\
\multicolumn{11}{l}{\emph{Assembly plants per model version}}\\
\quad EU & 1.76 & 1 & 1 & 2 & 3 & 12 & 66.6 & 92.3 & 3.22 & 6,134\\
\quad N. America & 1.64 & 1 & 1 & 2 & 3 & 16 & 71.0 & 91.4 & 2.83 & 4,010\\
\quad JPN-KOR & 1.57 & 1 & 1 & 1 & 3 & 12 & 75.1 & 94.4 & 2.77 & 6,535\\
\quad China & 1.27 & 1 & 1 & 1 & 2 & 14 & 86.5 & 96.6 & 2.20 & 9,843\\
\quad RoW & 1.50 & 1 & 1 & 2 & 3 & 12 & 74.2 & 92.5 & 2.44 & 8,582\\
\quad All & 1.51 & 1 & 1 & 1 & 3 & 16 & 76.1 & 93.9 & 2.71 & 35,104\\
\multicolumn{11}{l}{\emph{Countries per model version}}\\
\quad EU & 1.70 & 1 & 1 & 2 & 3 & 10 & 68.3 & 92.3 & 3.01 & 6,134\\
\quad N. America & 1.53 & 1 & 1 & 1 & 3 & 14 & 76.1 & 91.4 & 2.52 & 4,010\\
\quad JPN-KOR & 1.50 & 1 & 1 & 1 & 3 & 11 & 78.7 & 94.4 & 2.55 & 6,535\\
\quad China & 1.23 & 1 & 1 & 1 & 2 & 13 & 88.9 & 96.6 & 2.04 & 9,843\\
\quad RoW & 1.43 & 1 & 1 & 1 & 2 & 10 & 77.5 & 92.5 & 2.19 & 8,582\\
\quad All & 1.45 & 1 & 1 & 1 & 2 & 14 & 79.1 & 93.9 & 2.48 & 35,104\\
\bottomrule
\end{tabular}

\caption*{\textit{Notes:} Plant-years (first panel) and version-years (second and third panels). A plant makes a model version with 100 units or more in the year. \% one: share with a count of one. \% top: share of the plant's (version's) output from its largest version (plant). Region of the plant (of the version's largest plant).}
\end{table}

\begin{table}[tp]
\centering
\caption{Engine Programs and Engine Plants, 2000--2024} \label{tab:plants_e}
\footnotesize\setlength{\tabcolsep}{4pt}
\begin{tabular}{lcccccccccc}
\toprule
Zone & mean & p25 & median & p75 & p90 & max & \% one & \% top & units-weighted mean & N\\
\midrule
\multicolumn{11}{l}{\emph{Engine programs per engine plant}}\\
\quad EU & 6.55 & 2 & 4 & 8 & 16 & 39 & 17.6 & 62.0 & 11.28 & 1,305\\
\quad N. America & 3.71 & 2 & 3 & 5 & 8 & 15 & 21.6 & 69.9 & 5.08 & 1,021\\
\quad JPN-KOR & 8.80 & 4 & 8 & 12 & 18 & 37 & 7.7 & 51.9 & 14.36 & 981\\
\quad China & 5.37 & 2 & 4 & 7 & 11 & 33 & 16.3 & 63.5 & 7.46 & 1,952\\
\quad RoW & 3.90 & 2 & 3 & 5 & 8 & 24 & 24.2 & 67.6 & 6.55 & 2,040\\
\quad All & 5.40 & 2 & 4 & 7 & 12 & 39 & 18.3 & 63.7 & 9.26 & 7,299\\
\multicolumn{11}{l}{\emph{Engine plants per engine program}}\\
\quad EU & 1.29 & 1 & 1 & 1 & 2 & 10 & 80.5 & 95.4 & 2.38 & 6,983\\
\quad N. America & 1.68 & 1 & 1 & 2 & 3 & 14 & 66.8 & 90.2 & 2.89 & 2,584\\
\quad JPN-KOR & 1.51 & 1 & 1 & 2 & 3 & 12 & 72.4 & 91.9 & 2.96 & 6,008\\
\quad China & 1.26 & 1 & 1 & 1 & 2 & 13 & 88.0 & 96.2 & 3.80 & 8,568\\
\quad RoW & 1.42 & 1 & 1 & 1 & 2 & 12 & 75.5 & 92.7 & 2.29 & 4,361\\
\quad All & 1.38 & 1 & 1 & 1 & 2 & 14 & 79.0 & 94.0 & 2.91 & 28,504\\
\multicolumn{11}{l}{\emph{Countries per engine program}}\\
\quad EU & 1.25 & 1 & 1 & 1 & 2 & 7 & 82.1 & 95.4 & 2.12 & 6,983\\
\quad N. America & 1.52 & 1 & 1 & 2 & 2 & 12 & 69.7 & 90.2 & 2.36 & 2,584\\
\quad JPN-KOR & 1.34 & 1 & 1 & 1 & 2 & 10 & 79.7 & 91.9 & 2.30 & 6,008\\
\quad China & 1.18 & 1 & 1 & 1 & 1 & 11 & 91.1 & 96.2 & 2.87 & 8,568\\
\quad RoW & 1.38 & 1 & 1 & 1 & 2 & 10 & 76.2 & 92.7 & 2.12 & 4,361\\
\quad All & 1.29 & 1 & 1 & 1 & 2 & 12 & 82.3 & 94.0 & 2.37 & 28,504\\
\bottomrule
\end{tabular}

\caption*{\textit{Notes:} As Table~\ref{tab:plants_v}, for engine programs and engine plants.}
\end{table}

\begin{table}[tp]
\centering
\caption{Transmission Programs and Transmission Plants, 2000--2024} \label{tab:plants_t}
\footnotesize\setlength{\tabcolsep}{4pt}
\begin{tabular}{lcccccccccc}
\toprule
Zone & mean & p25 & median & p75 & p90 & max & \% one & \% top & units-weighted mean & N\\
\midrule
\multicolumn{11}{l}{\emph{Transmission programs per transmission plant}}\\
\quad EU & 4.23 & 2 & 3 & 5 & 11 & 23 & 23.4 & 71.8 & 7.49 & 1,145\\
\quad N. America & 2.88 & 1 & 2 & 4 & 6 & 11 & 29.1 & 75.2 & 3.86 & 649\\
\quad JPN-KOR & 7.82 & 3 & 5 & 10 & 17 & 40 & 9.6 & 56.0 & 15.65 & 1,095\\
\quad China & 4.20 & 1 & 3 & 6 & 9 & 26 & 27.9 & 71.5 & 7.32 & 1,404\\
\quad RoW & 2.77 & 1 & 2 & 4 & 5 & 10 & 27.9 & 74.8 & 3.73 & 1,310\\
\quad All & 4.43 & 2 & 3 & 5 & 10 & 40 & 23.6 & 69.7 & 8.77 & 5,603\\
\multicolumn{11}{l}{\emph{Transmission plants per transmission program}}\\
\quad EU & 1.33 & 1 & 1 & 1 & 2 & 8 & 80.1 & 94.6 & 2.41 & 3,961\\
\quad N. America & 1.69 & 1 & 1 & 2 & 3 & 7 & 63.7 & 89.2 & 2.63 & 1,285\\
\quad JPN-KOR & 1.36 & 1 & 1 & 1 & 2 & 7 & 77.0 & 93.7 & 2.20 & 6,379\\
\quad China & 1.24 & 1 & 1 & 1 & 2 & 9 & 84.5 & 96.0 & 2.21 & 4,482\\
\quad RoW & 1.55 & 1 & 1 & 2 & 3 & 8 & 73.3 & 91.0 & 2.70 & 1,999\\
\quad All & 1.37 & 1 & 1 & 1 & 2 & 9 & 78.2 & 93.8 & 2.38 & 18,106\\
\multicolumn{11}{l}{\emph{Countries per transmission program}}\\
\quad EU & 1.28 & 1 & 1 & 1 & 2 & 8 & 82.8 & 94.6 & 2.21 & 3,961\\
\quad N. America & 1.49 & 1 & 1 & 2 & 3 & 6 & 68.1 & 89.2 & 2.06 & 1,285\\
\quad JPN-KOR & 1.19 & 1 & 1 & 1 & 2 & 6 & 86.6 & 93.7 & 1.65 & 6,379\\
\quad China & 1.13 & 1 & 1 & 1 & 1 & 6 & 92.2 & 96.0 & 1.85 & 4,482\\
\quad RoW & 1.47 & 1 & 1 & 2 & 3 & 8 & 74.2 & 91.0 & 2.38 & 1,999\\
\quad All & 1.25 & 1 & 1 & 1 & 2 & 8 & 84.5 & 93.8 & 1.98 & 18,106\\
\bottomrule
\end{tabular}

\caption*{\textit{Notes:} As Table~\ref{tab:plants_v}, for transmission programs and transmission plants.}
\end{table}

\begin{table}[tp]
\centering
\caption{Links between Plants, 2000--2024} \label{tab:links}
\footnotesize\setlength{\tabcolsep}{3pt}
\begin{tabular}{>{\raggedright\arraybackslash\hangindent=1em\hangafter=1}p{4.4cm}cccccccccc}
\toprule
 & mean & p25 & median & p75 & p90 & max & \% one & \% top & \makecell[b]{units-weighted\\mean} & N\\
\midrule
Engines: model versions supplied per engine plant & 8.45 & 3 & 6 & 11 & 19 & 69 & 13.2 & 54.0 & 16.83 & 7,299\\
Engines: assembly plants supplied per engine plant & 5.66 & 2 & 4 & 8 & 13 & 40 & 22.8 & 67.1 & 10.77 & 7,301\\
Engines: carmaker groups supplied per engine plant & 2.58 & 1 & 2 & 3 & 6 & 30 & 48.6 & 91.6 & 3.67 & 7,301\\
Transmissions: model versions supplied per transmission plant & 12.10 & 3 & 8 & 15 & 27 & 135 & 11.5 & 49.5 & 27.21 & 5,601\\
Transmissions: assembly plants supplied per transmission plant & 8.68 & 2 & 6 & 11 & 18 & 118 & 16.7 & 58.4 & 19.16 & 5,601\\
Transmissions: carmaker groups supplied per transmission plant & 3.58 & 1 & 2 & 4 & 8 & 36 & 38.1 & 85.6 & 6.30 & 5,602\\
Vehicles: engine plants supplying an assembly plant & 2.55 & 1 & 2 & 3 & 5 & 21 & 31.7 & 77.8 & 2.94 & 16,197\\
Vehicles: transmission plants supplying an assembly plant & 3.01 & 2 & 3 & 4 & 5 & 24 & 22.6 & 74.2 & 3.70 & 16,161\\
\bottomrule
\end{tabular}

\caption*{\textit{Notes:} Plant-years, all regions. A link has 100 units or more in the year. \% top: share of the plant's output that goes to (comes from) its largest partner. Carmaker groups: manufacturing groups of the assembly plants supplied.}
\end{table}

\begin{table}[tp]
\centering
\caption{Engine Sources per Vehicle, 2000--2024} \label{tab:veh_e}
\footnotesize\setlength{\tabcolsep}{4pt}
\begin{tabular}{lcccccccc}
\toprule
Zone & \# sources mean & \% top & \% n$>$1 & \% n$>$2 & \% n$>$3 & max & \% units n$>$1 & choosers\\
\midrule
\multicolumn{9}{l}{\emph{Source = countries / Chooser = vehicle}}\\
\quad EU & 1.75 & 85.9 & 49.6 & 19.5 & 4.9 & 6 & 66.6 & 7,707\\
\quad N. America & 1.54 & 88.4 & 43.1 & 9.9 & 0.8 & 5 & 53.0 & 5,118\\
\quad JPN-KOR & 1.10 & 98.2 & 8.3 & 1.1 & 0.2 & 4 & 8.4 & 7,445\\
\quad China & 1.06 & 99.1 & 5.9 & 0.2 & 0.0 & 4 & 7.7 & 11,163\\
\quad RoW & 1.26 & 94.1 & 21.2 & 4.1 & 0.4 & 6 & 30.8 & 22,412\\
\quad All & 1.29 & 94.0 & 22.4 & 5.6 & 1.0 & 6 & 34.5 & 53,845\\
\multicolumn{9}{l}{\emph{Source = countries / Chooser = vehicle-engine spec}}\\
\quad EU & 1.00 & 99.9 & 0.2 & 0.0 & 0.0 & 2 & 0.6 & 46,125\\
\quad N. America & 1.02 & 99.4 & 1.6 & 0.1 & 0.0 & 3 & 3.3 & 14,187\\
\quad JPN-KOR & 1.00 & 100.0 & 0.1 & 0.0 & 0.0 & 2 & 0.2 & 22,689\\
\quad China & 1.00 & 99.9 & 0.2 & 0.0 & 0.0 & 3 & 0.4 & 27,023\\
\quad RoW & 1.01 & 99.9 & 0.5 & 0.0 & 0.0 & 3 & 1.1 & 48,275\\
\quad All & 1.00 & 99.9 & 0.4 & 0.0 & 0.0 & 3 & 1.1 & 158,299\\
\multicolumn{9}{l}{\emph{Source = plants / Chooser = vehicle}}\\
\quad EU & 2.13 & 80.3 & 65.6 & 30.9 & 11.8 & 7 & 85.0 & 7,707\\
\quad N. America & 1.84 & 83.6 & 55.0 & 21.3 & 6.0 & 6 & 63.7 & 5,118\\
\quad JPN-KOR & 1.52 & 90.1 & 37.3 & 11.0 & 3.1 & 5 & 45.8 & 7,445\\
\quad China & 1.53 & 89.9 & 33.8 & 10.1 & 4.5 & 9 & 36.0 & 11,163\\
\quad RoW & 1.38 & 91.4 & 29.4 & 6.9 & 1.4 & 6 & 39.7 & 22,412\\
\quad All & 1.58 & 88.6 & 39.0 & 12.9 & 4.2 & 9 & 54.9 & 53,845\\
\multicolumn{9}{l}{\emph{Source = plants / Chooser = vehicle-engine spec}}\\
\quad EU & 1.00 & 99.9 & 0.4 & 0.0 & 0.0 & 3 & 1.0 & 46,125\\
\quad N. America & 1.02 & 99.2 & 2.3 & 0.1 & 0.0 & 3 & 4.4 & 14,187\\
\quad JPN-KOR & 1.00 & 99.9 & 0.4 & 0.0 & 0.0 & 2 & 0.8 & 22,689\\
\quad China & 1.01 & 99.7 & 1.1 & 0.0 & 0.0 & 3 & 1.5 & 27,023\\
\quad RoW & 1.01 & 99.9 & 0.6 & 0.0 & 0.0 & 3 & 1.4 & 48,275\\
\quad All & 1.01 & 99.8 & 0.7 & 0.0 & 0.0 & 3 & 1.8 & 158,299\\
\bottomrule
\end{tabular}

\caption*{\textit{Notes:} Chooser-years by region of assembly. A plant or country is a source if it supplies any units. Columns as in \citet{HMM2022}, Tables 2 and 3, with the maximum and the share of units added.}
\end{table}

\begin{table}[tp]
\centering
\caption{Transmission Sources per Vehicle, 2000--2024} \label{tab:veh_t}
\footnotesize\setlength{\tabcolsep}{4pt}
\begin{tabular}{lcccccccc}
\toprule
Zone & \# sources mean & \% top & \% n$>$1 & \% n$>$2 & \% n$>$3 & max & \% units n$>$1 & choosers\\
\midrule
\multicolumn{9}{l}{\emph{Source = countries / Chooser = vehicle}}\\
\quad EU & 1.87 & 85.5 & 51.9 & 24.1 & 8.0 & 7 & 67.6 & 7,684\\
\quad N. America & 1.55 & 91.2 & 42.3 & 9.3 & 2.1 & 6 & 47.6 & 5,114\\
\quad JPN-KOR & 1.12 & 97.9 & 11.1 & 1.2 & 0.0 & 4 & 12.8 & 7,440\\
\quad China & 1.22 & 95.1 & 21.0 & 1.4 & 0.0 & 4 & 25.9 & 11,084\\
\quad RoW & 1.38 & 92.0 & 30.4 & 5.7 & 1.4 & 7 & 48.0 & 22,407\\
\quad All & 1.40 & 92.4 & 30.0 & 7.2 & 1.9 & 7 & 41.6 & 53,729\\
\multicolumn{9}{l}{\emph{Source = countries / Chooser = vehicle-transmission spec}}\\
\quad EU & 1.02 & 99.6 & 1.8 & 0.1 & 0.0 & 3 & 4.8 & 30,210\\
\quad N. America & 1.03 & 99.3 & 3.0 & 0.0 & 0.0 & 3 & 4.4 & 16,576\\
\quad JPN-KOR & 1.00 & 100.0 & 0.1 & 0.0 & 0.0 & 2 & 0.1 & 27,781\\
\quad China & 1.01 & 99.9 & 0.6 & 0.0 & 0.0 & 3 & 1.3 & 26,099\\
\quad RoW & 1.02 & 99.6 & 1.5 & 0.1 & 0.0 & 3 & 3.3 & 46,121\\
\quad All & 1.01 & 99.7 & 1.3 & 0.0 & 0.0 & 3 & 2.9 & 146,787\\
\multicolumn{9}{l}{\emph{Source = plants / Chooser = vehicle}}\\
\quad EU & 2.38 & 78.9 & 68.6 & 41.4 & 19.1 & 8 & 86.6 & 7,684\\
\quad N. America & 1.95 & 85.3 & 59.2 & 23.9 & 6.9 & 8 & 65.2 & 5,114\\
\quad JPN-KOR & 1.89 & 84.3 & 61.3 & 20.6 & 5.4 & 8 & 70.6 & 7,440\\
\quad China & 1.77 & 85.2 & 55.3 & 16.5 & 4.0 & 7 & 61.4 & 11,084\\
\quad RoW & 1.56 & 88.1 & 42.0 & 10.7 & 2.8 & 7 & 57.5 & 22,407\\
\quad All & 1.80 & 85.4 & 52.8 & 18.9 & 6.1 & 8 & 68.6 & 53,729\\
\multicolumn{9}{l}{\emph{Source = plants / Chooser = vehicle-transmission spec}}\\
\quad EU & 1.03 & 99.4 & 2.6 & 0.1 & 0.0 & 3 & 5.7 & 30,210\\
\quad N. America & 1.05 & 98.9 & 4.8 & 0.1 & 0.0 & 3 & 6.8 & 16,576\\
\quad JPN-KOR & 1.01 & 99.6 & 1.3 & 0.0 & 0.0 & 2 & 3.0 & 27,781\\
\quad China & 1.02 & 99.6 & 1.7 & 0.0 & 0.0 & 3 & 3.0 & 26,099\\
\quad RoW & 1.02 & 99.5 & 1.8 & 0.1 & 0.0 & 3 & 3.9 & 46,121\\
\quad All & 1.02 & 99.5 & 2.2 & 0.1 & 0.0 & 3 & 4.6 & 146,787\\
\bottomrule
\end{tabular}

\caption*{\textit{Notes:} As Table~\ref{tab:veh_e}, for transmissions.}
\end{table}

\begin{table}[tp]
\centering
\caption{Origin of the Engines Fitted in the Ten Largest Assembly Countries, 2000--2024} \label{tab:origin_e}
\footnotesize\setlength{\tabcolsep}{4pt}
\begin{tabular}{lcccccccccc}
\toprule
Origin & CHN & USA & JPN & DEU & KOR & IND & MEX & ESP & BRA & FRA\\
\midrule
JPN & 0.98 & 2.04 & 98.40 & 0.03 & 1.33 & 0.68 & 1.15 & 0.28 & 1.06 & 5.50\\
FRA & 0.02 & 0.00 & 0.50 & 1.31 & 0.36 & 0.01 & 0.05 & 28.75 & 1.23 & 72.98\\
KOR & 0.38 & 1.20 & 0.00 & 0.00 & 94.40 & 1.43 & 0.83 & 0.09 & 3.74 & 0.07\\
CHN & 97.34 & 0.00 & 0.35 & 0.00 & 0.37 & 0.34 & 0.11 & 0.00 & 0.31 & 0.20\\
USA & 0.13 & 67.50 & 0.23 & 0.12 & 0.20 & 0.00 & 25.88 & 0.09 & 0.48 & 0.00\\
IND & 0.00 & 0.00 & 0.00 & 0.00 & 0.00 & 93.60 & 0.00 & 0.00 & 0.43 & 0.10\\
BRA & 0.03 & 0.04 & 0.00 & 0.01 & 0.00 & 0.00 & 2.82 & 0.00 & 86.28 & 0.06\\
DEU & 0.22 & 4.04 & 0.13 & 44.72 & 0.48 & 0.80 & 3.65 & 21.37 & 0.49 & 1.39\\
MEX & 0.02 & 11.63 & 0.01 & 0.16 & 1.07 & 0.04 & 54.19 & 1.20 & 0.60 & 0.00\\
HUN & 0.33 & 0.04 & 0.00 & 22.32 & 0.18 & 0.29 & 2.60 & 7.75 & 0.33 & 0.48\\
ROW & 0.55 & 13.51 & 0.37 & 31.34 & 1.61 & 2.79 & 8.72 & 40.48 & 5.04 & 19.24\\
\midrule
Home & 97.34 & 67.50 & 98.40 & 44.72 & 94.40 & 93.60 & 54.19 & 14.80 & 86.28 & 72.98\\
\bottomrule
\end{tabular}

\caption*{\textit{Notes:} Columns: country of assembly. Rows: country where the engine was made, the ten origins with the largest average share and the rest of the world (ROW). Percent of the engines fitted, each column adds up to 100. Home: share made in the country of assembly. Spain is not among the ten engine origins, so that the engines made in Spain for Spanish cars (14.8\%) are in ROW.}
\end{table}

\begin{table}[tp]
\centering
\caption{Origin of the Transmissions Fitted in the Ten Largest Assembly Countries, 2000--2024} \label{tab:origin_t}
\footnotesize\setlength{\tabcolsep}{4pt}
\begin{tabular}{lcccccccccc}
\toprule
Origin & CHN & USA & JPN & DEU & KOR & IND & MEX & ESP & BRA & FRA\\
\midrule
JPN & 8.77 & 13.69 & 98.93 & 3.86 & 9.53 & 3.60 & 16.67 & 4.05 & 9.91 & 10.15\\
DEU & 3.76 & 5.11 & 0.10 & 71.25 & 1.13 & 0.93 & 8.98 & 14.71 & 0.75 & 3.12\\
FRA & 0.83 & 2.60 & 0.01 & 7.55 & 0.14 & 0.03 & 0.19 & 28.60 & 2.62 & 62.24\\
KOR & 1.40 & 1.60 & 0.00 & 0.01 & 85.93 & 3.17 & 0.86 & 0.13 & 6.46 & 0.00\\
USA & 0.73 & 62.49 & 0.64 & 0.52 & 0.37 & 0.07 & 30.38 & 1.60 & 1.69 & 0.01\\
IND & 0.00 & 0.00 & 0.00 & 0.00 & 0.00 & 86.61 & 0.00 & 1.26 & 0.49 & 0.15\\
CHN & 83.07 & 0.11 & 0.00 & 0.29 & 0.53 & 0.87 & 0.12 & 0.25 & 0.82 & 0.00\\
BRA & 0.01 & 0.17 & 0.00 & 0.01 & 0.00 & 0.05 & 1.76 & 0.00 & 51.02 & 0.00\\
MEX & 0.44 & 10.28 & 0.00 & 0.17 & 0.68 & 0.13 & 27.39 & 0.05 & 2.19 & 0.00\\
ESP & 0.05 & 0.00 & 0.05 & 2.61 & 0.16 & 0.58 & 4.67 & 21.94 & 0.06 & 7.87\\
ROW & 0.95 & 3.94 & 0.27 & 13.72 & 1.52 & 3.95 & 8.98 & 27.39 & 24.00 & 16.44\\
\midrule
Home & 83.07 & 62.49 & 98.93 & 71.25 & 85.93 & 86.61 & 27.39 & 21.94 & 51.02 & 62.24\\
\bottomrule
\end{tabular}

\caption*{\textit{Notes:} As Table~\ref{tab:origin_e}, for transmissions.}
\end{table}

\begin{table}[tp]
\centering
\caption{Vehicle Sources per Market, 2000--2023} \label{tab:markets}
\footnotesize\setlength{\tabcolsep}{4pt}
\begin{tabular}{lcccccccc}
\toprule
Zone of the market & \# sources mean & \% top & \% n$>$1 & \% n$>$2 & \% n$>$3 & max & \% units n$>$1 & choosers\\
\midrule
\multicolumn{9}{l}{\emph{Countries as markets: source = assembly plants / chooser = model $\times$ market}}\\
\quad EU & 1.08 & 97.5 & 7.6 & 0.6 & 0.0 & 5 & 23.4 & 125,387\\
\quad N. America & 1.11 & 97.3 & 8.9 & 1.5 & 0.3 & 5 & 34.3 & 21,045\\
\quad JPN-KOR & 1.05 & 98.5 & 5.1 & 0.4 & 0.0 & 5 & 10.3 & 12,687\\
\quad China & 1.06 & 99.0 & 5.0 & 0.5 & 0.0 & 4 & 14.7 & 13,216\\
\quad RoW & 1.05 & 98.4 & 4.5 & 0.3 & 0.0 & 6 & 13.9 & 127,860\\
\quad All & 1.07 & 98.0 & 6.1 & 0.5 & 0.0 & 6 & 21.0 & 300,195\\
\multicolumn{9}{l}{\emph{Countries as markets: source = origin countries / chooser = model $\times$ market}}\\
\quad EU & 1.06 & 98.3 & 5.4 & 0.4 & 0.0 & 4 & 16.8 & 125,587\\
\quad N. America & 1.06 & 98.6 & 5.6 & 0.6 & 0.0 & 4 & 22.7 & 21,050\\
\quad JPN-KOR & 1.02 & 99.5 & 2.0 & 0.0 & 0.0 & 3 & 1.2 & 12,691\\
\quad China & 1.03 & 99.7 & 2.6 & 0.2 & 0.0 & 3 & 6.2 & 13,217\\
\quad RoW & 1.03 & 98.9 & 3.3 & 0.1 & 0.0 & 4 & 6.0 & 128,043\\
\quad All & 1.04 & 98.7 & 4.2 & 0.2 & 0.0 & 4 & 12.2 & 300,588\\
\multicolumn{9}{l}{\emph{EU as one market: source = assembly plants / chooser = model $\times$ market}}\\
\quad EU & 1.16 & 96.9 & 13.4 & 2.3 & 0.3 & 7 & 37.3 & 13,066\\
\quad N. America & 1.11 & 97.3 & 8.9 & 1.5 & 0.3 & 5 & 34.3 & 21,045\\
\quad JPN-KOR & 1.05 & 98.5 & 5.1 & 0.4 & 0.0 & 5 & 10.3 & 12,687\\
\quad China & 1.06 & 99.0 & 5.0 & 0.5 & 0.0 & 4 & 14.7 & 13,216\\
\quad RoW & 1.05 & 98.4 & 4.5 & 0.3 & 0.0 & 6 & 13.9 & 127,860\\
\quad All & 1.06 & 98.2 & 5.7 & 0.6 & 0.1 & 7 & 24.0 & 187,874\\
\multicolumn{9}{l}{\emph{EU as one market: source = origin countries (EU as one) / chooser = model $\times$ market}}\\
\quad EU & 1.06 & 98.9 & 5.2 & 0.4 & 0.0 & 3 & 11.3 & 13,076\\
\quad N. America & 1.06 & 98.7 & 5.1 & 0.6 & 0.0 & 4 & 22.4 & 21,050\\
\quad JPN-KOR & 1.01 & 99.7 & 1.1 & 0.0 & 0.0 & 3 & 0.8 & 12,694\\
\quad China & 1.03 & 99.7 & 2.5 & 0.1 & 0.0 & 3 & 6.2 & 13,217\\
\quad RoW & 1.03 & 99.2 & 2.5 & 0.0 & 0.0 & 4 & 5.3 & 128,164\\
\quad All & 1.03 & 99.2 & 2.9 & 0.1 & 0.0 & 4 & 10.7 & 188,201\\
\bottomrule
\end{tabular}

\caption*{\textit{Notes:} Version-market-years by region of the market. A plant or origin is a source if it sells 100 units or more of the model version in the market in the year. With the EU as one market, the EU-28 destinations form one market and the EU-28 origins one origin.}
\end{table}

\begin{table}[tp]
\centering
\caption{The Largest Markets, 2000--2023} \label{tab:bigmarkets}
\footnotesize\setlength{\tabcolsep}{4pt}
\begin{tabular}{lcccccl}
\toprule
Market & Sales & Models & Origin plants & Origins & \% made in the market & Largest outside origins (\%)\\
\midrule
EU & 15.74 & 545 & 227 & 16.2 & 83.9 & JPN 4, TUR 4, KOR 3\\
USA & 15.62 & 349 & 169 & 10.9 & 57.5 & CAN 11, JPN 10, MEX 10\\
CHN & 15.08 & 551 & 230 & 8.0 & 95.7 & EU 2, JPN 1, USA 1\\
JPN & 5.11 & 375 & 129 & 10.0 & 94.3 & EU 4, THA 0, ZAF 0\\
IND & 2.53 & 114 & 41 & 5.8 & 99.8 & EU 0, KOR 0, THA 0\\
BRA & 2.31 & 165 & 90 & 10.4 & 85.4 & ARG 8, MEX 2, KOR 1\\
RUS & 1.74 & 255 & 136 & 13.1 & 71.5 & EU 12, JPN 7, KOR 3\\
CAN & 1.68 & 292 & 153 & 9.5 & 16.9 & USA 48, JPN 11, MEX 10\\
KOR & 1.52 & 154 & 72 & 7.1 & 90.3 & EU 6, USA 2, JPN 1\\
MEX & 1.11 & 237 & 151 & 13.2 & 41.1 & USA 16, BRA 10, EU 7\\
IRN & 1.04 & 74 & 46 & 8.8 & 89.7 & CHN 6, EU 1, KOR 1\\
AUS & 1.00 & 264 & 152 & 14.6 & 12.5 & JPN 33, THA 18, EU 16\\
\bottomrule
\end{tabular}

\caption*{\textit{Notes:} The twelve largest markets, the EU-28 as one. Sales: million vehicles a year. Models (model versions), origin plants and origins: number selling 100 units or more in the market, yearly averages, the EU-28 origins as one. \% made in the market: share of the period's sales made in the market. Largest outside origins: percent of the period's sales.}
\end{table}

\begin{table}[tp]
\centering
\caption{Markets Served per Assembly Plant and per Model Version, 2000--2023} \label{tab:served}
\footnotesize\setlength{\tabcolsep}{4pt}
\begin{tabular}{lcccccccccc}
\toprule
Zone & mean & p25 & median & p75 & p90 & max & \% one & \% top & units-weighted mean & N\\
\midrule
\multicolumn{11}{l}{\emph{Markets per assembly plant, countries as markets}}\\
\quad EU & 27.58 & 12 & 30 & 41 & 48 & 64 & 7.3 & 40.3 & 39.78 & 2,986\\
\quad N. America & 9.64 & 3 & 6 & 11 & 21 & 63 & 7.0 & 83.2 & 11.82 & 2,290\\
\quad JPN-KOR & 26.57 & 8 & 26 & 42 & 54 & 73 & 10.4 & 62.8 & 38.19 & 1,270\\
\quad China & 2.61 & 1 & 1 & 3 & 6 & 43 & 62.8 & 96.7 & 3.94 & 3,732\\
\quad RoW & 5.08 & 1 & 2 & 5 & 12 & 52 & 43.3 & 85.6 & 11.87 & 5,616\\
\quad All & 11.10 & 1 & 3 & 14 & 38 & 73 & 33.3 & 77.5 & 21.07 & 15,894\\
\multicolumn{11}{l}{\emph{Markets per model version, countries as markets}}\\
\quad EU & 16.25 & 2 & 12 & 27 & 37 & 67 & 18.2 & 52.9 & 36.54 & 8,430\\
\quad N. America & 6.79 & 2 & 3 & 7 & 17 & 69 & 21.9 & 82.8 & 14.32 & 5,316\\
\quad JPN-KOR & 5.87 & 1 & 1 & 5 & 18 & 70 & 51.7 & 81.2 & 16.80 & 10,575\\
\quad China & 2.12 & 1 & 1 & 1 & 3 & 66 & 76.4 & 94.5 & 6.34 & 11,051\\
\quad RoW & 3.89 & 1 & 1 & 3 & 9 & 57 & 55.3 & 85.1 & 10.66 & 10,996\\
\quad All & 6.50 & 1 & 2 & 6 & 22 & 70 & 48.9 & 80.3 & 18.03 & 46,368\\
\multicolumn{11}{l}{\emph{Markets per assembly plant, EU as one market}}\\
\quad EU & 12.42 & 4 & 11 & 18 & 27 & 41 & 9.7 & 79.1 & 18.43 & 3,000\\
\quad N. America & 7.96 & 3 & 6 & 11 & 16 & 38 & 7.2 & 83.4 & 9.75 & 2,292\\
\quad JPN-KOR & 17.70 & 6 & 19 & 27 & 33 & 48 & 10.3 & 63.6 & 24.64 & 1,270\\
\quad China & 2.46 & 1 & 1 & 3 & 6 & 25 & 62.8 & 96.7 & 3.59 & 3,733\\
\quad RoW & 3.86 & 1 & 2 & 5 & 10 & 36 & 43.3 & 86.8 & 8.05 & 5,616\\
\quad All & 6.84 & 1 & 3 & 9 & 19 & 48 & 33.7 & 85.3 & 12.68 & 15,911\\
\multicolumn{11}{l}{\emph{Markets per model version, EU as one market}}\\
\quad EU & 7.08 & 1 & 4 & 10 & 18 & 43 & 27.3 & 80.7 & 17.06 & 8,554\\
\quad N. America & 5.55 & 2 & 3 & 7 & 13 & 44 & 22.3 & 83.4 & 11.10 & 5,325\\
\quad JPN-KOR & 4.24 & 1 & 1 & 4 & 12 & 45 & 52.9 & 84.2 & 11.19 & 10,596\\
\quad China & 1.87 & 1 & 1 & 1 & 3 & 41 & 76.5 & 94.6 & 4.51 & 11,052\\
\quad RoW & 2.98 & 1 & 1 & 3 & 7 & 35 & 55.7 & 86.9 & 7.18 & 11,004\\
\quad All & 4.05 & 1 & 1 & 4 & 11 & 45 & 50.9 & 86.6 & 10.64 & 46,531\\
\bottomrule
\end{tabular}

\caption*{\textit{Notes:} A plant (model version) serves a market if it sells 100 units or more there in the year. Region of the plant (of the version's largest origin). \% top: share of the plant's (version's) sales in its largest market.}
\end{table}

\begin{table}[tp]
\centering
\caption{Documented Floods at the Plants, 2000--2024} \label{tab:floods_plants}
\footnotesize\setlength{\tabcolsep}{4pt}
\begin{tabular}{lcccccc}
\toprule
Plant & Plants & \makecell[b]{Flooded\\plants} & \makecell[b]{Flood\\plant-years} & \makecell[b]{Direct\\hits} & \makecell[b]{\% units,\\flooded} & \makecell[b]{\% units,\\direct hits}\\
\midrule
Assembly plant & 1,196 & 132 & 138 & 80 & 1.08 & 0.70\\
Engine plant & 443 & 47 & 50 & 29 & 0.78 & 0.51\\
Transmission plant & 350 & 30 & 32 & 16 & 0.58 & 0.30\\
\bottomrule
\end{tabular}

\caption*{\textit{Notes:} Plants: plants with positive output in 2000--2024. Flood plant-years: documented floods (direct hits and stoppages). \% units: share of all vehicles in 2000--2024 assembled at, or fitted with a component from, a plant in a year with a documented flood (direct hit) there.}
\end{table}

\begin{table}[tp]
\centering
\caption{Sources after a Documented Flood, 2000--2024} \label{tab:floods_choosers}
\footnotesize\setlength{\tabcolsep}{4pt}
\begin{tabular}{lcccccccc}
\toprule
Floods & Choosers & \makecell[b]{One source,\\$t-1$} & $t-1$ & $t$ & $t+1$ & $t+2$ & $t+3$ & \makecell[b]{Never\\flooded}\\
\midrule
\multicolumn{9}{l}{\emph{Engines: vehicle-engine spec / plants}}\\
\quad Documented floods & 1,041 & 0.97 & 0.03 & 0.02 & 0.01 & 0.01 & 0.01 & 0.007\\
\quad Direct hits & 645 & 0.96 & 0.04 & 0.03 & 0.02 & 0.01 & 0.01 & 0.007\\
\multicolumn{9}{l}{\emph{Engines: engine program / plants (100+ units)}}\\
\quad Documented floods & 216 & 0.48 & 0.52 & 0.54 & 0.57 & 0.57 & 0.59 & 0.172\\
\quad Direct hits & 147 & 0.42 & 0.58 & 0.57 & 0.58 & 0.56 & 0.55 & 0.182\\
\multicolumn{9}{l}{\emph{Transmissions: vehicle-transmission spec / plants}}\\
\quad Documented floods & 754 & 0.98 & 0.02 & 0.03 & 0.03 & 0.03 & 0.03 & 0.022\\
\quad Direct hits & 324 & 0.98 & 0.02 & 0.04 & 0.03 & 0.03 & 0.01 & 0.022\\
\multicolumn{9}{l}{\emph{Transmissions: transmission program / plants (100+ units)}}\\
\quad Documented floods & 121 & 0.53 & 0.47 & 0.48 & 0.46 & 0.48 & 0.45 & 0.190\\
\quad Direct hits & 57 & 0.47 & 0.53 & 0.54 & 0.47 & 0.48 & 0.50 & 0.209\\
\multicolumn{9}{l}{\emph{Model versions: model version / assembly plants (100+ units)}}\\
\quad Documented floods & 421 & 0.58 & 0.42 & 0.44 & 0.46 & 0.49 & 0.50 & 0.217\\
\quad Direct hits & 301 & 0.54 & 0.46 & 0.47 & 0.51 & 0.51 & 0.51 & 0.222\\
\bottomrule
\end{tabular}

\caption*{\textit{Notes:} Choosers whose source plant in $t-1$ has its first documented flood (direct hit) in $t$. One source, $t-1$: share supplied by a single plant in $t-1$. Columns $t-1$ to $t+3$: share supplied by two or more plants, among choosers still made. Never flooded: share of chooser-years with two or more sources among choosers never supplied by a flooded plant. For vehicle specifications a plant is a source if it supplies any units, for programs and models if it makes 100 units or more.}
\end{table}

\clearpage
\nextappendix{Flood Defences} \label{app:def}
This appendix supports the comparison of programs flooded with and without a flood defence nearby (panels C and D of Figure~\ref{fig:ext}). It shows the raw data behind the estimate, explains why the defended series rises after year 6, and reports the comparison within 5 km and under other definitions of protection. The design is that of Section~\ref{sec:ext}, i.e., flood signals, engine and transmission programs, and controls of the same country.

\begin{figure}[tp]
\centering
\caption{Flood Defences: The Flooded Programs and Their Controls} \label{fig:def_raw}
\includegraphics[width=0.85\textwidth]{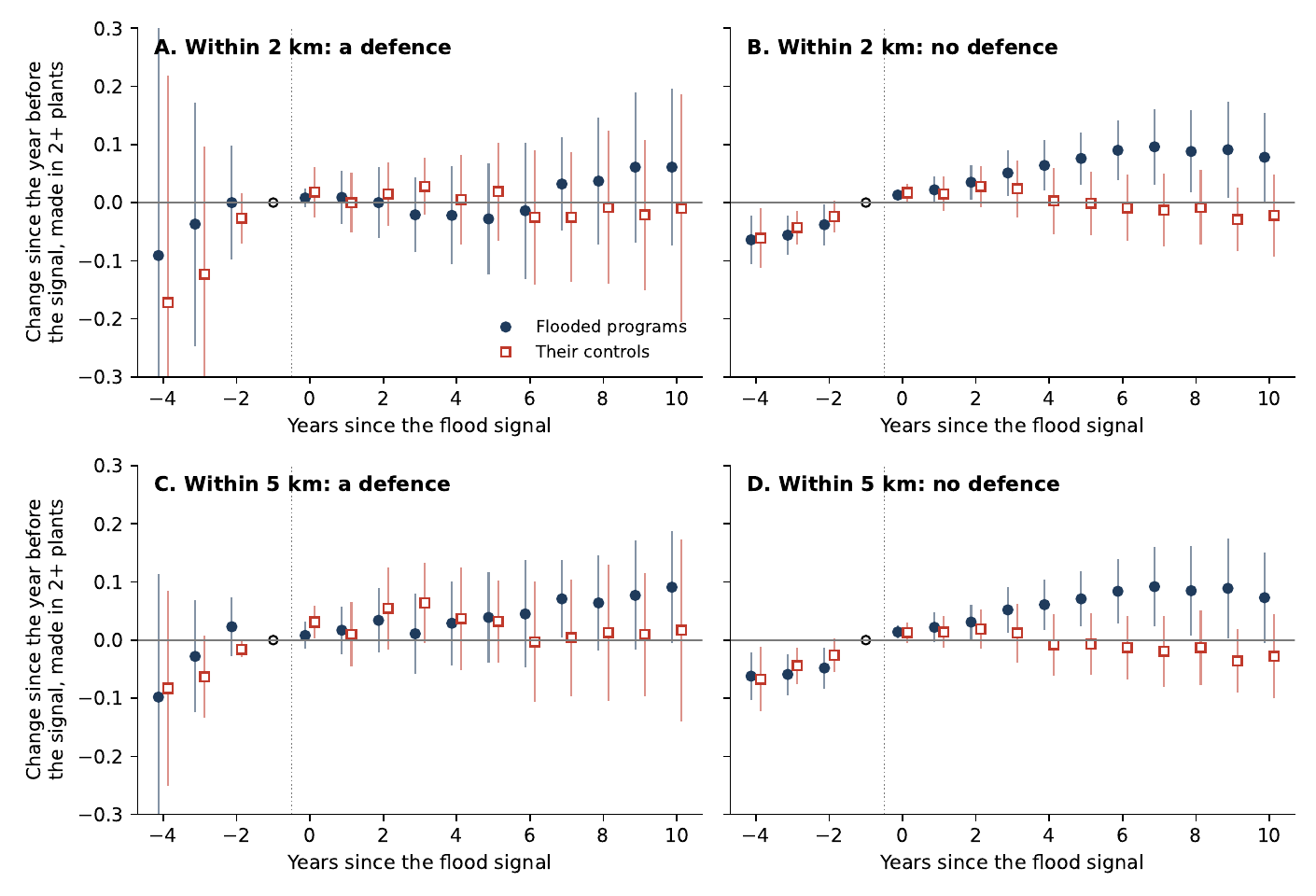}
\caption*{\textit{Notes:} Change in the probability of being made in two or more plants since the year before the flood signal, for the flooded programs and for their controls (the two terms of the equation above, whose difference is the estimate in Figure~\ref{fig:ext} and Figure~\ref{fig:def_5km}). A and B: a flood defence mapped within 2 km of the flooded plant, or none. C and D: within 5 km. Engine and transmission programs, controls of the same country. 95\% intervals, owner bootstrap. }
\end{figure}

\paragraph{{The raw data.}} Each point in panel C of Figure~\ref{fig:ext} is the flooded programs' change in the probability of being made in two or more plants since the year before the signal, minus the average change of their controls,
\begin{equation*}
\hat\theta^{G}_{q}=\frac{1}{N^{G}_{q}}\sum_{i\in G}\big(Y_{i,F_i+q}-Y_{i,F_i-1}\big)-\frac{1}{N^{G}_{q}}\sum_{i\in G}\frac{1}{|\mathcal C_i(q)|}\sum_{j\in\mathcal C_i(q)}\big(Y_{j,F_i+q}-Y_{j,F_i-1}\big).
\end{equation*}
Figure~\ref{fig:def_raw} plots the two terms separately. Within 2 km (panels A and B), the controls of both groups barely move after the signal and stay within 0.03 of zero. The gap therefore comes from the flooded programs. Those flooded at a plant with a defence are not made in more plants in the first six years, as their change lies between $-0.03$ and $+0.01$. Those flooded without one add plants steadily, from 0.035 in year 2 to 0.090 in year 6. The estimate does not rest on what the controls do.

\paragraph{{The late rise.}} After year 6, the programs flooded near a defence add some plants, as their change is between 0.03 and 0.06 in years 7--10, and the defended series in panel C of Figure~\ref{fig:ext} turns positive. Four features of the data account for this rise. First, it rests on few programs, from 63 in year 7 to 33 in year 10, so that each point is imprecise. Second, programs that stop being made drop out of the outcome, so that the late years rest on the programs that survive, which are more often made in several plants. When ended programs count as not made in two or more plants, as in panels B and C of Figure~\ref{fig:engines}, the defended series is $+0.025$ (s.e.\ 0.033) in years 5--10. Third, 81\% of the defended programs receive another flood signal at their plants within ten years, which may bring a late response. Fourth, the defences are not dated. For every defended plant, the defences within 2 km were first recorded in OpenStreetMap in or after the year of the signal, a median of 15 years later. Some of these plants may thus have been protected only after the flood, in which case they respond like unprotected plants. This misclassification biases the gap towards zero.

\begin{figure}[tp]
\centering
\caption{Flood Defences within 5 km} \label{fig:def_5km}
\includegraphics[width=0.85\textwidth]{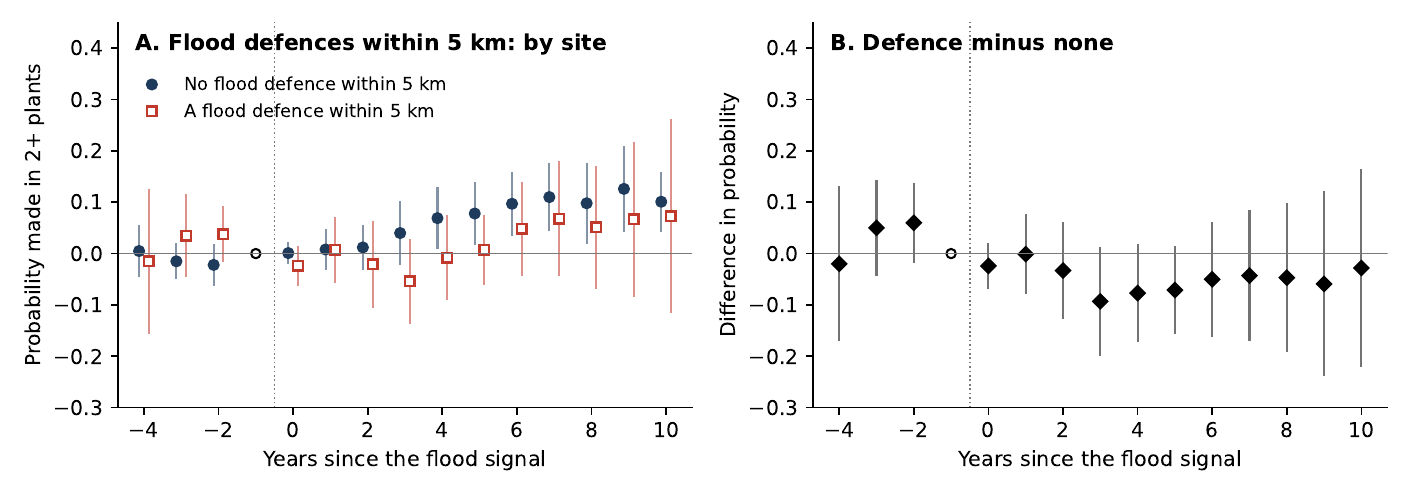}
\caption*{\textit{Notes:} As panels C and D of Figure~\ref{fig:ext}, with a flood defence mapped within 5 km of the flooded plant. A: engine and transmission programs made in two or more plants, by whether a defence is mapped within 5 km. B: their difference. 95\% intervals, owner bootstrap. Years $-4$ to 10. }
\end{figure}

\paragraph{{Within 5 km.}} Figure~\ref{fig:def_5km} shows the comparison within 5 km, as in earlier versions of this paper. The difference is $-0.046$ (s.e.\ 0.038) in years 0--4 and $-0.049$ (s.e.\ 0.060) in years 5--10. Within 5 km, the defended series dips below zero in years 2--4 and rises afterwards, and panels C and D of Figure~\ref{fig:def_raw} show why. The controls of the defended programs add plants in years 2 to 5, by 0.055 and 0.064 in years 2 and 3, and then stop, while the flooded programs add plants steadily. The rise of these controls is not an artefact of OpenStreetMap, which only decides which flooded programs, and hence which countries and launch cohorts, enter the comparison. Nor is it the same flood spreading to other programs of the country, since with controls from other countries of the same World Bank region the controls of the defended programs rise as much, by 0.055 in year 2 and 0.068 in year 3. It reflects what comparable programs did in the countries and years of the defended plants. A radius of 5 km also mixes plants protected by a defence with plants whose defence lies too far away to protect them. Plants whose nearest defence is 5--10 km away respond the most (0.197, s.e.\ 0.094, in years 5--10), more than plants with no defence within 25 km (0.060, s.e.\ 0.035).

\paragraph{{Other definitions.}} Table~\ref{tab:def_alt} reports the gap under other definitions. With controls from other countries of the same region, the gap within 2 km is larger ($-0.081$, s.e.\ 0.057, in years 5--10). Wider radii shrink it or reverse its sign, and so does the official river-flood protection standard of the plant's area (FLOPROS), as programs flooded where that standard is 100 years or more respond more than the others. One reading is that protection standards are highest in rich countries, where owners also have the plants abroad that make duplication cheap (Section~\ref{sec:cheap}), so that this comparison mixes protection with the cost of a second plant. The evidence on defences is therefore suggestive.

\begin{table}[tp]
\centering
\footnotesize
\caption{The Gap between Defended and Undefended Plants under Other Definitions} \label{tab:def_alt}
\begin{tabular}{lcc}
\toprule
Made in two or more plants, defended minus undefended & Years 0--4 & Years 5--10\\
\midrule
Within 2 km (Figure~\ref{fig:ext}) & $-0.038$ (0.030) & $-0.064$ (0.069)\\
Within 2 km, ended programs coded 0 & $-0.024$ (0.026) & $-0.026$ (0.038)\\
Within 2 km, controls from other countries of the region & $-0.062$ (0.040) & $-0.081$ (0.057)\\
Within 5 km (Figure~\ref{fig:def_5km}) & $-0.046$ (0.038) & $-0.049$ (0.060)\\
Within 5 km, controls from other countries of the region & $-0.026$ (0.034) & $-0.024$ (0.049)\\
Within 10 km & $-0.019$ (0.032) & $+0.013$ (0.058)\\
Within 25 km & $+0.008$ (0.029) & $+0.051$ (0.046)\\
Within 5 km, against no defence within 25 km & $-0.030$ (0.040) & $-0.008$ (0.059)\\
FLOPROS standard of 100 years or more, against less & $-0.008$ (0.032) & $+0.075$ (0.058)\\
\bottomrule
\end{tabular}
\caption*{\textit{Notes:} Flood signals, engine and transmission programs, controls of the same country unless stated. Defended: a dyke, levee, flood wall or seawall mapped in OpenStreetMap within the radius of the flooded plant. Owner-bootstrap standard errors in parentheses. Differences use the same draws. }
\end{table}

\clearpage
\nextappendix{Plants, Lines and Products} \label{app:plants}
Section~\ref{sec:engines} follows the flooded product. This appendix follows the plants. We distinguish three levels for each kind of plant: the product (engine program, transmission program, model version), the line, i.e., a product family at a plant with all the family's products made there (engine family line, transmission family line, platform assembly line), and the plant with all its products. A line is active in a year with 1,000 units or more, and a plant makes a product with 100 units or more. The output of engine and transmission plants is the number of units fitted in the light vehicles of the data. The owner of a plant is its manufacturing group. Standard errors come from a bootstrap over owners, with 499 draws.

\paragraph{{The flooded plant.}} The treatment is the first documented flood at a plant that was active the year before. Controls are plants not yet flooded, active the year before, in other countries and in the same cell of plant cohort (plants active in 2000--2001, otherwise the first active year in three-year bins) times the number of lines active the year before (1, 2, 3, 4 or more). The estimator is that of Section~\ref{sec:design}, with outcomes relative to the plant's output the year before the flood (Table~\ref{tab:plants_flooded}). For output over the demand index, the plant's output is divided by a shift-share index of what its markets bought: for engine and transmission plants, the vehicle output of the assembly countries the plant supplied the year before, weighted by their shares and net of the plant's own units. For assembly plants, the car sales of the destination countries of its cars, from S\&P's sales by plant and destination.

\paragraph{{The receiving plant.}} Each plant that starts a product after a flood, in the first such year $y_0$, is compared with plants that start only products that were never flooded in the same year, with the same number of active lines in $y_0-1$, in other countries. Plants with their own documented flood, or with another product added after a flood, within six years of $y_0$ are left out, and so are all the products that start at the plant in $y_0$ (Table~\ref{tab:plants_receiving}).

\paragraph{{Other estimators.}} Table~\ref{tab:plants_estimators} reports the change in plant output in the first years after the flood with the estimators of \citet{CH2024}, \citet{CS2021} and \citet{BJS2024}. These estimators cannot match on the number of lines in $F-1$, which changes over time, nor restrict the controls to other countries. With the same control rule, our estimator gives exactly the effects of \citet{CH2024}'s estimator.

\begin{table}[tp]
\centering
\caption{The Flooded Plant} \label{tab:plants_flooded}
\footnotesize\setlength{\tabcolsep}{4pt}
\begin{tabular}{>{\raggedright\arraybackslash\hangindent=1em\hangafter=1}p{4.6cm}cccccc}
\toprule
 & \multicolumn{2}{c}{Engine plants} & \multicolumn{2}{c}{Transmission plants} & \multicolumn{2}{c}{Assembly plants}\\
 & Years 0--4 & Years 5--10 & Years 0--4 & Years 5--10 & Years 0--4 & Years 5--10\\
\midrule
Plant output, flood year & \multicolumn{2}{c}{\makecell[t]{$-0.112$\\(0.061)}} & \multicolumn{2}{c}{\makecell[t]{$-0.066$\\(0.077)}} & \multicolumn{2}{c}{\makecell[t]{$-0.022$\\(0.049)}}\\
Plant output, year after the flood & \multicolumn{2}{c}{\makecell[t]{$+0.066$\\(0.070)}} & \multicolumn{2}{c}{\makecell[t]{$-0.014$\\(0.094)}} & \multicolumn{2}{c}{\makecell[t]{$+0.107$\\(0.074)}}\\
\addlinespace
Plant output & \makecell[t]{$+0.061$\\(0.061)} & \makecell[t]{$+0.123$\\(0.123)} & \makecell[t]{$-0.079$\\(0.095)} & \makecell[t]{$+0.068$\\(0.171)} & \makecell[t]{$+0.051$\\(0.068)} & \makecell[t]{$+0.095$\\(0.087)}\\
Output of the lines active in $F-1$ & \makecell[t]{$-0.042$\\(0.067)} & \makecell[t]{$-0.086$\\(0.113)} & \makecell[t]{$-0.063$\\(0.096)} & \makecell[t]{$+0.172$\\(0.168)} & \makecell[t]{$+0.074$\\(0.062)} & \makecell[t]{$+0.076$\\(0.096)}\\
Output of other lines & \makecell[t]{$+0.092$\\(0.059)} & \makecell[t]{$+0.191$\\(0.115)} & \makecell[t]{$-0.054$\\(0.070)} & \makecell[t]{$-0.172$\\(0.107)} & \makecell[t]{$-0.038$\\(0.046)} & \makecell[t]{$-0.003$\\(0.117)}\\
Lines of $F-1$ still active, share & \makecell[t]{$-0.012$\\(0.043)} & \makecell[t]{$-0.031$\\(0.092)} & \makecell[t]{$+0.064$\\(0.056)} & \makecell[t]{$+0.036$\\(0.083)} & \makecell[t]{$+0.044$\\(0.028)} & \makecell[t]{$+0.032$\\(0.047)}\\
New lines active, number & \makecell[t]{$+0.178$\\(0.136)} & \makecell[t]{$+0.229$\\(0.203)} & \makecell[t]{$+0.079$\\(0.157)} & \makecell[t]{$-0.057$\\(0.289)} & \makecell[t]{$-0.007$\\(0.080)} & \makecell[t]{$-0.002$\\(0.146)}\\
\addlinespace
Output of the products of $F-1$ & \makecell[t]{$+0.005$\\(0.061)} & \makecell[t]{$+0.028$\\(0.114)} & \makecell[t]{$-0.056$\\(0.078)} & \makecell[t]{$+0.147$\\(0.137)} & \makecell[t]{$+0.004$\\(0.066)} & \makecell[t]{$-0.037$\\(0.070)}\\
Products of $F-1$ still made there, share & \makecell[t]{$+0.029$\\(0.035)} & \makecell[t]{$+0.030$\\(0.062)} & \makecell[t]{$+0.045$\\(0.048)} & \makecell[t]{$+0.008$\\(0.074)} & \makecell[t]{$-0.016$\\(0.031)} & \makecell[t]{$-0.037$\\(0.037)}\\
Products of $F-1$ moved away, share & \makecell[t]{$+0.024$\\(0.030)} & \makecell[t]{$+0.107$\\(0.054)} & \makecell[t]{$+0.012$\\(0.048)} & \makecell[t]{$-0.000$\\(0.063)} & \makecell[t]{$+0.008$\\(0.016)} & \makecell[t]{$+0.020$\\(0.034)}\\
Products of $F-1$ ended, share & \makecell[t]{$-0.052$\\(0.042)} & \makecell[t]{$-0.136$\\(0.055)} & \makecell[t]{$-0.056$\\(0.035)} & \makecell[t]{$-0.008$\\(0.065)} & \makecell[t]{$+0.009$\\(0.031)} & \makecell[t]{$+0.018$\\(0.049)}\\
\addlinespace
Plant output over the demand index & \makecell[t]{$-0.048$\\(0.092)} & \makecell[t]{$-0.017$\\(0.134)} & \makecell[t]{$-0.166$\\(0.097)} & \makecell[t]{$-0.152$\\(0.102)} & \makecell[t]{$-0.039$\\(0.061)} & \makecell[t]{$-0.032$\\(0.083)}\\
\midrule
Flooded plants with controls, year 0; year 5 [owners] & 43 & 31 [13] & 22 & 16 [12] & 112 & 67 [20]\\
\bottomrule
\end{tabular}

\caption*{\textit{Notes:} Plants with a first documented flood in $F$, relative to their output in $F-1$, each ratio capped at 3, so that the outputs of the lines of $F-1$ and of other lines need not add up to plant output. Controls: plants not yet flooded, active in $F-1$, in other countries, of the same cohort and with the same number of active lines in $F-1$. Lines: engine family lines, transmission family lines and platform assembly lines. Products: engine programs, transmission programs and model versions. ``Moved away'': no longer made at the plant but made at another plant. ``Ended'': made nowhere. Shares of lines and products are shares of those of $F-1$. Output of transmission plants already rises faster than at the controls before the flood. Owner bootstrap, 499 draws. }
\end{table}

\begin{table}[tp]
\centering
\caption{The Receiving Plant} \label{tab:plants_receiving}
\scriptsize\setlength{\tabcolsep}{4pt}
\begin{tabular}{>{\raggedright\arraybackslash\hangindent=1em\hangafter=1}p{4.0cm}ccccccccc}
\toprule
 & \multicolumn{3}{c}{Engine plants} & \multicolumn{3}{c}{Transmission plants} & \multicolumn{3}{c}{Assembly plants}\\
 & $-4$ to $-2$ & 0--2 & 3--6 & $-4$ to $-2$ & 0--2 & 3--6 & $-4$ to $-2$ & 0--2 & 3--6\\
\midrule
Plant output, new products left out & \makecell[t]{$+0.126$\\(0.143)} & \makecell[t]{$+0.021$\\(0.125)} & \makecell[t]{$+0.055$\\(0.180)} & \makecell[t]{$+0.078$\\(0.138)} & \makecell[t]{$-0.404$\\(0.174)} & \makecell[t]{$-0.355$\\(0.243)} & \makecell[t]{$+0.052$\\(0.105)} & \makecell[t]{$+0.031$\\(0.121)} & \makecell[t]{$+0.205$\\(0.160)}\\
Receiving lines, new products left out & \makecell[t]{$+0.210$\\(0.124)} & \makecell[t]{$-0.179$\\(0.264)} & \makecell[t]{$-0.124$\\(0.446)} & \makecell[t]{$+0.237$\\(0.202)} & \makecell[t]{$-0.421$\\(0.398)} & \makecell[t]{$+0.153$\\(0.524)} & \makecell[t]{$+0.187$\\(0.183)} & \makecell[t]{$-0.146$\\(0.204)} & \makecell[t]{$+0.095$\\(0.215)}\\
The plant's other lines & \makecell[t]{$+0.124$\\(0.190)} & \makecell[t]{$-0.006$\\(0.175)} & \makecell[t]{$+0.002$\\(0.241)} & \makecell[t]{$-0.068$\\(0.189)} & \makecell[t]{$-0.225$\\(0.218)} & \makecell[t]{$-0.306$\\(0.267)} & \makecell[t]{$-0.005$\\(0.122)} & \makecell[t]{$+0.129$\\(0.148)} & \makecell[t]{$+0.307$\\(0.186)}\\
Plant output over the demand index & \makecell[t]{$+0.080$\\(0.130)} & \makecell[t]{$+0.005$\\(0.138)} & \makecell[t]{$+0.029$\\(0.205)} & \makecell[t]{$+0.094$\\(0.170)} & \makecell[t]{$-0.329$\\(0.170)} & \makecell[t]{$-0.319$\\(0.237)} & \makecell[t]{$+0.106$\\(0.113)} & \makecell[t]{$+0.057$\\(0.146)} & \makecell[t]{$+0.267$\\(0.171)}\\
\addlinespace
Products of $y_0-1$ still made there, share & n.a. & \makecell[t]{$+0.092$\\(0.041)} & \makecell[t]{$+0.197$\\(0.078)} & n.a. & \makecell[t]{$-0.065$\\(0.060)} & \makecell[t]{$-0.096$\\(0.110)} & n.a. & \makecell[t]{$-0.030$\\(0.042)} & \makecell[t]{$-0.018$\\(0.059)}\\
Products of $y_0-1$ moved away, share & n.a. & \makecell[t]{$-0.004$\\(0.032)} & \makecell[t]{$+0.050$\\(0.053)} & n.a. & \makecell[t]{$+0.031$\\(0.034)} & \makecell[t]{$+0.101$\\(0.052)} & n.a. & \makecell[t]{$-0.013$\\(0.032)} & \makecell[t]{$-0.011$\\(0.038)}\\
Products of $y_0-1$ ended, share & n.a. & \makecell[t]{$-0.087$\\(0.035)} & \makecell[t]{$-0.247$\\(0.059)} & n.a. & \makecell[t]{$+0.034$\\(0.061)} & \makecell[t]{$-0.005$\\(0.095)} & n.a. & \makecell[t]{$+0.043$\\(0.032)} & \makecell[t]{$+0.029$\\(0.062)}\\
\midrule
Receiving plants (owners); ordinary plant-years & \multicolumn{3}{c}{38 (13); 527} & \multicolumn{3}{c}{21 (11); 308} & \multicolumn{3}{c}{51 (26); 1,133}\\
\bottomrule
\end{tabular}

\caption*{\textit{Notes:} Plants that start a product after a flood in $y_0$, against plants that start only never-flooded products in the same year, with the same number of active lines in $y_0-1$, in other countries. Outcomes relative to $y_0-1$, the products that start in $y_0$ left out. Receiving lines: the families of the new products at the plant, if active in $y_0-1$. Shares of products are shares of those made at the plant in $y_0-1$, reported from $y_0$ on (n.a. before). Owner bootstrap, 499 draws. }
\end{table}

\begin{table}[tp]
\centering
\caption{The Flooded Plant's Output with Other Estimators} \label{tab:plants_estimators}
\footnotesize\setlength{\tabcolsep}{4pt}
\begin{tabular}{>{\raggedright\arraybackslash\hangindent=1em\hangafter=1}p{7.6cm}ccc}
\toprule
 & \makecell[b]{Flood\\year} & Year 1 & Year 2\\
\midrule
\multicolumn{4}{l}{\emph{Engine plants}}\\
\quad Matched & \makecell[t]{$-0.112$\\(0.061)} & \makecell[t]{$+0.066$\\(0.070)} & \makecell[t]{$+0.106$\\(0.082)}\\
\quad de Chaisemartin--D'Haultf\oe uille, no cells & \makecell[t]{$-0.157$\\(0.059)} & \makecell[t]{$-0.033$\\(0.095)} & \makecell[t]{$+0.061$\\(0.048)}\\
\quad de Chaisemartin--D'Haultf\oe uille, cohort & \makecell[t]{$-0.126$\\(0.061)} & \makecell[t]{$+0.002$\\(0.094)} & \makecell[t]{$+0.089$\\(0.050)}\\
\quad Callaway--Sant'Anna & \makecell[t]{$-0.157$\\(0.083)} & \makecell[t]{$-0.033$\\(0.114)} & \makecell[t]{$+0.061$\\(0.065)}\\
\quad Borusyak--Jaravel--Spiess & \makecell[t]{$-0.114$\\(0.101)} & \makecell[t]{$+0.016$\\(0.127)} & \makecell[t]{$+0.170$\\(0.089)}\\
\addlinespace
\multicolumn{4}{l}{\emph{Transmission plants}}\\
\quad Matched & \makecell[t]{$-0.066$\\(0.077)} & \makecell[t]{$-0.014$\\(0.094)} & \makecell[t]{$-0.141$\\(0.117)}\\
\quad de Chaisemartin--D'Haultf\oe uille, no cells & \makecell[t]{$-0.082$\\(0.035)} & \makecell[t]{$+0.011$\\(0.032)} & \makecell[t]{$-0.053$\\(0.083)}\\
\quad de Chaisemartin--D'Haultf\oe uille, cohort & \makecell[t]{$-0.032$\\(0.038)} & \makecell[t]{$+0.062$\\(0.042)} & \makecell[t]{$+0.028$\\(0.099)}\\
\quad Callaway--Sant'Anna & \makecell[t]{$-0.082$\\(0.050)} & \makecell[t]{$+0.011$\\(0.036)} & \makecell[t]{$-0.053$\\(0.071)}\\
\quad Borusyak--Jaravel--Spiess & \makecell[t]{$-0.083$\\(0.072)} & \makecell[t]{$+0.005$\\(0.087)} & \makecell[t]{$+0.022$\\(0.101)}\\
\addlinespace
\multicolumn{4}{l}{\emph{Assembly plants}}\\
\quad Matched & \makecell[t]{$-0.022$\\(0.049)} & \makecell[t]{$+0.107$\\(0.074)} & \makecell[t]{$+0.088$\\(0.098)}\\
\quad de Chaisemartin--D'Haultf\oe uille, no cells & \makecell[t]{$-0.064$\\(0.037)} & \makecell[t]{$+0.061$\\(0.041)} & \makecell[t]{$+0.053$\\(0.049)}\\
\quad de Chaisemartin--D'Haultf\oe uille, cohort & \makecell[t]{$-0.039$\\(0.038)} & \makecell[t]{$+0.094$\\(0.046)} & \makecell[t]{$+0.087$\\(0.054)}\\
\quad Callaway--Sant'Anna & \makecell[t]{$-0.064$\\(0.045)} & \makecell[t]{$+0.061$\\(0.059)} & \makecell[t]{$+0.053$\\(0.073)}\\
\quad Borusyak--Jaravel--Spiess & \makecell[t]{$-0.005$\\(0.051)} & \makecell[t]{$+0.196$\\(0.039)} & \makecell[t]{$+0.197$\\(0.037)}\\
\bottomrule
\end{tabular}

\caption*{\textit{Notes:} Effect of the first documented flood on plant output in the flood year and the two years after. Matched: the estimator of Section~\ref{sec:design} with cells of plant cohort times active lines in $F-1$ and controls from other countries, output relative to $F-1$, owner bootstrap. Other rows: log plant output, not-yet-treated controls from any country, errors clustered by owner. ``No cells'' is \citet{CH2024}'s estimator without options, ``cohort'' with the plant cohort as \texttt{trends\_nonparam}. \citet{CH2024}'s package also reports a joint placebo test over five pre-flood years: $p$ is 0.10 without cells and 0.01 with the cohort for engine plants, 0.80 and 0.18 for transmission plants, and 0.02 and 0.31 for assembly plants. }
\end{table}

\clearpage
\nextappendix{Recipient Plants: Full Estimates} \label{app:recv}

\begin{table}[!htp]
\centering
\caption{The Receiving Plant against the Owner's Other Plants} \label{tab:recv2}
\scriptsize\setlength{\tabcolsep}{4pt}
\begin{tabular}{>{\raggedright\arraybackslash\hangindent=1em\hangafter=1}p{4.0cm}ccccccccc}
\toprule
 & \multicolumn{3}{c}{Engine programs} & \multicolumn{3}{c}{Transmission programs} & \multicolumn{3}{c}{Vehicle models}\\
\cmidrule(lr){2-4}\cmidrule(lr){5-7}\cmidrule(lr){8-10}
 & All & \makecell[b]{After a\\flood} & Ordinary & All & \makecell[b]{After a\\flood} & Ordinary & All & \makecell[b]{After a\\flood} & Ordinary\\
\midrule
Plant slack, lifetime peaks (Fact 3) & \makecell[t]{$+0.001$\\(0.019)} & \makecell[t]{$+0.002$\\(0.042)} & \makecell[t]{$+0.000$\\(0.019)} & \makecell[t]{$+0.022$\\(0.024)} & \makecell[t]{$-0.013$\\(0.069)} & \makecell[t]{$+0.025$\\(0.024)} & \makecell[t]{$+0.063$\\(0.011)} & \makecell[t]{$+0.085$\\(0.034)} & \makecell[t]{$+0.062$\\(0.012)}\\
Plant slack, peaks to date & \makecell[t]{$-0.035$\\(0.015)} & \makecell[t]{$-0.040$\\(0.052)} & \makecell[t]{$-0.035$\\(0.015)} & \makecell[t]{$-0.063$\\(0.016)} & \makecell[t]{$-0.034$\\(0.102)} & \makecell[t]{$-0.066$\\(0.016)} & \makecell[t]{$+0.004$\\(0.012)} & \makecell[t]{$-0.010$\\(0.039)} & \makecell[t]{$+0.005$\\(0.011)}\\
Slack of the platform's line, lifetime peak (plants with it) & \makecell[t]{$+0.018$\\(0.017)} & \makecell[t]{$+0.022$\\(0.066)} & \makecell[t]{$+0.017$\\(0.018)} & \makecell[t]{$+0.046$\\(0.045)} & \makecell[t]{$-0.119$\\(0.177)} & \makecell[t]{$+0.063$\\(0.044)} & \makecell[t]{$+0.108$\\(0.014)} & \makecell[t]{$+0.152$\\(0.064)} & \makecell[t]{$+0.104$\\(0.017)}\\
Slack of the platform's line, peak to date (plants with it) & \makecell[t]{$-0.074$\\(0.022)} & \makecell[t]{$-0.057$\\(0.053)} & \makecell[t]{$-0.076$\\(0.024)} & \makecell[t]{$-0.093$\\(0.023)} & \makecell[t]{$-0.211$\\(0.120)} & \makecell[t]{$-0.081$\\(0.024)} & \makecell[t]{$-0.028$\\(0.009)} & \makecell[t]{$+0.011$\\(0.032)} & \makecell[t]{$-0.031$\\(0.009)}\\
Output growth, $b-3$ to $b$ & \makecell[t]{$+0.074$\\(0.031)} & \makecell[t]{$+0.117$\\(0.101)} & \makecell[t]{$+0.070$\\(0.033)} & \makecell[t]{$+0.182$\\(0.057)} & \makecell[t]{$-0.184$\\(0.210)} & \makecell[t]{$+0.218$\\(0.053)} & \makecell[t]{$+0.166$\\(0.061)} & \makecell[t]{$+0.159$\\(0.139)} & \makecell[t]{$+0.166$\\(0.061)}\\
Output growth, $b$ to $b+3$, without the product & \makecell[t]{$+0.115$\\(0.052)} & \makecell[t]{$+0.083$\\(0.148)} & \makecell[t]{$+0.118$\\(0.054)} & \makecell[t]{$+0.252$\\(0.101)} & \makecell[t]{$-0.215$\\(0.187)} & \makecell[t]{$+0.299$\\(0.103)} & \makecell[t]{$+0.019$\\(0.033)} & \makecell[t]{$+0.194$\\(0.112)} & \makecell[t]{$+0.007$\\(0.036)}\\
Platform line growth, $b$ to $b+3$, without the product (plants with it) & \makecell[t]{$+0.296$\\(0.084)} & \makecell[t]{$+0.277$\\(0.177)} & \makecell[t]{$+0.298$\\(0.093)} & \makecell[t]{$+0.204$\\(0.154)} & \makecell[t]{$-0.338$\\(0.072)} & \makecell[t]{$+0.257$\\(0.155)} & \makecell[t]{$+0.036$\\(0.033)} & \makecell[t]{$+0.397$\\(0.283)} & \makecell[t]{$+0.013$\\(0.035)}\\
Makes the product's platform & \makecell[t]{$+0.439$\\(0.065)} & \makecell[t]{$+0.360$\\(0.131)} & \makecell[t]{$+0.448$\\(0.067)} & \makecell[t]{$+0.428$\\(0.048)} & \makecell[t]{$+0.359$\\(0.096)} & \makecell[t]{$+0.434$\\(0.051)} & \makecell[t]{$+0.326$\\(0.021)} & \makecell[t]{$+0.303$\\(0.055)} & \makecell[t]{$+0.328$\\(0.020)}\\
Makes its type, another platform & \makecell[t]{$-0.155$\\(0.027)} & \makecell[t]{$-0.190$\\(0.074)} & \makecell[t]{$-0.152$\\(0.030)} & \makecell[t]{$+0.018$\\(0.024)} & \makecell[t]{$+0.109$\\(0.090)} & \makecell[t]{$+0.010$\\(0.026)} & \makecell[t]{$+0.073$\\(0.017)} & \makecell[t]{$+0.003$\\(0.036)} & \makecell[t]{$+0.078$\\(0.017)}\\
Plant of the next stage on the site & \makecell[t]{$+0.020$\\(0.026)} & \makecell[t]{$+0.175$\\(0.088)} & \makecell[t]{$+0.003$\\(0.030)} & \makecell[t]{$-0.062$\\(0.056)} & \makecell[t]{$+0.018$\\(0.040)} & \makecell[t]{$-0.070$\\(0.062)} & \makecell[t]{$+0.013$\\(0.032)} & \makecell[t]{$+0.063$\\(0.116)} & \makecell[t]{$+0.009$\\(0.029)}\\
Residual flood hazard 1\%+ a year & \makecell[t]{$+0.025$\\(0.020)} & \makecell[t]{$+0.062$\\(0.040)} & \makecell[t]{$+0.021$\\(0.021)} & \makecell[t]{$-0.033$\\(0.021)} & \makecell[t]{$-0.065$\\(0.114)} & \makecell[t]{$-0.029$\\(0.022)} & \makecell[t]{$-0.012$\\(0.023)} & \makecell[t]{$+0.016$\\(0.069)} & \makecell[t]{$-0.014$\\(0.023)}\\
Country where the product is not made & \makecell[t]{$-0.049$\\(0.031)} & \makecell[t]{$-0.047$\\(0.048)} & \makecell[t]{$-0.049$\\(0.034)} & \makecell[t]{$-0.043$\\(0.034)} & \makecell[t]{$-0.016$\\(0.087)} & \makecell[t]{$-0.046$\\(0.035)} & \makecell[t]{$+0.030$\\(0.042)} & \makecell[t]{$+0.072$\\(0.103)} & \makecell[t]{$+0.027$\\(0.039)}\\
Km to the product's markets & \makecell[t]{$-998$\\(342)} & \makecell[t]{$-109$\\(707)} & \makecell[t]{$-1,094$\\(363)} & \makecell[t]{$-766$\\(284)} & \makecell[t]{$-432$\\(763)} & \makecell[t]{$-799$\\(312)} & \makecell[t]{$-250$\\(403)} & \makecell[t]{$+257$\\(816)} & \makecell[t]{$-285$\\(397)}\\
Share of the market in the plant's country & \makecell[t]{$+0.049$\\(0.017)} & \makecell[t]{$+0.026$\\(0.042)} & \makecell[t]{$+0.051$\\(0.019)} & \makecell[t]{$+0.050$\\(0.021)} & \makecell[t]{$-0.022$\\(0.082)} & \makecell[t]{$+0.057$\\(0.022)} & \makecell[t]{$-0.003$\\(0.025)} & \makecell[t]{$+0.024$\\(0.073)} & \makecell[t]{$-0.005$\\(0.023)}\\
Share in contiguous countries & \makecell[t]{$+0.025$\\(0.019)} & \makecell[t]{$-0.006$\\(0.018)} & \makecell[t]{$+0.029$\\(0.020)} & \makecell[t]{$+0.026$\\(0.023)} & \makecell[t]{$+0.018$\\(0.044)} & \makecell[t]{$+0.027$\\(0.025)} & \makecell[t]{$+0.022$\\(0.013)} & \makecell[t]{$+0.013$\\(0.015)} & \makecell[t]{$+0.022$\\(0.014)}\\
Share in countries with a common official language & \makecell[t]{$+0.003$\\(0.016)} & \makecell[t]{$+0.009$\\(0.028)} & \makecell[t]{$+0.003$\\(0.017)} & \makecell[t]{$-0.012$\\(0.006)} & \makecell[t]{$+0.006$\\(0.043)} & \makecell[t]{$-0.014$\\(0.007)} & \makecell[t]{$+0.012$\\(0.008)} & \makecell[t]{$-0.009$\\(0.019)} & \makecell[t]{$+0.014$\\(0.008)}\\
Share in countries with a trade agreement & \makecell[t]{$+0.048$\\(0.024)} & \makecell[t]{$+0.118$\\(0.045)} & \makecell[t]{$+0.040$\\(0.027)} & \makecell[t]{$+0.042$\\(0.034)} & \makecell[t]{$-0.006$\\(0.090)} & \makecell[t]{$+0.047$\\(0.037)} & \makecell[t]{$-0.035$\\(0.027)} & \makecell[t]{$-0.134$\\(0.046)} & \makecell[t]{$-0.028$\\(0.029)}\\
Tariff to the product's markets, ln(1+t) & \makecell[t]{$-0.007$\\(0.001)} & \makecell[t]{$-0.007$\\(0.005)} & \makecell[t]{$-0.007$\\(0.001)} & \makecell[t]{$-0.005$\\(0.002)} & \makecell[t]{$-0.003$\\(0.003)} & \makecell[t]{$-0.005$\\(0.002)} & \makecell[t]{$-0.011$\\(0.004)} & \makecell[t]{$-0.011$\\(0.008)} & \makecell[t]{$-0.011$\\(0.005)}\\
Log output & \makecell[t]{$+0.352$\\(0.070)} & \makecell[t]{$+0.265$\\(0.262)} & \makecell[t]{$+0.361$\\(0.082)} & \makecell[t]{$+0.299$\\(0.086)} & \makecell[t]{$-0.034$\\(0.511)} & \makecell[t]{$+0.332$\\(0.094)} & \makecell[t]{$-0.452$\\(0.107)} & \makecell[t]{$-0.930$\\(0.345)} & \makecell[t]{$-0.420$\\(0.112)}\\
Lines & 633 & 62 & 571 & 324 & 29 & 295 & 1,188 & 75 & 1,113\\
\bottomrule
\end{tabular}

\caption*{\textit{Notes:} The receiving plant minus the mean of the other plants of its choice set (the owner's plants of the same kind that make its products the year before the addition, $b$, but not this product), mean over additions, owner-bootstrap standard errors in parentheses. Owner: design parent for engine and transmission programs, manufacturer group for model versions. Slack and markets as in Section~\ref{sec:receiving}, with peaks over 2000--2024 (lifetime) or over 2000--$b$ (to date). Growth: symmetric rate, $(x_1-x_0)$ over their mean, and after the addition without the product's own units. Type: fuel times cylinder layout for engines, design times drive for transmissions, and segment times size class for vehicles. Next stage on the site: within 2 km, an assembly plant for engines and transmissions, an engine plant for vehicles. Residual flood hazard: net of public protection. Tariff: applied, preferential where one applies, 0 at home. Additions in 2004--2024, plants with every trait of Table~\ref{tab:clogit} measured. }
\end{table}

\begin{table}[tp]
\centering
\caption{Which of the Owner's Plants Gets the Program: Conditional Logit} \label{tab:clogit}
\scriptsize\setlength{\tabcolsep}{4pt}
\begin{tabular}{>{\raggedright\arraybackslash\hangindent=1em\hangafter=1}p{4.0cm}ccccccccc}
\toprule
 & \multicolumn{3}{c}{Engine programs} & \multicolumn{3}{c}{Transmission programs} & \multicolumn{3}{c}{Vehicle models}\\
\cmidrule(lr){2-4}\cmidrule(lr){5-7}\cmidrule(lr){8-10}
 & All & \makecell[b]{After a\\flood} & Ordinary & All & \makecell[b]{After a\\flood} & Ordinary & All & \makecell[b]{After a\\flood} & Ordinary\\
\midrule
Plant slack, lifetime peaks (Fact 3), per 10 points & \makecell[t]{$-0.008$\\(0.035)} & \makecell[t]{$-0.020$\\(0.154)} & \makecell[t]{$-0.008$\\(0.036)} & \makecell[t]{$+0.059$\\(0.056)} & \makecell[t]{$-0.072$\\(0.129)} & \makecell[t]{$+0.073$\\(0.057)} & \makecell[t]{$+0.018$\\(0.020)} & \makecell[t]{$+0.038$\\(0.103)} & \makecell[t]{$+0.016$\\(0.021)}\\
Slack of the platform's line, lifetime peak (Fact 3), per 10 points & \makecell[t]{$+0.014$\\(0.020)} & \makecell[t]{$+0.006$\\(0.093)} & \makecell[t]{$+0.014$\\(0.023)} & \makecell[t]{$+0.037$\\(0.029)} & \makecell[t]{$-0.283$\\(0.117)} & \makecell[t]{$+0.054$\\(0.030)} & \makecell[t]{$+0.091$\\(0.014)} & \makecell[t]{$+0.070$\\(0.075)} & \makecell[t]{$+0.093$\\(0.017)}\\
Makes the product's platform & \makecell[t]{$+3.353$\\(0.455)} & \makecell[t]{$+2.849$\\(1.560)} & \makecell[t]{$+3.432$\\(0.472)} & \makecell[t]{$+3.050$\\(0.370)} & \makecell[t]{$+7.011$\\(1.749)} & \makecell[t]{$+2.843$\\(0.375)} & \makecell[t]{$+2.095$\\(0.188)} & \makecell[t]{$+2.124$\\(0.617)} & \makecell[t]{$+2.102$\\(0.198)}\\
Makes its type, another platform & \makecell[t]{$+1.169$\\(0.247)} & \makecell[t]{$+0.197$\\(0.986)} & \makecell[t]{$+1.275$\\(0.290)} & \makecell[t]{$+1.515$\\(0.271)} & \makecell[t]{$+2.834$\\(1.156)} & \makecell[t]{$+1.436$\\(0.296)} & \makecell[t]{$+1.672$\\(0.174)} & \makecell[t]{$+1.150$\\(0.386)} & \makecell[t]{$+1.717$\\(0.180)}\\
Output growth over the three years before, per 10 points & \makecell[t]{$+0.016$\\(0.005)} & \makecell[t]{$+0.034$\\(0.026)} & \makecell[t]{$+0.015$\\(0.006)} & \makecell[t]{$+0.026$\\(0.011)} & \makecell[t]{$-0.076$\\(0.031)} & \makecell[t]{$+0.032$\\(0.011)} & \makecell[t]{$+0.025$\\(0.007)} & \makecell[t]{$+0.018$\\(0.026)} & \makecell[t]{$+0.026$\\(0.007)}\\
Plant of the next stage on the site & \makecell[t]{$+0.034$\\(0.112)} & \makecell[t]{$+0.442$\\(0.457)} & \makecell[t]{$-0.029$\\(0.119)} & \makecell[t]{$-0.267$\\(0.244)} & \makecell[t]{$-0.507$\\(0.562)} & \makecell[t]{$-0.284$\\(0.256)} & \makecell[t]{$+0.137$\\(0.185)} & \makecell[t]{$+0.493$\\(0.771)} & \makecell[t]{$+0.115$\\(0.175)}\\
Residual flood hazard 1\%+ a year & \makecell[t]{$+0.186$\\(0.095)} & \makecell[t]{$+0.330$\\(0.501)} & \makecell[t]{$+0.174$\\(0.121)} & \makecell[t]{$-0.687$\\(0.208)} & \makecell[t]{$-1.594$\\(0.514)} & \makecell[t]{$-0.638$\\(0.253)} & \makecell[t]{$-0.156$\\(0.151)} & \makecell[t]{$+0.206$\\(0.504)} & \makecell[t]{$-0.179$\\(0.157)}\\
Country where the product is not made & \makecell[t]{$-0.163$\\(0.233)} & \makecell[t]{$-0.638$\\(0.577)} & \makecell[t]{$-0.085$\\(0.261)} & \makecell[t]{$-0.200$\\(0.300)} & \makecell[t]{$-0.777$\\(1.456)} & \makecell[t]{$-0.140$\\(0.290)} & \makecell[t]{$+0.234$\\(0.267)} & \makecell[t]{$+1.381$\\(1.381)} & \makecell[t]{$+0.147$\\(0.248)}\\
Log km to the product's markets & \makecell[t]{$-0.176$\\(0.051)} & \makecell[t]{$-0.093$\\(0.234)} & \makecell[t]{$-0.191$\\(0.053)} & \makecell[t]{$-0.104$\\(0.108)} & \makecell[t]{$+0.023$\\(0.263)} & \makecell[t]{$-0.109$\\(0.123)} & \makecell[t]{$-0.147$\\(0.065)} & \makecell[t]{$+0.163$\\(0.368)} & \makecell[t]{$-0.157$\\(0.065)}\\
Market in the plant's country, per 10 points & \makecell[t]{$-0.055$\\(0.028)} & \makecell[t]{$+0.069$\\(0.112)} & \makecell[t]{$-0.069$\\(0.031)} & \makecell[t]{$-0.061$\\(0.048)} & \makecell[t]{$-0.289$\\(0.156)} & \makecell[t]{$-0.042$\\(0.050)} & \makecell[t]{$-0.014$\\(0.033)} & \makecell[t]{$+0.268$\\(0.248)} & \makecell[t]{$-0.030$\\(0.030)}\\
Market in contiguous countries, per 10 points & \makecell[t]{$+0.002$\\(0.020)} & \makecell[t]{$-0.102$\\(0.121)} & \makecell[t]{$+0.014$\\(0.017)} & \makecell[t]{$+0.059$\\(0.048)} & \makecell[t]{$+0.021$\\(0.449)} & \makecell[t]{$+0.052$\\(0.053)} & \makecell[t]{$+0.062$\\(0.026)} & \makecell[t]{$+0.198$\\(0.141)} & \makecell[t]{$+0.053$\\(0.026)}\\
Market in countries with a common language, per 10 points & \makecell[t]{$-0.006$\\(0.035)} & \makecell[t]{$+0.114$\\(0.296)} & \makecell[t]{$-0.016$\\(0.038)} & \makecell[t]{$-0.045$\\(0.055)} & \makecell[t]{$+0.037$\\(0.364)} & \makecell[t]{$-0.060$\\(0.061)} & \makecell[t]{$+0.045$\\(0.022)} & \makecell[t]{$+0.008$\\(0.156)} & \makecell[t]{$+0.045$\\(0.022)}\\
Market in countries with a trade agreement, per 10 points & \makecell[t]{$+0.004$\\(0.023)} & \makecell[t]{$+0.092$\\(0.054)} & \makecell[t]{$-0.011$\\(0.025)} & \makecell[t]{$-0.001$\\(0.030)} & \makecell[t]{$-0.072$\\(0.105)} & \makecell[t]{$+0.012$\\(0.032)} & \makecell[t]{$-0.075$\\(0.023)} & \makecell[t]{$-0.144$\\(0.084)} & \makecell[t]{$-0.068$\\(0.025)}\\
Tariff to the product's markets, ln(1+t), per 10 points & \makecell[t]{$-1.341$\\(0.269)} & \makecell[t]{$-0.768$\\(0.902)} & \makecell[t]{$-1.462$\\(0.264)} & \makecell[t]{$-1.715$\\(0.564)} & \makecell[t]{$-6.697$\\(3.722)} & \makecell[t]{$-1.498$\\(0.560)} & \makecell[t]{$-0.312$\\(0.089)} & \makecell[t]{$-0.409$\\(0.317)} & \makecell[t]{$-0.318$\\(0.091)}\\
Log output & \makecell[t]{$+0.056$\\(0.043)} & \makecell[t]{$+0.027$\\(0.187)} & \makecell[t]{$+0.059$\\(0.048)} & \makecell[t]{$+0.097$\\(0.082)} & \makecell[t]{$-0.144$\\(0.265)} & \makecell[t]{$+0.143$\\(0.084)} & \makecell[t]{$-0.182$\\(0.032)} & \makecell[t]{$-0.352$\\(0.111)} & \makecell[t]{$-0.173$\\(0.033)}\\
\midrule
Pseudo $R^2$ & 0.244 & 0.217 & 0.253 & 0.292 & 0.414 & 0.296 & 0.173 & 0.228 & 0.173\\
Hit rate (random pick) & \makecell[t]{0.33\\(0.09)} & \makecell[t]{0.39\\(0.08)} & \makecell[t]{0.33\\(0.09)} & \makecell[t]{0.42\\(0.11)} & \makecell[t]{0.52\\(0.12)} & \makecell[t]{0.44\\(0.11)} & \makecell[t]{0.20\\(0.06)} & \makecell[t]{0.23\\(0.05)} & \makecell[t]{0.21\\(0.06)}\\
Lines & 633 & 62 & 571 & 324 & 29 & 295 & 1,188 & 75 & 1,113\\
Plants compared & 12,059 & 1,144 & 10,915 & 3,919 & 292 & 3,627 & 38,899 & 2,274 & 36,625\\
Owners & 23 & 10 & 23 & 23 & 5 & 23 & 43 & 12 & 42\\
Draws that converge (of 499) & 499 & 468 & 499 & 499 & 364 & 499 & 499 & 489 & 499\\
\bottomrule
\end{tabular}

\caption*{\textit{Notes:} Conditional logit \citep{mcfadden1974} over the plants of the addition's choice set, the year before the addition (Table~\ref{tab:recv2}). Coefficients, owner-bootstrap standard errors in parentheses (499 draws, draws that do not converge are left out). Slack, growth and market shares per 10 points. Platform-line slack is 0 at plants without the platform. Km: market-weighted mean of log(1 + km). Hit rate: share of additions whose receiving plant has the highest fitted probability, against a random pick in parentheses. For all lines, the pseudo $R^2$ is 0.228, 0.273 and 0.162 (engine programs, transmission programs, vehicle models) with plant traits only (capacity, platform, type, growth, site, hazard, country and size), and 0.034, 0.031 and 0.027 with gravity only (the market terms and size). }
\end{table}

\clearpage
\nextappendix{Estimates behind the Figures} \label{app:est}

\begin{table}[!htp]
\centering
\footnotesize
\caption{Estimates behind the Figures} \label{tab:est}
\setlength{\tabcolsep}{4pt}
\begin{tabularx}{\textwidth}{@{}Lcccc@{}}
\toprule
 & \makecell[b]{Years\\$-4$ to $-2$} & \makecell[b]{Years\\0--4} & \makecell[b]{Years\\5--10} & Products\\
\midrule
\multicolumn{5}{@{}l}{\emph{Documented floods, controls from other countries}}\\
Engine programs & $+0.024$ (0.024) & $+0.085$ (0.027) & $+0.194$ (0.061) & 203, 99\\
\quad plants of the year before still making it (share) & $+0.011$ (0.013) & $+0.022$ (0.032) & $+0.036$ (0.055) & 216, 160\\
\quad output at those plants & $+0.031$ (0.075) & $+0.052$ (0.046) & $+0.087$ (0.066) & 201, 153\\
\quad made only at added plants & $-0.001$ (0.003) & $-0.006$ (0.002) & $-0.013$ (0.010) & 203, 99\\
\quad made in one plant, a plant abroad makes the family & $+0.000$ (0.036) & $+0.146$ (0.069) & $+0.359$ (0.105) & 56, 20\\
\quad made in one plant, no plant abroad does & $+0.019$ (0.025) & $-0.018$ (0.007) & $-0.030$ (0.018) & 39, 10\\
Model versions & $-0.008$ (0.012) & $+0.041$ (0.019) & $-0.057$ (0.061) & 376, 94\\
\quad years 3--4 & n.a. & $+0.096$ (0.028) & n.a. & 160\\
\quad made in one plant, a plant abroad builds the platform & $-0.005$ (0.026) & $+0.044$ (0.059) & $-0.026$ (0.031) & 122, 24\\
\quad made in one plant, no plant abroad does & $-0.040$ (0.021) & $+0.010$ (0.041) & $+0.045$ (0.092) & 90, 23\\
Transmission programs & $+0.050$ (0.023) & $+0.029$ (0.031) & $+0.026$ (0.066) & 111, 65\\
\quad manual gearboxes & $+0.022$ (0.029) & $-0.002$ (0.068) & $+0.112$ (0.094) & 48, 32\\
\quad automatic, CVT and dual-clutch & $+0.073$ (0.065) & $+0.021$ (0.046) & $-0.076$ (0.067) & 41, 26\\
\midrule
\multicolumn{5}{@{}l}{\emph{Documented floods, controls from other countries shipped the same way (Figure~\ref{fig:stock})}}\\
Engine programs, fitted nearby & $+0.050$ (0.037) & $+0.088$ (0.042) & $+0.336$ (0.081) & 102, 52\\
\quad shipped & $-0.003$ (0.049) & $+0.071$ (0.057) & $+0.088$ (0.109) & 95, 44\\
Transmission programs, fitted nearby & $+0.052$ (0.052) & $+0.010$ (0.042) & $+0.181$ (0.149) & 40, 26\\
\quad shipped & $+0.057$ (0.022) & $+0.039$ (0.055) & $+0.006$ (0.090) & 63, 34\\
\quad manual gearboxes, fitted nearby & $-0.025$ (0.019) & $-0.025$ (0.071) & $+0.333$ (0.159) & 16, 12\\
\quad manual gearboxes, shipped & $+0.045$ (0.039) & $+0.053$ (0.115) & $+0.136$ (0.145) & 29, 18\\
\quad automatic, CVT and dual-clutch, fitted nearby & $-0.020$ (0.196) & $+0.044$ (0.147) & $-0.036$ (0.020) & 14, 3\\
\quad automatic, CVT and dual-clutch, shipped & $+0.089$ (0.075) & $+0.025$ (0.078) & $-0.151$ (0.057) & 25, 13\\
The two groups averaged: engine programs & $+0.024$ (0.028) & $+0.079$ (0.033) & $+0.215$ (0.064) & 197, 96\\
\quad transmission programs & $+0.055$ (0.027) & $+0.027$ (0.039) & $+0.077$ (0.094) & 103, 60\\
\quad transmission programs, the engines' shares & $+0.055$ (0.031) & $+0.024$ (0.037) & $+0.096$ (0.101) & 103, 60\\
\midrule
\multicolumn{5}{@{}l}{\emph{Flood signals, controls of the same country: engine and transmission programs}}\\
No flood defence within 2 km & $-0.010$ (0.016) & $+0.020$ (0.019) & $+0.100$ (0.029) & 1086, 537\\
A flood defence within 2 km & $+0.065$ (0.072) & $-0.018$ (0.023) & $+0.037$ (0.061) & 124, 72\\
\midrule
\multicolumn{5}{@{}l}{\emph{Flood signals, controls of the same region: engine programs, regression of program effects}}\\
High hazard $\times$ insurer concentration, per ten points ($b_3$) & $-0.011$ (0.011) & $+0.014$ (0.004) & $+0.046$ (0.017) & 925, 502\\
\bottomrule
\end{tabularx}
\caption*{\textit{Notes:} Outcome: made in two or more plants unless stated. Owner-bootstrap standard errors in parentheses, except the last row (clustered by country). Products: flooded products with controls in year 0 and in year 5 (years 3--4: in year 3; the two groups averaged: both groups; last row: programs in the regressions). The row of years 3--4 is a window within years 0--4, so the other windows are n.a. Model versions: brand $\times$ nameplate $\times$ S\&P vehicle program. }
\end{table}

\begin{table}[tp]
\centering
\scriptsize
\caption{The Design: What Is Matched, and Alternatives} \label{tab:design}
\setlength{\tabcolsep}{3pt}
\begin{tabularx}{\textwidth}{@{}Lcccccc@{}}
\toprule
 & \multicolumn{3}{c}{Engine programs} & \multicolumn{3}{c}{Model versions}\\
\cmidrule(lr){2-4}\cmidrule(lr){5-7}
 & \multicolumn{2}{c}{Made in 2+ plants} & Plants & \multicolumn{2}{c}{Made in 2+ plants} & Plants\\
 & \makecell[b]{Years\\$-4$ to $-2$} & \makecell[b]{Years\\5--10} & \makecell[b]{Years\\$-4$ to $-2$} & \makecell[b]{Years\\$-4$ to $-2$} & \makecell[b]{Years\\3--4} & \makecell[b]{Years\\$-4$ to $-2$}\\
\midrule
\multicolumn{7}{@{}l}{\emph{(A) Owner bootstrap}}\\
Main: cohort $\times$ plants in $F-1$, other countries & $+0.022$ (0.024) & $+0.194$ (0.061) & $-0.013$ (0.063) & $-0.009$ (0.012) & $+0.096$ (0.028) & $-0.095$ (0.058)\\
Main $\times$ export markets in $F-1$ & $+0.015$ (0.030) & $+0.242$ (0.084) & $-0.002$ (0.087) & $-0.006$ (0.017) & $+0.076$ (0.036) & $-0.026$ (0.055)\\
\multicolumn{7}{@{}l}{\emph{(B) Stacked regression, errors clustered by owner}}\\
Main: cohort $\times$ plants in $F-1$, other countries & $+0.022$ (0.015) & $+0.194$ (0.027) & $-0.013$ (0.042) & $-0.009$ (0.007) & $+0.096$ (0.014) & $-0.095$ (0.030)\\
Controls of any country & $+0.022$ (0.014) & $+0.180$ (0.029) & $-0.009$ (0.041) & $-0.008$ (0.007) & $+0.098$ (0.014) & $-0.093$ (0.030)\\
Controls of other owners only & $+0.025$ (0.014) & $+0.208$ (0.030) & $-0.010$ (0.043) & $-0.006$ (0.007) & $+0.095$ (0.014) & $-0.091$ (0.030)\\
Controls of the same owner only & $-0.056$ (0.017) & $+0.171$ (0.127) & $-0.091$ (0.058) & $-0.055$ (0.015) & $+0.099$ (0.047) & $-0.064$ (0.067)\\
Exact launch year $\times$ plants in $F-1$ & $+0.023$ (0.018) & $+0.210$ (0.044) & $+0.026$ (0.045) & $-0.008$ (0.008) & $+0.094$ (0.017) & $-0.027$ (0.039)\\
Main $\times$ export markets in $F-1$ & $+0.015$ (0.020) & $+0.242$ (0.035) & $-0.002$ (0.049) & $-0.006$ (0.012) & $+0.076$ (0.021) & $-0.026$ (0.037)\\
Exact launch year only (age) & $-0.022$ (0.015) & $+0.026$ (0.033) & $-0.168$ (0.030) & $-0.041$ (0.008) & $+0.017$ (0.026) & $-0.293$ (0.040)\\
Launch cohort only, any country & $-0.023$ (0.012) & $-0.000$ (0.039) & $-0.159$ (0.026) & $-0.046$ (0.007) & $+0.016$ (0.023) & $-0.312$ (0.040)\\
No cells, any country & $-0.017$ (0.011) & $-0.050$ (0.038) & $-0.146$ (0.025) & $-0.050$ (0.008) & $+0.006$ (0.023) & $-0.321$ (0.039)\\
Two-way fixed effects (product, year) & $-0.002$ (0.019) & $-0.053$ (0.069) & $-0.114$ (0.063) & $-0.027$ (0.015) & $-0.014$ (0.040) & $-0.193$ (0.074)\\
\bottomrule
\end{tabularx}
\caption*{\textit{Notes:} $\hat\theta_q$ of equation~\eqref{eq:theta}, averaged over the years shown, under alternative control sets. Main: launch cohort (three-year bins) times plants in $F-1$ (1, 2, 3, 4+), controls with their largest plant in another country whose plants of $F-1$ stay dry through $F+q$. Export markets in $F-1$: the cell also matches the number of export markets served with 100 units or more (0, 1--2, 3--5, 6--10, 11+, no sales). For engine programs, a market is an assembly country supplied with 100 units or more, an export market one other than the plant's. Controls of the same (other) owner: the main cell, with controls of the product's own owner (of other owners) only. Exact launch year: the first year of output instead of the cohort. Launch cohort only and no cells: any not-yet-treated product of any country, as in \texttt{did\_multiplegt\_dyn} with and without \texttt{trends\_nonparam} on the cohort. Two-way fixed effects: event study of the level on event-time dummies (binned at $-5$ and $10$) with product and year effects, no matching. Plants: number of plants making the product. New assembly (destination) countries: countries entered since $F-1$ over the countries of $F-1$. Stacked regression: long differences of flooded products and their controls on event-time dummies, a fixed effect per flooded product and horizon, controls weighted by one over the cell size. Its point estimates equal those of panels A and C for the same control set. Flooded products with controls in year 0: 203 engine programs of 14 owners and 376 model versions of 28 owners (same owner only: 116 of 10 and 273 of 16). Years $-4$ to $-2$ leave out the years before a product's first output; Table~\ref{tab:est} and the figures count them as not made in two or more plants, which gives $+0.024$ and $-0.008$ for the main rows. }
\end{table}

\begin{table}[tp]
\ContinuedFloat
\centering
\scriptsize
\caption{The Design: What Is Matched, and Alternatives (continued)}
\setlength{\tabcolsep}{3pt}
\begin{tabularx}{\textwidth}{@{}Lcccc@{}}
\toprule
 & \multicolumn{2}{c}{Engine programs} & \multicolumn{2}{c}{Model versions}\\
\cmidrule(lr){2-3}\cmidrule(lr){4-5}
 & \multicolumn{2}{c}{New assembly countries supplied} & \multicolumn{2}{c}{New destination countries}\\
 & \makecell[b]{Years\\$-4$ to $-2$} & \makecell[b]{Years\\3--4} & \makecell[b]{Years\\$-4$ to $-2$} & \makecell[b]{Years\\0--2}\\
\midrule
\multicolumn{5}{@{}l}{\emph{(C) Owner bootstrap}}\\
Main: cohort $\times$ plants in $F-1$, other countries & $-0.033$ (0.051) & $+0.214$ (0.076) & $-0.090$ (0.097) & $-0.164$ (0.056)\\
Main $\times$ export markets in $F-1$ & $-0.025$ (0.058) & $+0.216$ (0.115) & $-0.081$ (0.120) & $-0.401$ (0.198)\\
\multicolumn{5}{@{}l}{\emph{(D) Stacked regression, errors clustered by owner}}\\
Main: cohort $\times$ plants in $F-1$, other countries & $-0.033$ (0.037) & $+0.214$ (0.042) & $-0.090$ (0.066) & $-0.164$ (0.033)\\
Main $\times$ export markets in $F-1$ & $-0.025$ (0.035) & $+0.216$ (0.057) & $-0.081$ (0.081) & $-0.401$ (0.152)\\
Controls of any country & $-0.028$ (0.036) & $+0.207$ (0.042) & $-0.084$ (0.063) & $-0.155$ (0.036)\\
Controls of other owners only & $-0.035$ (0.037) & $+0.215$ (0.041) & $-0.089$ (0.065) & $-0.166$ (0.034)\\
Controls of the same owner only & $-0.056$ (0.047) & $+0.311$ (0.149) & $-0.055$ (0.117) & $-0.093$ (0.085)\\
Exact launch year $\times$ plants in $F-1$ & $-0.014$ (0.036) & $+0.215$ (0.044) & $-0.056$ (0.049) & $-0.153$ (0.039)\\
Exact launch year only (age) & $-0.050$ (0.032) & $+0.179$ (0.039) & $-0.325$ (0.082) & $-0.119$ (0.027)\\
Launch cohort only, any country & $-0.033$ (0.033) & $+0.210$ (0.042) & $-0.292$ (0.076) & $-0.153$ (0.030)\\
No cells, any country & $-0.016$ (0.028) & $+0.144$ (0.054) & $-0.350$ (0.085) & $-0.215$ (0.033)\\
\bottomrule
\end{tabularx}
\caption*{\textit{Notes:} As in the first part of the table. New assembly (destination) countries: countries entered since $F-1$ over the countries of $F-1$. }
\end{table}

\clearpage
\nextappendix{A Stricter Definition of Direct Hits} \label{app:strict}
In the record, a direct hit is a closure of the plant because of a flood or storm at the plant or in its own area, whether or not water entered the plant (Section~\ref{sec:floods}). As a robustness check, we restrict direct hits to floods that put water in the plant. We went back to every documented flood that other public reports describe differently, 111 plant-years of 78 events, and classified each plant-year by what happened at the plant: water entered it; it stayed dry and stopped because a flood cut off its parts, logistics or access; it closed as a precaution; the event was not a flood, mostly the earthquake of March 2011; the flood hit another site of the firm, such as an R\&D centre, a vehicle yard or a former plant, or another plant; or no source documents an event at the plant in that year. The stricter definition keeps as direct hits the floods that put water in the plant and the direct hits that no other source questions, counts every other closure of a dry plant as a stoppage, and drops the events that were not floods at the plant. It drops about a fifth of the flood plant-years and turns about two thirds of the direct hits into stoppages or drops them. Flooded engine programs fall from 216 to 185, and those whose first documented flood is a direct hit from 137 to 50, of seven owners. Because the stricter definition shrinks the sample substantially, above all the direct hits, we keep it as a robustness check in this appendix and do not pursue it for the main results.

\begin{table}[tp]
\centering
\footnotesize
\caption{Engine Programs under the Stricter Definition of Direct Hits} \label{tab:strict}
\setlength{\tabcolsep}{4pt}
\begin{tabularx}{\textwidth}{@{}Lcccc@{}}
\toprule
Made in two or more plants & Years $-4$ to $-2$ & Years 0--4 & Years 5--10 & Programs [owners]\\
\midrule
\multicolumn{5}{@{}l}{\emph{All flooded programs}}\\
Main-text definition (Figure~\ref{fig:engines}) & $+0.024$ (0.024) & $+0.085$ (0.027) & $+0.194$ (0.061) & 203; 99 [11]\\
Stricter definition & $+0.028$ (0.023) & $+0.090$ (0.029) & $+0.171$ (0.084) & 177; 85 [9]\\
\midrule
\multicolumn{5}{@{}l}{\emph{By the first documented flood, main-text definition}}\\
Direct hit & $+0.033$ (0.025) & $+0.054$ (0.037) & $+0.214$ (0.087) & 128; 61 [10]\\
Stoppage, plant dry & $+0.008$ (0.047) & $+0.141$ (0.061) & $+0.177$ (0.096) & 75; 38 [4]\\
Stoppage minus direct hit & $-0.026$ (0.047) & $+0.087$ (0.075) & $-0.037$ (0.140) & 203; 99 [11]\\
\midrule
\multicolumn{5}{@{}l}{\emph{By the first documented flood, stricter definition}}\\
Direct hit (water in the plant) & $+0.001$ (0.017) & $+0.024$ (0.044) & $+0.257$ (0.102) & 49; 27 [6]\\
Stoppage, plant dry & $+0.040$ (0.045) & $+0.128$ (0.036) & $+0.154$ (0.105) & 128; 58 [6]\\
Stoppage minus direct hit & $+0.039$ (0.048) & $+0.104$ (0.045) & $-0.103$ (0.147) & 177; 85 [9]\\
\bottomrule
\end{tabularx}
\caption*{\textit{Notes:} $\hat\theta_q$ of equation~\eqref{eq:theta}, averaged over the years shown, with controls from other countries and cells of launch cohort times plants in $F-1$. Programs: flooded programs with controls in year 0 and in year 5 (differences: both groups). A program's first documented flood is a direct hit if a plant that made the program in $F-1$ has a direct hit in $F$, and a stoppage otherwise. Owners in year 5 in brackets. Owner-bootstrap standard errors in parentheses (499 draws); differences use the same draws. }
\end{table}

The main estimate is similar under the stricter definition, 0.171 (s.e.\ 0.084) in years 5--10 against 0.194 (s.e.\ 0.061), with a wider interval, as nine owners remain in year 5 instead of eleven. Dropping the precautionary closures as well gives 0.175 (s.e.\ 0.083). Both kinds of flood are still followed by a second plant. In years 5--10, flooded programs are made in two or more plants more often than their controls by 0.257 (s.e.\ 0.102) after a flood that put water in the plant and by 0.154 (s.e.\ 0.105) after a stoppage with the plant dry, and the two cannot be told apart. After a stoppage, the second plant comes sooner, by 0.104 (s.e.\ 0.045) in years 0--4. Only the plant-years that other sources describe differently were re-checked, and the remaining ones keep their classification.

\end{document}